\documentclass[opre,nonblindrev]{informs3_hide} %

\OneAndAHalfSpacedXI %

\usepackage{endnotes}
\let\footnote=\endnote
\let\enotesize=\normalsize
\def\notesname{Endnotes}%
\def\makeenmark{$^{\theenmark}$}
\def\enoteformat{\rightskip0pt\leftskip0pt\parindent=1.75em
  \leavevmode\llap{\theenmark.\enskip}}

\usepackage{amsmath,amsfonts,amssymb}
\usepackage{mathtools}
\usepackage{bm}
\usepackage[mathscr]{euscript}
\usepackage{enumitem}
\usepackage{xcolor}
\usepackage{graphicx}
\usepackage{multirow}
\usepackage{booktabs}
\usepackage{microtype}
\usepackage{fix-cm}
\usepackage[ruled,vlined,linesnumbered]{algorithm2e}
\usepackage{physics}
\usepackage{xfrac}
\usepackage{nicematrix}

\DeclareMathOperator{\pr}{\mathbb{P}}
\DeclareMathOperator{\E}{\mathbb{E}}
\DeclareMathOperator{\Var}{\mathrm{Var}}
\DeclareMathOperator{\Cov}{\mathrm{Cov}}

\DeclareMathOperator{\MSE}{\mathrm{MSE}}
\DeclareMathOperator{\nnz}{\mathrm{nnz}}
\DeclareMathOperator{\ind}{\mathbb{I}}
\DeclareMathOperator{\supp}{\mathrm{supp}}

\newcommand{\Real}{\mathbb R}

\newcommand{\NatInt}{\mathbb N}
\newcommand{\SFc}{\mathsf c}
\newcommand{\SFp}{\mathsf p}
\newcommand{\SFq}{\mathsf q}

\newcommand{\SFz}{\mathsf z}
\newcommand{\SFB}{\mathsf B}
\newcommand{\SFM}{\mathsf M}
\newcommand{\SFY}{\mathsf Y}

\newcommand{\CalA}{\mathcal A}
\newcommand{\CalD}{\mathcal D}

\newcommand{\CalL}{\mathcal L}

\newcommand{\CalO}{\mathcal O}
\newcommand{\CalP}{\mathcal P}
\newcommand{\CalR}{\mathcal R}
\newcommand{\CalS}{\mathcal S}

\newcommand{\CalX}{\mathcal X}

\newcommand{\ScrF}{\mathscr F}
\newcommand{\ScrG}{\mathscr G}
\newcommand{\ScrH}{\mathscr H}
\newcommand{\ScrI}{\mathscr I}

\newcommand{\ScrX}{\mathscr X}

\usepackage{natbib}
 \bibpunct[, ]{(}{)}{,}{a}{}{,}%
 \def\bibfont{\small}%
 \def\BIBand{and}%

\TheoremsNumberedThrough     %
\ECRepeatTheorems

\EquationsNumberedThrough    %

\MANUSCRIPTNO{OPRE-2020-10-697.R4} %

\begin{document}

\RUNAUTHOR{Ding and Zhang}

\RUNTITLE{Sample and Computationally Efficient Stochastic Kriging in High Dimensions}

\TITLE{Sample and Computationally Efficient Stochastic Kriging in High Dimensions}

\ARTICLEAUTHORS{%
\AUTHOR{Liang Ding}
\AFF{School of Data Science, Fudan University, Shanghai 200433, China, \EMAIL{dingl1990@gmail.com}}
\AUTHOR{Xiaowei Zhang}
\AFF{Faculty of Business and Economics, The University of Hong Kong, Pokfulam Road, Hong Kong SAR, \EMAIL{xiaoweiz@hku.hk} }
} %

\ABSTRACT{%
Stochastic kriging has been widely employed for simulation metamodeling to predict the response surface of complex simulation models.
However, its use is limited to cases where the design space is low-dimensional
because, in general, the sample complexity (i.e., the number of design points required for stochastic kriging to produce an accurate prediction) grows exponentially in the dimensionality of the design space.
The large sample size results in both a prohibitive sample cost for running the simulation model and a severe computational challenge due to the need to invert large covariance matrices.
Based on tensor Markov kernels and sparse grid experimental designs, we develop a novel methodology that dramatically alleviates the curse of dimensionality.
We show that the sample complexity of the proposed methodology grows only slightly in the dimensionality,
even under model misspecification.
We also develop fast algorithms that compute stochastic kriging in its exact form without any approximation schemes.
We demonstrate via extensive numerical experiments that our methodology can
handle problems with a design space of more than 10,000 dimensions, improving both prediction accuracy and computational efficiency by orders of magnitude relative to typical alternative methods in practice.
}%

\KEYWORDS{simulation metamodeling; stochastic kriging; Gaussian process; tensor Markov kernel; sparse grid; experimental design; matrix inversion; high-dimensional inputs}
\maketitle

\section{Introduction}

Simulation is a general-purpose decision-making tool that has been widely used in operations research and management science.
It allows one to model the dynamics of a complex system and to assess an arbitrary performance measure of interest.
However, simulation models are often computationally intensive to execute.
It is costly to collect sufficient simulation samples to facilitate informed decisions.
Simulation metamodeling is a methodology developed to address this issue and has attracted substantial research interest from the simulation community; see \cite{Barton15} and \cite{Kleijnen17} for recent reviews.
The basic idea is to postulate a statistical model for the response surface, that is, the mean performance as a (deterministic) function of the design variables.
This statistical model is fast to run and is called the metamodel.
The simulation model is then executed at a set of judiciously selected design points.
After calibration using the simulation samples,
the metamodel is used to predict the value of the response surface, so that no further simulation will be needed.

Kriging is a popular metamodeling technique, as it imposes minimal structural assumptions on the response surface and can provide robust maps of the entire design space.
Originating in geostatistics \citep{Matheron63}, the kriging technique was introduced  into the design and analysis of computer experiments by \cite{sacks1989} to model the response surface of a deterministic simulation model and has been highly effective in this area.
In the present paper, we focus on stochastic kriging \citep{AnkenmanNelsonStaum10}, which generalizes kriging to stochastic simulation that usually involves heteroscedastic simulation noise.

Stochastic kriging assumes that the response surface is a sample path of a Gaussian process (GP).
Given a sampling budget, its prediction accuracy depends critically on the concrete specification of two aspects:
(i) the kernel function (i.e., the covariance function) of the GP
and (ii) the experimental design (i.e., the choice of the design points).
Upon collecting samples following the experimental design, predicting the response surface with the chosen kernel function
reduces to numerical linear algebra.
In particular, it involves inverting the sample covariance matrix.
Due to its nonparametric nature and analytical tractability, stochastic kriging has been widely used in recent years in the simulation field, including
output analysis \citep{BartonNelsonXie14,XieNelsonBarton14} and simulation optimization \citep{sun2014,JalaliNieuwenhuysePicheny17,SalemiSongNelsonStaum19}; see  \cite{gramacy2020surrogates} for a general exposition.

Despite its popularity,
the use of stochastic kriging has mostly been  limited to problems with a low-dimensional design space.
This is mainly due to two issues; one is related to sample complexity, and the other is related to computational complexity.
The former issue is notable in high-dimensional settings, as simulation samples are expensive to acquire and, in general, the sample size required for stochastic kriging to achieve the prescribed level of prediction accuracy grows exponentially in the dimensionality (see Section~\ref{sec:CoD}).
It is known that the curse of dimensionality on the sample complexity can be mitigated if the response surface is ``highly smooth'' \citep{Stone80}.
For example, sample paths of GPs with a Mat\'ern kernel have a higher degree of differentiability as the smoothness parameter of the kernel increases.
Intuitively, this is because a space of smoother functions is smaller, making it easier to identify the true function in the space.
In the extreme case, if the response surface is infinitely differentiable, which is true for sample paths of GPs with a Gaussian kernel, then the sample complexity can be independent of the dimensionality of the design space \citep{VaartZanten11}.
However, in practice response surfaces of interest may not possess such a high level of smoothness \citep{SalemiStaumNelson19}.
Using a kernel with a high degree of smoothness in stochastic kriging when the response surface is in fact less smooth leads to model misspecification.
It may further exacerbate the curse of dimensionality on sample complexity---at least from the perspective of the rate of convergence---if the misspecification is significant \citep{TuoWang20}.

In the present paper, we develop a novel methodology that does not rely on strong assumptions about the smoothness of the response surface to solve the sample complexity issue in high dimensions.
Instead, we impose a certain tensor structure on the response surface using tensor Markov (TM) kernels, which in fact induce non-differentiable sample paths (see Section~\ref{sec:non-diff}).
TM kernels are of product form, constructed by multiplying the kernel functions of Gauss--Markov processes that are one-dimensional and include Brownian motion and the Ornstein--Uhlenbeck process.
\cite{SalemiStaumNelson19} propose a new class of GPs called generalized integrated Brownian fields (GIBFs) that exhibit excellent prediction capability. A simple example of a GIBF is a Brownian field, the kernel of which is the product of Brownian kernels, thus making it a TM kernel.

However,
the convergence rate of stochastic kriging with TM kernels may still suffer from the curse of dimensionality if a proper experimental design is not chosen (see Section~\ref{sec:deter-sim}).
Based on sparse grid (SG) experimental designs \citep{bungartz_griebel_2004}, we propose a new experimental design called \emph{truncated sparse grids} (TSGs).
The joint use of TM kernels and TSGs allow the convergence rate of stochastic kriging to be  only slightly affected by the dimensionality of the design space.

Another manifestation of the curse of dimensionality is the prohibitive computational complexity.
The matrix inversion in stochastic kriging normally incurs a computational cost that is cubic in the sample size, which quickly surpasses the processing capability of typical modern computers as the sample size increases.
There is a vast and fast-growing literature on approximations to reduce the computational complexity of stochastic kriging and Gaussian process regression (GPR), a closely related method in machine learning.
The basic idea is to identify a suitable subset of the samples and then use it to construct a low-rank approximation of the covariance matrix, which is substantially easier to invert.
Such approximations are usually developed for general kernel functions and general experimental designs.
We refer to \citet[Chapter~8]{RasmussenWilliams06} and \citet[Chapter~9]{gramacy2020surrogates} for overviews of the approach and to \cite{GramacyApley15}, \cite{LuRudiBorgonovoRosasco20}, \cite{BurtRasmussenWilk20}, and references therein for recent advances in this regard.

We do not seek to develop new approximation strategies in the present paper.
Instead, we take advantage of properties that are specific to TM kernels and TSG designs to develop fast algorithms for \emph{exact} computation of stochastic kriging.
We compare our approach with those based on matrix approximations in terms of both prediction accuracy and computational efficiency through extensive numerical experiments.

Another important issue related to matrix inversion is the
numerical instability due to limitations of floating point arithmetic and the fact that large covariance matrices tend to be ill-conditioned \citep{HaalandQian11}.
This issue is particularly serious when Gaussian kernels are used \citep{AbabouBagtzoglouWood94}.
One may apply the framework in \cite{HaalandWangMaheshwari18} and \cite{WangHaaland19} to analyze such numerical errors.
We do not pursue this direction in the present paper, although numerical experiments show that covariance matrices induced by TM kernels and TSG designs do not appear to suffer from this issue.

\subsection{Main Contributions}

Our foremost contribution is to provide a novel solution to two fundamental issues that preclude stochastic kriging from being used for high-dimensional simulation metamodeling, that is, high sample complexity and prohibitive computational cost.
The solution is based on two key concepts---TM kernels and TSG designs.
The former appeared in a study of the multidimensional Markov properties of random fields \citep{CarnalWalsh91}, but as far as we know, no prior theoretical analysis exists  on the use of TM kernels in stochastic kriging.
The TSG designs are indeed new, and they allows us to
derive error bounds on predictions of stochastic kriging with TM kernels and to
devise fast algorithms that leverage properties that are specific to TM kernels.

First, we establish an asymptotic upper bound on the mean squared prediction error of stochastic kriging with TM kernels and TSG designs under the premise that the underlying model is correctly specified.
That is, the kernel used in stochastic kriging is identical to the kernel that drives the GP from which the response surface is realized.
The upper bound is uniform in the design variable, and it deteriorates slightly in the dimensionality $d$, with $d$ appearing in the exponent of $\log n$ as opposed to the usual case of appearing in the exponent of $n$, where $n$ is the number of design points.
Thus, the curse of dimensionality on sample complexity is  substantially mitigated.

Second, in light of the fact that the GP sample paths induced by a TM kernel are non-differentiable everywhere, whereas response surfaces in practice are generally smoother, we extend the theoretical analysis to cope with the model misspecification.
That is, one uses a TM kernel in stochastic kriging, but the response surface is actually realized from a GP induced by a tensor product kernel with a higher degree of smoothness.
Surprisingly, the upper rate of convergence under model misspecification is substantially faster than that under correct specification.
Intuitively, this may be attributed to the fact that the function space induced by a smoother tensor product kernel is ``smaller'' than one induced by a TM kernel.
This theoretical result provides the kind of robustness that is reassuring and encouraging to potential users of our methodology.

Third, typical SG designs are not flexible to use because they are not defined directly via the number of design points $n$ but via the so-called level parameter $\tau$.
The existing algorithms in the literature for the computation of kriging with SG designs are very rigid,
requiring that $n$ should be identical to the size of a level-$\tau$ SG design for some $\tau$.
The computational efficiency would break down otherwise.
We lift such restriction by proposing the TSG design and develop algorithms that allow any arbitrary number of design points.
The key driving force is an explicit characterization of the sparse structure of the inverse kernel matrix, permitting
the algorithms to perform fast, exact computation of stochastic kriging with TM kernels without any approximation strategies.

Fourth, the combination of the upper rates of convergence and the fast algorithms implies that our methodology is a viable option for high-dimensional metamodeling.
Through extensive numerical experiments with problems having a design space of dimension as high as 16675,
we demonstrate that our methodology significantly outperforms competing approaches in both prediction accuracy and computational efficiency.

Last, as earlier studies predominantly focused on Mat\'ern kernels and Gaussian kernels, the existing approaches to theoretical analysis used there may not apply to our setting.
We therefore develop a variety of new technical results, particularly an orthogonal expansion of TM kernels and related properties (see the e-companion).
These are of independent interest and may be used to facilitate future research that involves TM kernels and TSG designs.

\subsection{Related Work}

There is a vast body of literature on the convergence rates of kriging and similar methods, such as GPR and kernel ridge regression (KRR) in machine learning.
The assumptions in these three areas are distinctive due to the differences in application context.
In general, kriging focuses on noiseless samples and fixed designs, whereas both GPR and KRR focus on homoscedastic noise and random designs; meanwhile, both kriging and GPR assume that the function to estimate is a GP sample path, whereas KRR assumes that it lies in the reproducing kernel Hilbert space induced by the kernel.
Despite the differences in assumptions and mathematical tools, the theoretical results in the three areas are comparable.
An incomplete list of recent papers include those by \cite{WangTuoWu20} and
\cite{TuoWang20} regarding  kriging;
 \cite{VaartZanten11} and \cite{PatiBhattacharyaCheng15} regarding GPR; and \cite{CaponnettoVito07} and  \cite{TuoWangWu20} regarding KRR.
Convergence rate analysis for stochastic kriging, which assumes heteroscedastic noise, is scarce, although the results for GPR may conceivably be generalized to this setting.
However, most of the existing work in the literature is derived for Mat\'ern kernels or Gaussian kernels.
To our knowledge, the present paper is the first work that establishes error bounds for stochastic kriging with TM kernels.

Another related strand of research relates SG methods.
SG experimental designs were introduced in \cite{Zenger91} to address the curse of dimensionality in standard grid-based numerical methods for partial differential equations: a $d$-dimensional grid consists of $M^d$ points if the discretization scheme in each dimension employs $M$ points.
SGs can reduce the number of points from $\CalO(M^d)$ to $\CalO(M(\log M)^{d-1})$ with a minor sacrifice of accuracy, thereby greatly alleviating the curse of dimensionality.
We refer to \cite{bungartz_griebel_2004} for a survey of the fundamentals of SGs.
Since their inception, SG methods have been applied in a great variety of research fields, including statistics, economics, and financial engineering \citep{GarckeGriebel13}.
Recently, \cite{Plumlee14} used SGs to enhance the computational efficiency of kriging with tensor product kernels.
However, the algorithms developed there do not apply to stochastic kriging.

\smallskip

The remainder of the paper is organized as follows.
Section~\ref{sec:SK} reviews stochastic kriging, including typical choices of kernels and experimental designs in practice, and discusses the curse of dimensionality that comes with such choices.
Section~\ref{sec:TMK} and Section~\ref{sec:SG} introduce TM kernels and TSG designs, respectively, both of which  are the key ingredients of our methodology.
Section~\ref{sec:sample} analyzes the convergence rate of stochastic kriging with TM kernels and TSG designs.
The issue of model misspecification is also discussed.
Section~\ref{sec:computation} develops fast algorithms for implementing stochastic kriging with TM kernels and TSG designs.
Section~\ref{sec:numerical} presents extensive numerical experiments that focus on high-dimensional problems.
Section~\ref{sec:conclusions} concludes with a brief discussion on possible extensions.
All the proofs are collected in the e-companion to this paper.

\section{Stochastic Kriging}\label{sec:SK}

We denote the $d$-dimensional design variable by $\BFx=(x_1,\ldots, x_d)^\intercal$ and the design space by $\ScrX\subseteq \Real^d$.
In (deterministic) kriging, the unknown response surface $y:\ScrX\mapsto \Real$ is modeled as a realized sample path of the GP
\[
\SFY(\cdot) = \mathrm{GP}(\mu(\cdot), k(\cdot, \cdot) ),
\]
where $\mu(\cdot)$ is the mean function and $k(\cdot,\cdot)$ is the covariance function, which is also called the \emph{kernel} of $\SFY$.
In particular, $\mu(\BFx)= \E[\SFY(\BFx)]$ and $k(\BFx,\BFx') = \Cov[\SFY(\BFx), \SFY(\BFx')] $ for all $\BFx,\BFx'\in\ScrX$.

The mean function $\mu(\cdot)$ characterizes the ``trend'' of the response surface.
It is often modeled as $\mu(\BFx)=\BFf^\intercal(\BFx)\BFbeta$,  where $\BFf(\cdot)$ is a vector of known functions, which may encode the modeler's prior knowledge regarding the shape of the response surface, and $\BFbeta$ is a vector of unknown parameters of compatible dimensions that need to be estimated.
Throughout the paper, however, we assume that $\mu(\cdot)\equiv 0$ for simplicity.
Our theoretical results can be generalized to the nonzero trend case.

Suppose that the response surface $y(\cdot)$ is sampled at design points $\{\BFx_1,\ldots,\BFx_n\}$.
For deterministic simulation models, $y(\BFx_i)$ is observed  without noise for each $i$.
Kriging then serves as an interpolation method; that is, the prediction at each design point equals exactly the simulation output there.

For stochastic simulation models, however, the response surface is observed with random noise.
Suppose that the simulation model is executed for $m_i$ replications at each $\BFx_i$, resulting in random outputs $Y_r(\BFx_i)=y(\BFx_i)+\epsilon_r(\BFx_i)$, $r=1,\ldots,m_i$,
where $\epsilon_r(\BFx_i)$ is the simulation noise at $\BFx_i$ for the $r$-th replication, which is an independent Gaussian random variable with mean zero and variance $\sigma^2(\BFx_i)$.
Assume that $\sup_{\BFx\in\ScrX} \sigma^2(\BFx) <\infty$.
In stochastic kriging, the simulation output $Y_r(\BFx_i)$ is modeled as a realization of the random variable
$\SFY(\BFx_i) + \epsilon_r(\BFx_i)$.
Throughout the present paper, we assume that the simulation model is executed independently at different design points, that is, $\epsilon_r(\BFx_i)$ and $\epsilon_r(\BFx_j)$ are independent for all $r$ and $i\neq j$.
For stochastic kriging with the use of common random numbers, which would introduce dependence between $\epsilon_r(\BFx_i)$ and $\epsilon_r(\BFx_j)$, we refer to \cite{PearcePoloczekBranke22} and references therein.

We denote the sample mean at $\BFx_i$ by
$\bar{Y}(\BFx_i) \coloneqq  m_i^{-1} \sum_{r=1}^{m_i} Y_r(\BFx_i)$.
Let $\bar{\BFY}\coloneqq (\bar{Y}(\BFx_1),\ldots,\bar{Y}(\BFx_n))^\intercal$
and  $\BFk(\BFx)\coloneqq (k(\BFx, \BFx_1),\ldots, k(\BFx, \BFx_n))^\intercal$.
Let $\BFK\in\Real^{n\times n}$ be the covariance matrix, also referred to as the \emph{kernel matrix}, with the $(i,j)$-th entry $\Cov[\SFY(\BFx_i), \SFY(\BFx_j)]=k(\BFx_i,\BFx_j)$,  $1\leq i,j\leq n$.
Let $\BFSigma $ be the covariance matrix with the $(i,j)$-th entry $\Cov[m_i^{-1}\sum_{r=1}^{m_i}\epsilon_r(\BFx_i), m_j^{-1}\sum_{s=1}^{m_j}\epsilon_s(\BFx_j)] =  \sigma^2(\BFx_i)/m_i$ for $i=j$ and 0 otherwise.
Stochastic kriging is concerned with predicting the response surface at an arbitrary $\BFx\in\ScrX$.
Specifically, the prediction is given by
\begin{equation}\label{eq:SK-predictor}
    \widehat \SFY(\BFx) = \BFk^\intercal(\BFx) (\BFK + \BFSigma)^{-1}\bar{\BFY}.
\end{equation}
The mean squared error (MSE) of the prediction is
\begin{equation}\label{eq:SK-MSE}
    \MSE[\widehat \SFY(\BFx)] = \E[(\widehat \SFY(\BFx) - \SFY(\BFx))^2] = k(\BFx, \BFx) - \BFk^\intercal(\BFx)  (\BFK + \BFSigma)^{-1} \BFk(\BFx).
\end{equation}

In stochastic kriging, the simulation noise is called the intrinsic uncertainty, because it reflects the nature of the stochastic simulation model; the GP process $\SFY$ is called the extrinsic uncertainty, because it is imposed (not inherent in the stochastic simulation model) as a metamodel of the response surface \citep{AnkenmanNelsonStaum10}.
The MSE \eqref{eq:SK-MSE} accounts for both kinds of uncertainty;
$\BFSigma$ represents the (intrinsic) variance of the simulation noise at the sampled design points, while $k(\BFx,\BFx)$ is the (extrinsic) variance of $\SFY(\BFx)$.

Without loss of generality,
henceforth we assume for simplicity that the variance $\sigma^2(\BFx_i)$ is known and $m_i\equiv 1$ for all $i=1,\ldots,n$, unless explicitly specified otherwise.
Hence, the terms ``sample size'' and ``number of design points'' will be used interchangeably, both referring to $n$.
When $\sigma^2(\BFx_i)$ is unknown in practice, one usually runs multiple replications (i.e., $m_i >1$) to calculate the sample variance
$\hat\sigma^2(\BFx_i) \coloneqq  (m_i-1)^{-1} \sum_{r=1}^{m_i} (Y_r(\BFx_i) - \bar{Y}(\BFx_i))^2$ and uses it in lieu of $\sigma^2(\BFx)$.

Despite a flexible metamodeling approach that ensures global convergence in general,
the prediction capability of stochastic kriging may vary substantially depending on the choice of the kernel and the experimental design.
To a large extent, they collectively determine both the sample complexity and the computational complexity of stochastic kriging.
Before introducing the new methodology, we briefly overview typical examples of kernels and experimental designs that are adopted in practice.

\subsection{Typical Examples of Kernels}\label{sec:kernels}

We introduce typical examples of kernels and refer to \citet[Chapter~4]{RasmussenWilliams06} for more discussion.
\begin{example}[Gaussian Kernel]
Let $\BFtheta\in\Real^d$. The Gaussian kernel is defined by
\[
k_{\mathsf{Gauss}}(\BFx,\BFx';\BFtheta) \coloneqq \exp\left(-\norm{\BFtheta^\intercal (\BFx-\BFx')}^2\right), \quad \BFx,\BFx'\in\Real^d,
\]
where $\norm{\cdot}$ denotes the Euclidean norm.
\end{example}

\begin{example}[Mat\'{e}rn Kernel]
Let $\BFtheta\in\Real^d$ and $\nu>0$. The Mat\'{e}rn$(\nu)$ kernel is defined by
\[
k_{\mathsf{Mat\acute{e}rn}(\nu)}(\BFx,\BFx';\BFtheta) \coloneqq \frac{1}{2^{\nu-1}\Gamma(\nu)}\left(\sqrt{2\nu}\norm{\BFtheta^\intercal (\BFx-\BFx')}\right)^\nu K_\nu\left(\sqrt{2\nu}\norm{\BFtheta^\intercal (\BFx-\BFx')}\right),\quad \BFx,\BFx'\in\Real^d,
\]
where $\Gamma(\cdot)$ is the gamma function and $K_\nu(\cdot)$ is the modified Bessel function of the second kind of order $\nu$.
The Mat\'ern kernel has a simplified form if $\nu$ is half-integer: $\nu=p+\sfrac{1}{2}$ for some non-negative integer $p$.
For instance,
\begin{align*}
    k_{\mathsf{Mat\acute{e}rn}(\sfrac{1}{2})}(\BFx, \BFx';\BFtheta) \coloneqq{}& \exp\left(-\norm{\BFtheta^\intercal(\BFx-\BFx')}\right), \\
    k_{\mathsf{Mat\acute{e}rn}(\sfrac{3}{2})}(\BFx, \BFx';\BFtheta) \coloneqq{}& \left(1+\sqrt{3}\norm{\BFtheta^\intercal(\BFx-\BFx')}\right) \exp\left(-\sqrt{3}\norm{\BFtheta^\intercal(\BFx-\BFx')}\right).
\end{align*}
\end{example}

The parameter $\nu$ in the Mat\'ern kernel determines the smoothness of the GP sample paths;
a larger $\nu$ leads to sample paths having a higher degree of differentiability.
Therefore, the Mat\'ern kernel is often used for modeling response surfaces with a finite degree of differentiability.
This is in contrast to the Gaussian kernel that induces infinitely differentiable GP sample paths.
In fact, the Gaussian kernel can be obtained as a limit of the Mat\'ern kernel as $\nu\to\infty$ \cite[page~192]{gramacy2020surrogates}.

Another approach to constructing multidimensional kernels is via tensor products.
If $k_j(\cdot,\cdot)$ is a kernel defined on $\ScrX_j\subseteq \Real$ for each $j=1,\ldots,d$, then
\begin{equation}\label{eq:tensor-prod-kernel}
    k(\BFx,\BFx') = \prod_{j=1}^d k_j(x_j,  x'_j)
\end{equation}
is a kernel defined on the product space $\ScrX=\bigtimes_{j=1}^d\ScrX_j$.
Tensor product kernels imply that the dependence structure of the associated GP across each dimension is separable. (But it does not mean that the sample paths are separable functions.)
A particular advantage of the tensor product form is that it may induce substantial computational savings (Section~\ref{sec:exp-design}).

Note that the Gaussian kernel has a product form, whereas the Mat\'ern kernel does not.
Recently, \cite{SalemiStaumNelson19}  proposed a new class of tensor product kernels and numerically demonstrated  that the corresponding GPs, called GIBFs, have excellent prediction capability.

\begin{example}[GIBF Kernel]\label{example:GIBF}
Let $\ell_j\geq 0$ be an integer and $\BFtheta_j = (\theta_{j,0},\theta_{j,1},\ldots,\theta_{j,\ell_j+1})\in\Real^{\ell_j+2}_{>0}$ for each  $j=1,\ldots,d$.
The GIBF kernel is defined by the form \eqref{eq:tensor-prod-kernel}, where
\[k_j(x_j,  x'_j;\BFtheta_j) = \sum_{i=0}^{\ell_j} \theta_{j, i}\frac{(x_j  x'_j)^i}{(i!)^2} + \theta_{j, \ell_j+1}
\int_0^\infty \frac{(x_j-u)_+^{\ell_j}( x'_j-u)_+^{\ell_j}}{(\ell_j!)^2}\dd{u},
\quad \BFx,\BFx'\in\Real_{\geq 0}^d,
\]
and $(x)_+$ denotes $\max(x, 0)$.
\end{example}

\subsection{Typical Examples of Experimental Designs}\label{sec:exp-design}

There are two typical choices of experimental design---\emph{space-filling} designs and \emph{lattice} designs.
We discuss them briefly here and refer to \cite{santner2003} for more extensive exposition on the subject.
Space-filling designs are widely used for computer experiments.
They are developed following the principle that design points should be spread evenly to provide information about all portions of the design space,
as the response surface is unknown \emph{a priori} and its interesting features are just as likely to appear in one part of the design space as another.

One class of space-filling designs are Latin hypercube designs (LHDs) developed in \cite{McKay_LatinHypercube79} and their variations.
The popularity of LHDs stems from their ease of use and the fact that their projections onto one-dimensional subspaces are evenly dispersed.
They have been shown to perform well in predicting response surfaces.
However, the superiority of LHDs relative to, say, random designs that are formed by generating design points independently from some distribution usually holds only for large sample sizes, particularly when the design space is high-dimensional \citep{zhang_distance-distributed_2021}.
The large sample size makes it difficult to compute the stochastic kriging predictor \eqref{eq:SK-predictor}, as it involves inverting a large matrix (see Section~\ref{sec:CoD} for more discussion).

Lattice designs are formed by the Cartesian product of one-dimensional designs, that is, $\CalX = \bigtimes_{j=1}^d \CalX_j = \{(x_1,\ldots,x_d):x_j \in \CalX_j, j=1,\ldots,d\}$, where $\CalX_j$ is a one-dimensional design of size $n_j$ in the $j$-th dimension of the design space.
Lattice designs can lead to significant computational savings in kriging when used in conjunction with kernels of the product form \eqref{eq:tensor-prod-kernel}.
Specifically, the inverse kernel matrix can be expressed as a Kronecker product of smaller matrices,
that is,
$\BFK^{-1} = \bigotimes_{j=1}^d \BFK_j^{-1}$.
Here, $\BFK_j$ denotes the kernel matrix corresponding to the kernel $k_j$ on the $j$-th dimension and its associated one-dimensional design $\CalX_j$; that is,
entries of $\BFK_j$ are of the form $k_j(x, x')$ for $x, x'\in\CalX_j$.
Computing $\BFK^{-1}$  via $\BFK^{-1} = \bigotimes_{j=1}^d \BFK_j^{-1}$ is massively faster than direct inversion of $\BFK$ because it reduces inverting one large matrix of size $\prod_{j=1}^d n_j\times \prod_{j=1}^d n_j$ to inverting multiple small matrices, each of size $n_j\times n_j$.

Despite the considerable computational advantage, lattice designs are not practical for high-dimensional problems because they would result in an enormous sample size.
For example, when $d=20$ a lattice design has at least $2^d > 10^6$ design points, implying an excessive sample cost.

\subsection{Curse of Dimensionality}\label{sec:CoD}

The curse of dimensionality affects both  sample complexity and computational complexity.

\subsubsection{Sample Complexity}\label{sec:CoD-sample}

Under homoscedasticity,
the convergence rate of stochastic kriging with the Mat\'ern kernel  has been well studied.
It is shown in \cite{VaartZanten11} that under mild conditions, if
(i) the response surface is a GP sample path induced by the Mat\'ern($\alpha$) kernel
and (ii) the Mat\'ern($\nu$) kernel and a random design are used in stochastic kriging,
then the predictor \eqref{eq:SK-predictor} converges at the rate $n^{-\min(\alpha,\nu)/(2\nu+d)}$, assuming a single observation at each design point.
It is easy to see that the fastest rate $n^{-\alpha/(2\alpha+d)}$ is attained at $\nu=\alpha$;
that is, the model is correctly specified.
In practice, the convergence rate of stochastic kriging is likely to be worse, as $\alpha$ is usually unknown and therefore prone to misspecification.

Note that the rate $n^{-\alpha/(2\alpha+d)}$ coincides with the \emph{minimax-optimal} rate for estimating with noisy samples an unknown function in a given space of ``smooth'' functions having a parameter $\alpha$ that measures the degree of smoothness \citep{Stone80}.
The minimax-optimal rate basically specifies a fundamental lower bound on the error of any estimator of an unknown function in such a space \citep{Tsybakov09}.
Hence, if the response surface is a GP sample path induced by the Mat\'ern($\alpha$) kernel,
then
regardless of the experimental design,
the sample complexity for the predictor \eqref{eq:SK-predictor} to achieve a small error $\delta$ is at least of the order of magnitude $(1/\delta)^{2 + d/\alpha}$,
which grows exponentially in $d$, thus manifesting the curse of dimensionality.

We introduce TM kernels in Section~\ref{sec:TMK} and  TSG designs in Section~\ref{sec:SG}.
In Section~\ref{sec:sample},
we show that their joint use can  to some extent circumvent the curse of dimensionality on the convergence rate of stochastic kriging.

\subsubsection{Computational Complexity}

The main computational burden of stochastic kriging is to invert the matrix $(\BFK + \BFSigma)$ of size $n\times n$, which requires $\CalO(n^3)$ arithmetic operations in general.
The matrix inversion is involved in the computation of the predictor and  its MSE  (see equations~\eqref{eq:SK-predictor} and \eqref{eq:SK-MSE}, respectively).
As previously discussed, the sample size $n$ is usually large for high-dimensional problems, and it in turn makes the matrix inversion computationally difficult.
For example, when $n>10^5$, both storing $(\BFK+\BFSigma)$ and computing its inverse are costly or even prohibitive on typical modern computers, although the limitation can be somewhat lifted by using high-performance computing infrastructures.
Existing research has focused on approximation methods that replace the matrix of interest with one that is easier to invert while preserving the prediction accuracy as much as possible (see Section~\ref{sec:apprx} for a brief overview).
However, for an approximation method to perform well it often requires fine tuning of certain parameters involved to determine the proper trade-off between the approximation accuracy and the computational requirement.
Moreover, using approximations in lieu of the exact inverse matrix  may yield spurious estimates of the MSE of the prediction \citep{ShahriariSwerskyWangAdamsdeFreitas16}.

We do not pursue approximation methods in the present paper.
In Section~\ref{sec:computation}, we show  that the joint use of TSG designs and TM kernels
permits us to perform exact computation of stochastic kriging
with dramatically lower computational complexity.

\subsubsection{Approximation Methods}\label{sec:apprx}

Stochastic kriging is closely related to kernel methods in machine-learning literature \citep{ScholkopfSmola02}.
A plethora of approximation methods for computing $(\BFK+\BFSigma)^{-1}$ on a large dataset have been developed; see \cite{LiuOngShenCai20} for a recent survey.
Among the most popular are the Nystr\"om method and the random features method.
Both methods seek to construct
another kernel function, say $\tilde{k}$, whose kernel matrix $\tilde{\BFK}$ is of rank $\ell<n$.
By applying the Woodbury matrix identity \cite[page~19]{HornJohnson12}, the approximation $(\tilde{\BFK}+\BFSigma)^{-1}$ can normally be computed with complexity $\CalO(\ell^2n)$.

The Nystr\"om method approximates the first $\ell$ eigenvectors of $\BFK$
by selecting a subset of size $\ell$ of the  $n$ design points \citep{smola2000sparse,williams2001using}.
Let $\tilde{\BFx}_1,\ldots,\tilde{\BFx}_\ell$ denote the selected points.
Then, an approximate kernel function is constructed by $\tilde{k}(\BFx,\BFx')=\BFk^\intercal_\ell(\BFx)\BFK_{\ell,\ell}^{-1}\BFk_\ell(\BFx')$, where $\BFk_\ell(\BFx)=(k(\tilde{\BFx}_1,\BFx),\ldots, k(\tilde{\BFx}_\ell, \BFx))^\intercal$ and $\BFK_{\ell,\ell}=(k(\tilde{\BFx}_i, \tilde{\BFx}_j))_{i,j=1}^{\ell}$.
See \cite{rudi2016generalization} and \cite{LuRudiBorgonovoRosasco20} for recent developments of this method.

The random features method originates from the seminal work by \cite{rahimi2007random} and has attracted substantial interest in recent years \citep{LiuHuangChenSuykens22}.
In contrast to the Nystr\"om method, for which the approximate kernels are dependent on the given design points, the random features method constructs \emph{data-independent} approximations using Fourier transforms.
In particular, this method uses $\tilde{k}(\BFx,\BFx')=\sum_{i=1}^\ell \psi_i(\BFx)\psi_i(\BFx')$ to approximate $k(\BFx,\BFx')$, where $\psi_1(\BFx), \ldots,\psi_\ell(\BFx)$ are basis functions---also known as features in machine-learning literature---that are constructed based on random samples drawn from the spectral density (i.e., the Fourier transform) of the kernel function $k$. (For instance, the spectral density of a Gaussian kernel is the probability density of a multivariate normal distribution.)

These approximation methods can be implemented for general kernel functions. (One exception is that the random features method may not be suitable for nonstationary kernels.)
However, we stress that they focus on fast computations.
The statistical properties, such as the rate of convergence of the predictor to the true response surface, also depend mainly on the choice of the kernel and need to be analyzed separately.
In Section~\ref{sec:numerical}, we numerically compare both the Nystr\"om method and the random features method with the approach developed in the present paper.

\section{Tensor Markov Kernels}\label{sec:TMK}

Let us begin with one-dimensional Gauss--Markov processes.
Their kernels share a common functional form,  characterized in the following lemma.
This is a well-known result; see Lemma 5.1.8 and Lemma 5.1.9 in \cite{MarcusRosen06} for the proof.

\begin{lemma}\label{lemma:PD}
Let $\ScrI\subseteq \Real$ be an interval (open or closed).
Let $\SFY=\{\SFY(x):x\in \ScrI\}$ be a zero mean GP with continuous positive definite kernel $k(\cdot,\cdot)$.
Then, $\SFY$ is a Gauss--Markov process if and only if there exist positive functions $p$ and $q$ on $\ScrI$ with $p/q$ strictly increasing such that
\begin{equation}\label{eq:Gaus-Markov}
    k(x,  x') = p(x \wedge  x') q(x \vee  x'),\quad x, x' \in \ScrI,
\end{equation}
where $x\wedge  x'=\min(x, x')$ and $x\vee  x' = \max(x, x')$.
\end{lemma}

Following are some typical examples of Gauss--Markov processes and their kernels; see \citet[Chapter~5.6]{KaratzasShreve91} for more discussion on their properties.

\begin{example}[Brownian Motion Kernel]\label{example:BM}
Brownian motion is a prominent example of Gauss--Markov processes.
The kernel of the standard Brownian motion is
\[k_{\mathsf{BM}}(x, x') = x\wedge  x',\quad x, x'\in\Real_{\geq 0}.\]
Then, $k_{\mathsf{BM}}$ satisfies the form \eqref{eq:Gaus-Markov} with $p(x) = x$ and $q(x) = 1$ for $x\in\Real_{\geq 0}$.
\end{example}

\begin{example}[Brownian Bridge Kernel]\label{example:BB}
The kernel of a Brownian bridge on $[0,T]$ is given by
\[k_{\mathsf{BB}}(x, x') = (x\wedge  x')\left(1- \frac{x \vee  x'}{T}\right),\quad x, x'\in [0,T].\]
Then, $k_{\mathsf{BB}}$ has the form \eqref{eq:Gaus-Markov} with $p(x) = x$ and $q(x) = 1-x/T$ for $x\in[0,T]$.
\end{example}

\begin{example}[Laplace Kernel]\label{example:Laplace}
Let $\theta\in\Real_{>0}$. The Laplace kernel is defined by
\[k_{\mathsf{Laplace}}(x, x';\theta) = \exp(-\theta |x- x'|),\quad x, x'\in\Real.\]
It is the same as the Mat\'ern(\sfrac{1}{2}) kernel.
Then, $k_{\mathsf{Laplace}}$ has the form \eqref{eq:Gaus-Markov} with
$p(x) = e^{\theta x}$ and $q(x)=e^{-\theta x}$ for $x\in\Real$.
The corresponding Gauss--Markov process is the stationary Ornstein--Uhlenbeck process.
\end{example}

We now introduce TM kernels for $d$-dimensional GPs based on the characterization on the kernels of one-dimensional Gauss--Markov processes in Lemma~\ref{lemma:PD}.

\begin{definition}[Tensor Markov Kernel]\label{def:TMK}
For each $j=1,\ldots,d$, let $\ScrX_j\subseteq \Real$ be an interval (open or closed), and let
$p_j$ and $q_j$ be positive functions on $\ScrX_j$ with $p_j/q_j$ strictly increasing.
Then,
\begin{equation}\label{eq:TMK}
    k(\BFx,\BFx') = \prod_{j=1}^d p_j(x_j\wedge x'_j) q_j(x_j\vee x'_j),\quad \BFx,\BFx'\in \ScrX
\end{equation}
is called a  \emph{tensor Markov (TM) kernel} on $\ScrX = \bigtimes_{j=1}^d \ScrX_j$.
We also call the corresponding GP a \emph{tensor Markov Gaussian process} (TMGP).
\end{definition}

Note that each component kernel $k_j$ may be different in general and that the theory is developed with such generality.
However, for simplicity, we focus on the case where the component kernels are identical in the numerical experiments in Section~\ref{sec:numerical}.

The GIBF kernel (Example \ref{example:GIBF}) developed in \cite{SalemiStaumNelson19} is a TM kernel if its order is $(\ell_1,\ldots,\ell_j)=\BFzero$.
In this case, the GIBF kernel is reduced to
\[k_{\mathsf{GIBF}(\BFzero)}(\BFx,\BFx') = \prod_{j=1}^d [ \theta_{j,0} +  \theta_{j,1} (x_j \wedge  x'_j)],\quad \BFx,\BFx' \in \Real_{\geq 0}^d.\]
The form \eqref{eq:TMK} is verified by setting $p_j(x) = \theta_{j,0} +  \theta_{j,1} (x)$ and $q_j(x)=1$ for $x\in\Real$.
The corresponding GP is a Brownian field.

\subsection{A Connection to Ordinary Differential Equations}

The theory we develop in the present paper relies on a connection between Gauss--Markov processes and linear ordinary differential equations (ODEs).
This is well known in the probability and statistics literature.
We provide a heuristic, exemplary introduction below and refer interested readers to
\cite{DolphWoodbury52}, \cite{Ylvisaker87}, and \cite{MouraGoswami97}   for a rigorous treatment, which is beyond the scope of the present paper.

Suppose that $\{\SFY(x):x\in\Real\}$ is the stationary Ornstein--Uhlenbeck process and $k(x,x')$ is the Laplace kernel with parameter $\theta=1$ (Example~\ref{example:Laplace}) that is associated with it.
For each $i\in \mathbb{N}$,
let $x_i=ih$ for some constant $h>0$.
Then, the discretized process $\{\SFY(x_i):i\in\mathbb{N}\}$ forms a Markov chain, and its transition dynamics can be written as
\begin{equation*}
    \SFY(x_{i+1})=e^{-h}\SFY(x_{i})+\varepsilon_i\sqrt{1-e^{-2h}},
\end{equation*}
where $\varepsilon_i$'s are independent and identically distributed (i.i.d.) standard normal random variables.

We can show via a straightforward calculation that the joint distribution of $\{\SFY(x_i):i\in\mathbb{N}\}$ satisfies
\begin{align*}
    \pr\bigl(\{\SFY(x_i):i\in\mathbb{N} \}\bigr) \propto{} & \exp\biggl(-\frac{1}{1-e^{-2h}}\sum_{i\in\mathbb{N}}\bigl(\SFY(x_{i})-e^{-h}\SFY(x_{i-1})\bigr)^2\biggr)\\
    ={}&\exp\biggl(-\frac{1}{1-e^{-2h}}\sum_{i\in\mathbb{N}}(1+e^{-2h})\SFY^2(x_{i})-e^{-h}\SFY(x_{i})\SFY(x_{i+1})-e^{-h}\SFY(x_{i})\SFY(x_{i-1})\biggr)\\
    ={}& \exp\biggl(-\sum_{i,j\in\mathbb{N}}L_{i,j}\SFY(x_i)\SFY(x_j)\biggr),
\end{align*}
where $L_{i,i}=\frac{1+e^{-2h}}{1-e^{-2h}}$,  $L_{i-1,i}=L_{i,i+1}=\frac{-e^{-h}}{1-e^{-2h}}$, and $L_{i,j}=0$ for all $i,j\in\mathbb{N}$ such that $|i-j|\geq 2$.
Hence, the joint distribution of $\{\SFY(x_i):i\in\mathbb{N}\}$ is determined by $\{L_{i,j}:i,j\in\mathbb{N}\}$.
Further, note that
\begin{align*}
\sum_{i,j\in\mathbb{N}}L_{i,j}\SFY(x_i)\SFY(x_j)={}&
\frac{h}{2} \sum_{i\in\mathbb{N}}\SFY(x_i)\biggl(-\frac{\SFY(x_{i+1})-2\SFY(x_i)+\SFY(x_{i-1})}{h^2}
+\SFY(x_i)\biggr)+\CalO(h)\\
\to {} & \frac{1}{2}\int \SFY(x)(\CalL \SFY)(x)\dd{x},
\end{align*}
as $h\to 0$, where $\CalL$ is a differential operator defined as $(\CalL f)(x)\coloneqq -\frac{\dd^2 }{\dd x^2}f(x)+ f(x)$.
Hence, the joint distribution of $\{\SFY(x):x\in\Real\}$ is determined by the operator $\CalL$, which can be interpreted as a continuous analog of $\{L_{i,j}:i,j\in\mathbb{N}\}$.
Moreover,
let $p(x)=e^x$ and $q(x)=e^{-x}$ so that $k(x,x') = p(x\wedge x') q(x\vee x')$, as in Example~\ref{example:Laplace}.
One can easily verify that
$(\CalL p)(x) = (\CalL q)(x) = 0$.

In general, the distribution of a Gauss--Markov process can be fully characterized by a second-order linear differential operator such as $\CalL$.
As all second-order linear ODEs can be recast to a Sturm--Liouville equation of form \eqref{eq:diff-op-L} below by a simple transformation \citep[Chapter 8]{ArfkenWeberHarris12},
we impose the following regularity condition on such an ODE.

\begin{assumption}\label{assump:SL}
Let $k(\BFx,\BFx') = \prod_{j=1}^d p_j(x_j\wedge  x'_j)q_j(x_j \vee  x'_j)$ be a TM kernel for $\BFx,\BFx' \in (0,1)^d$.
For each $j=1,\ldots,d$, both $p_j$ and $q_j$ are continuously differentiable and satisfy the ODE $(\CalL_j f)(x)=0$ for all $x\in(0,1)$, where
\begin{equation}\label{eq:diff-op-L}
    (\CalL_j f)(x) = \frac{\dd }{\dd x}\left(u_j(x) \frac{\dd f}{\dd x} (x)\right) + v_j(x) f(x),
\end{equation}
where $u_j$ is a continuously differentiable function on $(0,1)$ and $v_j$ is a continuous function on $(0,1)$.
\end{assumption}
It is straightforward to verify that Examples~\ref{example:BM}--\ref{example:Laplace} all satisfy Assumption~\ref{assump:SL}, so we omit the details.

\subsection{Markov Property}

Let $\SFY$ be a TMGP defined on $\ScrX=\bigtimes_{j=1}^d \ScrX_j$, with each $\ScrX_j\subseteq \Real$ an interval.
It follows from
Corollary~3.3 in \cite{CarnalWalsh91} that $\SFY$
has the \emph{sharp Markov property with respect to rectangular sets}, which is equivalent to the $\BFzero$-order Markov property introduced in \cite{SalemiStaumNelson19}.
Specifically, let $\ScrI = \bigtimes_{j=1}^d \ScrI_j$ be a rectangular set in $\ScrX$, where each $\ScrI_j\subseteq \ScrX_j$ is an interval.
Then,
$\varsigma(\{\SFY(\BFx): \BFx\in \ScrI\})$
is conditionally independent of $\varsigma(\{\SFY(\BFx): \BFx\in \ScrX\setminus \ScrI\})$
given $\varsigma(\{\SFY(\BFx): \BFx\in \partial(\ScrI)\})$, where $\varsigma(\mathcal S)$ denotes the sigma-algebra generated by $\mathcal S$, and $\partial(\ScrI)$ denotes the boundary of  $\ScrI$.
Basically, the $\BFzero$-order Markov property indicates that given  information about $\SFY$ on the boundary of a rectangular region,
information about $\SFY$ outside the region provides no additional value for predicting $\SFY$ inside the region.

It is nontrivial to generalize the usual Markov property for stochastic processes on $\Real$ to random fields on $\Real^d$.
A particular challenge is to properly define  the ``past'', ``present'', and ``future'' on a multidimensional space, analogously to the real line.
Indeed, there are several different Markovian notions other than the sharp Markov property for random fields; see, for example, the Appendix in \cite{Adler81}. It is beyond the scope of the present paper to study these properties for TMGPs in detail.
Our sample complexity analysis and development of computational algorithms do not directly rely on the multidimensional Markovian notions.
Instead, it is the usual Markov property on the real line that is of great significance for us in developing relevant theory.
In particular, for $d=1$, by virtue of the Markov property of the underlying Gauss--Markov process, we show below that $\BFK^{-1}$ is a tridiagonal matrix and that its nonzero entries can be calculated analytically.

\begin{proposition}\label{prop:inverse}
Let $k(x, x') = p(x\wedge  x')q(x\vee x')$ be a TM kernel in one dimension.
Let $n\geq 3$, $x_0=-\infty$, $x_{n+1}=\infty$, and $\{x_1,\ldots,x_n\}$ be an increasing sequence.
Let $\SFp_0=\SFq_{n+1}=0$, $\SFp_{n+1}=\SFq_0=1$,
$\SFp_i = \SFp(x_i)$, and $\SFq_i=\SFq(x_i)$ for $i=1,\ldots,n$.
Then, $\BFK^{-1}$ and $\BFK^{-1}\BFk(x)$ are specified as follows.
\begin{enumerate}[label=(\roman*)]
    \item
$\BFK^{-1}$ is a tridiagonal matrix, that is, $(\BFK^{-1})_{i,i+2}= (\BFK^{-1})_{i+2,i} =0 $ for all $i=1,\ldots,n-2$;
moreover,
\begin{equation}\label{eq:K-inverse}
\begin{array}{ll}
    (\BFK^{-1})_{i,i} = \displaystyle \frac{\SFp_{i+1}\SFq_{i-1}-\SFp_{i-1}\SFq_{i+1}}{(\SFp_i\SFq_{i-1}-\SFp_{i-1}\SFq_i)(\SFp_{i+1}\SFq_i-\SFp_i\SFq_{i+1})}, &\quad i = 1, \ldots, n, \\[2ex]
    (\BFK^{-1})_{i,i+1} = (\BFK^{-1})_{i+1,i} = \displaystyle \frac{-1}{\SFp_{i+1} \SFq_{i}-\SFp_{i}\SFq_{i+1}},&\quad  i = 1, \ldots, n-1.
\end{array}
\end{equation}
\item Let $i^*=0,1,\ldots,n$ such that $x\in [x_{i^*}, x_{i^*+1})$. Then,
\begin{equation}\label{eq:K-inverse-times-k}
(\BFK^{-1}\BFk(x))_i = \left\{
\begin{array}{ll}
\displaystyle\frac{\SFp_{i^*+1}q(x) - p(x) \SFq_{i^*+1} }{\SFp_{i^*+1}\SFq_{i^*} - \SFp_{i^*} \SFq_{i^*+1} },     &  \mbox{ if } i = i^*,\\[2ex]
\displaystyle\frac{p(x)\SFq_{i^*} - \SFp_{i^*} q(x) }{\SFp_{i^*+1}\SFq_{i^*} - \SFp_{i^*} \SFq_{i^*+1} },     & \mbox{ if } i=i^*+1, \\[2ex]
0, & \mbox{ otherwise}.
\end{array}
\right.
\end{equation}
\end{enumerate}

\end{proposition}

\begin{remark}
One should not confuse TMGPs with Gaussian Markov random fields (GMRFs).
The former are GPs defined on a continuous domain in $\Real^d$, whereas the latter are GPs defined on a discrete set, usually a graph with nodes representing the index and edges indicating the dependence structure. We refer to \cite{RueHeld05} for a general introduction to GMRFs.
\end{remark}

\subsection{Non-differentiability}\label{sec:non-diff}

It can be shown that TMGPs can be represented by a time-changed Brownian field after proper possibly nonlinear scaling; see,  for example, Theorem~3.2 in \cite{CarnalWalsh91} and Remark~5.1.11 in \cite{MarcusRosen06}.
In particular, if $\SFY=\{\SFY(\BFx):\BFx\in\ScrX\}$ is a TMGP with kernel \eqref{eq:TMK}, then $\SFY$ has the same distribution as \[\left\{\left(\prod_{j=1}^d q_j(x_j) \right) \SFB\bigl(p_1(x_1)/q_1(x_1), \ldots, p_d(x_d)/q_d(x_d)\bigr) :\BFx\in\ScrX\right\},\]
where $\SFB$ is a standard Brownian field on $\Real_{\geq 0}^d$.
This representation suggests that, like Brownian fields, the sample paths of TMGPs are continuous but nowhere differentiable.

From a modeling viewpoint, the nowhere differentiability seems rather restrictive.
After all, in practice most response surfaces of interest do exhibit some degree of smoothness; non-differentiability often appears in some but not all parts of the design space.
In Section~\ref{sec:misspec}, we discuss the prediction capability of TMGPs
under such model misspecification.
Somewhat surprisingly, we prove that the convergence rate of stochastic kriging with TM kernels
for predicting smooth surfaces does not suffer much from the curse of dimensionality under mild conditions,
provided that the experimental design is chosen to be a TSG.

\section{Sparse Grid Experimental Designs}\label{sec:SG}

This section proceeds in two main steps.
First, we introduce classical SGs.
Then, we propose TSG designs that are much more flexible with regard to specifying the sampling budget.

\subsection{Classical Sparse Grids}

Recall that a lattice design is defined as the Cartesian product $\CalX=\bigtimes_{j=1}^d\CalX_j$,
where $\CalX_j$ is a one-dimensional
\emph{component design} in the $j$-th dimension, having $n_j$ points.
Then, the kernel matrix of size $\prod_{j=1}^d n_j\times \prod_{j=1}^d n_j$  can be expressed as a Kronecker product of much smaller matrices, each of size $n_j\times n_j$, provided that the kernel is of the product form \eqref{eq:tensor-prod-kernel}.

SG designs extend the idea of lattice designs,
retaining the computational advantage induced by the tensor structure while dramatically reducing the grid size in high dimensions.
A key notion in constructing SG designs is the \emph{level} that we denote by $\tau \geq 1$.
Given $\tau$, we specify, for each dimension, a \emph{nested} sequence of one-dimensional designs $\emptyset = \CalX_{j,0} \subseteq \CalX_{j,1} \subseteq \CalX_{j,2}\subseteq \cdots \subseteq \CalX_{j,\tau}$.
There are multiple ways to choose the nested sequences to construct an SG; see \cite{bungartz_griebel_2004}.
We adopt the ``classical'' approach in the present paper.
Suppose that the design space is a hypercube $\ScrX=(0,1)^d$.
We construct the nested sequence for each dimension by recursive dyadic partitioning of the interval $(0,1)$.
That is, for each $j=1,\ldots,d$,
\begin{equation}\label{eq:dyadic}
\begin{aligned}
      \CalX_{j, 1} ={}& \{\sfrac{1}{2}\}, \\
      \CalX_{j, 2} ={}& \{\sfrac{1}{4},\; \sfrac{1}{2},\; \sfrac{3}{4}\}, \\
    \vdots& \\
    \CalX_{j, \tau} ={}& \{1\cdot 2^{-\tau},\; 2\cdot  2^{-\tau},\ldots, \; (2^\tau-1)\cdot 2^{-\tau} \}.
\end{aligned}
\end{equation}

The parameter $\tau$ thus indicates the refinement level of the nested sequence of each dimension.
Then, an SG of level $\tau$ is defined as
\begin{equation}\label{eq:SG}
    \CalX^{\mathsf{SG}}_\tau = \bigcup_{\abs{\BFl}\leq \tau+d-1} \CalX_{1,l_1} \times \cdots \times \CalX_{d,l_d},
\end{equation}
where $\BFl=(l_1,\ldots,l_d)\in \NatInt^d$ and $\abs{\BFl}=\sum_{j=1}^d l_j$.
Namely, an SG is the union of multiple lattices.
We refer to the SGs constructed using the nested sequences in \eqref{eq:dyadic} as the classical SGs.

\begin{remark}
The ``full grid'' that uses the same refinement level $\tau$ for each dimension can be expressed in a form similar to \eqref{eq:SG}:
\begin{equation}\label{eq:fullgrid}
\bigtimes_{j=1}^d \CalX_{j,\tau} =  \bigcup_{\norm{\BFl}_\infty\leq \tau} \CalX_{1,l_1} \times \cdots \times \CalX_{d,l_d},
\end{equation}
where $\norm{\BFl}_\infty = \max(l_1,\ldots,l_d)$.
Contrasting \eqref{eq:SG} with \eqref{eq:fullgrid} reveals  a key reason why SG designs have a much smaller grid size relative to  lattice designs of the same refinement level.
In the SG design \eqref{eq:SG}, the size of each constituent lattice is small because
the refinement level of each dimension cannot be large simultaneously due to the condition $\sum_{j=1}^d l_j \leq \tau+d-1$.
For instance, if there exists a dimension in which $l_j=\tau$, then $l_{j'}=1$ for all $j'\neq j$.
\end{remark}

\begin{figure}[ht]
\FIGURE
{
\includegraphics[width=\textwidth]{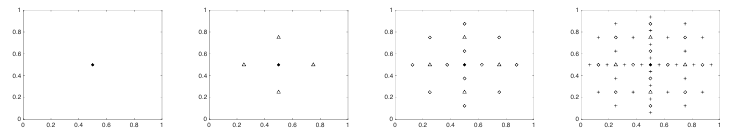}
}
{Classical SGs of Levels 1 to 4 in Two Dimensions. \label{fig:sgg}}
{The new points added as the level increases are denoted by a different symbol.}
\end{figure}

\begin{example}[Two-Dimensional Classical Sparse Grids]\label{example:SG}
Assume $d=2$ and $\tau=2$.
By the definition \eqref{eq:SG}, the level multi-index satisfies $\abs{\BFl}\leq 3$, so the two-dimensional classical SG of level 2 is a union of three smaller lattices with level indices $\BFl \in \{(1,1), (1, 2), (2, 1)\}$; that is,
\begin{align*}
\CalX^{\mathsf{SG}}_2 ={}& \big( \CalX_{1,1} \times \CalX_{2,1} \big) \cup \big( \CalX_{1,1} \times \CalX_{2,2} \big) \cup  \big( \CalX_{1,2} \times \CalX_{2,1} \big) \\
={}& \big( \{\sfrac{1}{2}\} \times \{\sfrac{1}{2}\} \big) \cup \big( \{\sfrac{1}{2}\} \times \{\sfrac{1}{4},\; \sfrac{1}{2},\; \sfrac{3}{4}\} \big) \cup  \big( \{\sfrac{1}{4},\; \sfrac{1}{2},\; \sfrac{3}{4}\} \times \{\sfrac{1}{2}\} \big)   \\
={}& \{ (\sfrac{1}{2}, \sfrac{1}{4}),\; (\sfrac{1}{2}, \sfrac{1}{2}),\; (\sfrac{1}{2}, \sfrac{3}{4}),\; (\sfrac{1}{4}, \sfrac{1}{2}),\; (\sfrac{3}{4},\sfrac{1}{2}) \}.
\end{align*}
Figure~\ref{fig:sgg} shows the two-dimensional classical SGs of levels 1 to 4.
\end{example}

Note that the constituent lattices of an SG have common points.
For instance,  in Example~\ref{example:SG}
\[\big( \CalX_{1,1} \times \CalX_{2,1} \big)
\cap ( \CalX_{1,1} \times \CalX_{2,2}) \cap  ( \CalX_{1,2} \times \CalX_{2,1} ) = \{(\sfrac{1}{2}, \sfrac{1}{2})\}.
\]
To facilitate theoretical analysis in later sections,
we represent classical SGs as a union of non-overlapping sets of design points by defining the following notations.
Let $\SFc_{l, i}\coloneqq i\cdot 2^{-l}$ for $l\geq 1$ and $i=1,\ldots,2^l -1$,
and let
$\BFc_{\BFl, \BFi}\coloneqq (\SFc_{l_1, i_1},\ldots, \SFc_{l_d, i_d})$.
For any level multi-index $\BFl$, we define a set for the  multi-index $\BFi$ as follows,
\begin{equation}\label{eq:index-set-rho}
\rho(\BFl) \coloneqq \bigtimes_{j=1}^d \{i_j: i_j\mbox{ is an odd number between 1 and }2^{l_j} \}  = \bigtimes_{j=1}^d  \{1, 3, 5, \ldots, 2^{l_j}-1 \}.
\end{equation}
Then, we may write the component designs in \eqref{eq:dyadic} as $\CalX_{j, l} = \{\SFc_{l, i}: i=1,\ldots,2^l-1 \}$ for all $j=1,\ldots,d$ and $l=1,\ldots,\tau$.
Furthermore, it is easy to show that due to the nested structure, the classical SG can be represented as
\begin{equation}\label{eq:disjoint-rep}
    \CalX^{\mathsf{SG}}_\tau = \bigcup_{\abs{\BFl}\leq \tau+d-1} \{\BFc_{\BFl,\BFi}: \BFi\in\rho(\BFl)\}.
\end{equation}
Moreover, when we augment the classical SG from level $\tau-1$ to level $\tau$, the new added points are exactly
$\{\BFc_{\BFl,\BFi}: \abs{\BFl}=\tau+d-1, \BFi\in\rho(\BFl)\}$; see Figure~\ref{fig:sgg} for an illustration in two dimensions.
Thus, \eqref{eq:disjoint-rep} represents the classical SG as a union of non-overlapping sets of design points.

Lemma~3.6 in \cite{bungartz_griebel_2004} stipulates that the sample size of $\CalX_{\tau}^{\mathsf{SG}}$ is given by
\begin{equation}\label{eq:SSGNumpt}
    \abs{\CalX_{\tau}^{\mathsf{SG}}}=\sum_{\ell=0}^{\tau-1}2^\ell \binom{\ell+d-1}{d-1} = 2^\tau\cdot\left(\frac{\tau^{d-1}}{(d-1)!}+\CalO(\tau^{d-2})\right)\asymp 2^\tau \tau^{d-1},
\end{equation}
where $a_n\asymp b_n$ denotes the relation that $\limsup_{n\to\infty}a_n/b_n<\infty$ and $\limsup_{n\to\infty}b_n/a_n<\infty$  for two sequences $\{a_n\}$ and $\{b_n\}$.
In contrast, the sample size of a full grid of the same refinement level is $\abs{\CalX_{j, \tau}}^d = (2^\tau-1)^d \asymp 2^{\tau d}$, a dramatically larger number; see
Table~\ref{tab:Lattice-SG} for the stark contrast between full grids and classical SGs in terms of the sample size, particularly  in high dimensions.
In addition, Table~\ref{tab:SSG_num} shows the sample sizes of classical SGs of different levels and dimensions.

\begin{table}[ht]
\TABLE
{Full Grids versus Classical Sparse Grids. \label{tab:Lattice-SG}}
{
    \begin{tabular}{ c c r c  r }
    \toprule
    Dimension $d$ && Full Grid && Sparse Grid of Level 4 \\
    \midrule
    1 && 15 && 15\\
    2 && 225 && 49\\
    5 && 759,375 && 351\\
    10 && $5.77\times10^{11}$ && 2,001\\
    20 && $3.33\times10^{23}$ && 13,201\\
    50 && $6.38\times10^{58}$ && 182,001\\
    \bottomrule
    \end{tabular}
}
{
The refinement level of each full grid is $\tau=4$, so its sample size is $(2^\tau-1)^d = 15^d$.
The size of a classical SG is calculated by \eqref{eq:SSGNumpt}.}
\end{table}

\begin{table}[ht]
\TABLE
{Classical Sparse Grids of Different Levels and Dimensions. \label{tab:SSG_num}}
{
    \begin{tabular}{c  c r c r  c r c r c r}
    \toprule
    Level $\tau$  && $d=2$ && $d=5$ && $d=10$  && $d=20$ && $d=50$\\
    \midrule
    2 && 5 && 11 && 21 && 41 && 101\\
    3 && 17 && 71 && 241 && 881 && 5,201\\
    4 && 49 && 351 && 2,001 && 13,201 && 182,001\\
    5 && 129 && 1,471 && 13,441  && 154,881 && 4,867,201\\
    \bottomrule
    \end{tabular}
}
{The size of a classical SG is calculated by \eqref{eq:SSGNumpt}.}
\end{table}

\subsection{Truncated Sparse Grids}

\begin{figure}[ht]
\FIGURE{
\includegraphics[width=\textwidth]{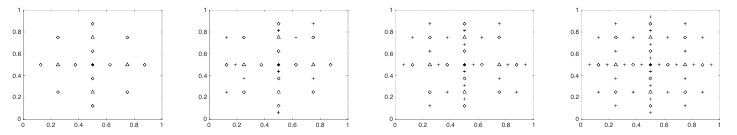}
}
{Examples of TSGs in Two Dimensions. \label{fig:TSG}}
{(i) $\tau=3$, $\Tilde{n}=0$; (ii) $\tau=3$, $\Tilde{n}=10$;  (iii) $\tau=3$, $\Tilde{n}=20$; and (iv) $\tau=4$, $\Tilde{n}=0$. The first and last are classical SGs.}
\end{figure}

Despite the huge reduction in sample size and the significant computational savings that follow (see Section~\ref{sec:computation}), SG designs lack sufficient flexibility that would allow them to be widely adopted in practice.
This is because SGs are not defined directly via the sample size $n$ but via the level $\tau$.
If $n$ lies between $\abs{\CalX_{\tau}^{\mathsf{SG}}}$ and $\abs{\CalX_{\tau+1}^{\mathsf{SG}}}$ for some $\tau$, then we are forced to choose between the two SGs.
However, choosing $\CalX_{\tau}^{\mathsf{SG}}$ would mean wasting the budget, while choosing  $\CalX_{\tau+1}^{\mathsf{SG}}$ would mean that one has to increase the budget, which may not be possible.
This issue becomes particularly serious in high dimensions
because the sample size of the classical SG increases very quickly as $\tau$ increases;
see Table~\ref{tab:SSG_num}.
To resolve the issue, we propose a new experimental design, as follows; see
Figure~\ref{fig:TSG} for examples of TSGs in two dimensions.

\begin{definition}[Truncated Sparse Grid]\label{def:TSG}
Given a sampling budget $n$, there exists $\tau\geq 1$ such that $\abs{\CalX_{\tau}^{\mathsf{SG}}} \leq n <  \abs{\CalX_{\tau+1}^{\mathsf{SG}}}$, which can be calculated via
formula \eqref{eq:SSGNumpt}.
Let $\Tilde{n} = n-\abs{\CalX_{\tau}^{\mathsf{SG}}}$ and  $\CalD_{\tau+1}\coloneqq \CalX_{\tau+1}^{\mathsf{SG}} \setminus \CalX_{\tau}^{\mathsf{SG}}$.
We construct $\CalA_{\Tilde{n}}$, a set of points of size $\Tilde{n}$, by arbitrarily selecting the design points in $\CalD_{\tau+1}$. Then, a TSG design of size $n$, denoted by $\CalX_{n}^{\mathsf{TSG}}$, is the union of $\CalX_{\tau}^{\mathsf{SG}}$ and $\CalA_{\Tilde{n}}$, that is,
$\CalX_n^{\mathsf{TSG}} = \CalX_{\tau}^{\mathsf{SG}} \cup \CalA_{\Tilde{n}}$.
Further, if $\CalA_{\Tilde{n}}$ is constructed by selecting the design points in $\CalD_{\tau+1}$ uniformly at random without replacement,
then the resulting design is called a \emph{random truncated sparse grid} (RTSG) design.
\end{definition}

A natural question follows the simple construction of TSGs.
Being ``incomplete'' SGs, can TSGs retain the computational tractability of classical SGs, yielding fast computation of kriging for tensor product kernels \eqref{eq:tensor-prod-kernel} as shown in \cite{Plumlee14}?
This is a nontrivial question, and the answer is negative in general.
However,
we show later in Theorem~\ref{theo:K-inverse} that the inverse kernel matrices
induced by TM kernels and TSG designs possess a nice sparse structure, which
facilitates fast computation of stochastic kriging.
The proof depends critically on a deep connection between classical SGs and an orthogonal basis expansion of TM kernels that may not exist for general tensor product kernels.

\section{Sample Efficiency} \label{sec:sample}

In this section, we analyze the convergence rate of stochastic kriging with TM kernels in terms of MSE.
We aim to answer a fundamental question regarding how many simulation samples are needed for stochastic kriging to produce good predictions of a high-dimensional response surface.
Note that this section does not concern the computational issue of large matrix inversion that is involved in stochastic kriging. (One could use generic matrix inversion algorithms. It would not change the sample efficiency but would result in a time complexity of $\CalO(n^3)$.)
We present algorithms for fast computation of stochastic kriging with TM kernels and TSG designs in Section~\ref{sec:computation}.

Our  analysis of the sample efficiency comprises three parts, as follows.
\begin{enumerate}[label=(\roman*)]
    \item We first focus on the case of deterministic simulation. Assuming that the response surface is indeed a realization of a TMGP, we establish an upper bound on the convergence rate, if the design points form a TSG.
    \item We then consider model misspecification; the response surface is realized from a GP induced by a tensor product kernel having a higher degree of smoothness than the TM kernel.
    We prove that the curse of dimensionality on the convergence rate of the ``misspecified'' predictor
    is significantly alleviated relative to typical stochastic kriging.

    \item Third, we generalize the above results to the stochastic setting.
\end{enumerate}

\subsection{Deterministic Simulation}\label{sec:deter-sim}

In this section, we study the convergence rate of (deterministic) kriging with TM kernels and TSGs.
Let $\widehat \SFY_n(\BFx)$ denote the kriging predictor with kernel $k$ and design points $\{\BFx_1,\ldots,\BFx_n\}$,
\begin{equation}\label{eq:Kriging}
    \widehat \SFY_n(\BFx) =  \BFk^\intercal (\BFx) \BFK^{-1}\BFy,
\end{equation}
where $\BFy = (\SFY(\BFx_1),\ldots, \SFY(\BFx_n))^\intercal $.
Our first main result establishes an upper bound on the MSE \eqref{eq:SK-MSE} uniformly in $\BFx$ in the absence of simulation noise, assuming that the response surface is a realization of a TMGP and the experimental design forms a TSG.

\begin{theorem}\label{theo:determ-TSG}
Let $\{\SFY(\BFx):\BFx\in(0,1)^d\}$ be a zero mean TMGP with kernel $k$  that satisfies Assumption~\ref{assump:SL}.
Suppose that the true response surface is a realization of $\SFY$, the simulation has no noise, and the design points $\{\BFx_1,\ldots,\BFx_n\}$  form a TSG.
Let $\widehat \SFY_n(\BFx)$ be the kriging predictor with kernel $k$ and the TSG design.
Then, as $n\to\infty$,
\begin{equation}\label{eq:upper-bound-determ-TSG}
    \sup_{\BFx\in(0,1)^d}\E[(\widehat \SFY_n(\BFx) - \SFY(\BFx))^2] = \CalO\left(n^{-1}(\log n)^{2(d-1)}\right).
\end{equation}
\end{theorem}

Note that in the upper bound \eqref{eq:upper-bound-determ-TSG}, the dimensionality $d$ does not appear in the exponent of $n$ but only in the exponent of $\log n$.
This suggests that for a fixed $d$, the MSE of the kriging predictor  with TM kernels and TSGs can be upper bounded by a rate arbitrarily close to $n^{-1}$ as $n\to\infty$, almost overcoming the curse of dimensionality on the sample complexity.
Admittedly, the setting of Theorem~\ref{theo:determ-TSG} is somewhat restrictive;
the simulation model is deterministic and the response surface is realized from a TMGP, which suggests nowhere differentiability.
However, these restrictions will be relaxed shortly, and the curse of dimensionality on the convergence rate will still not be substantial.

Let us now contrast Theorem~\ref{theo:determ-TSG} with existing results.
\cite{TuoWang20} thoroughly investigate the convergence rate of deterministic kriging for the Mat\'ern  kernels.
The authors prove that if the response surface $y$ is realized from a GP with  Mat\'ern($\alpha$) kernel, then for \emph{any} experimental design, the kriging predictor with the Mat\'ern($\nu$) kernel, where $\nu$ is allowed to be different from $\alpha$, has the following lower bound on its MSE:
\begin{equation}\label{eq:lower-bound-Matern}
    \sup_{\BFx\in(0,1)^d}\E[(\widehat \SFY_n(\BFx) - \SFY(\BFx))^2] \geq  c n^{-2\alpha/d}
\end{equation}
for some constant $c>0$ and all sufficiently large $n$.
Unlike Theorem~\ref{theo:determ-TSG}, in the above lower bound the dimensionality $d$ appears in the exponent of $n$ and decreases the lower rate rapidly, manifesting the severity of the curse of dimensionality.

A fundamental reason behind the stark difference between the decay rates in \eqref{eq:upper-bound-determ-TSG} and  \eqref{eq:lower-bound-Matern} is that TM kernels are not isotropic but are defined via a tensor structure,
which can rule out some irregular functions that would be included in the function space that are formed by the GP sample paths induced by the Mat\'ern kernel.
We refer to \citet[page~260]{Evans10} for one such irregular function that possesses certain smoothness properties but is unbounded on each open subset of the design space.

Meanwhile, the function space formed by TMGP sample paths is still broad in the sense that if the experimental design is not well chosen, the decay rate of the MSE of the kriging predictor with a TM kernel will suffer severely from the curse of dimensionality, as shown  below.
\begin{proposition}
\label{prop:FG-convergence}
Let $\{\SFY(\BFx):\BFx\in(0,1)^d\}$ be a zero mean TMGP with kernel $k$  that satisfies Assumption~\ref{assump:SL}.
Suppose that the true response surface is a realization of $\SFY$, the simulation has no noise, and the design points $\{\BFx_1,\ldots,\BFx_n\}$ form a lattice
$\bigtimes_{j=1}^d \CalX_{j,\tau} $ for some $\tau\geq 1$, where $\CalX_{j,\tau} = \{i\cdot 2^{-\tau}:1\leq i\leq 2^\tau-1\}$.
Let $\widehat \SFY_n(\BFx)$ be the kriging predictor with kernel $k$ and the lattice design.
Then, there exists a constant $c>0$ such that for all sufficiently large $n$,
\[
\sup_{\BFx\in(0,1)^d}\E[(\widehat \SFY_n(\BFx) - \SFY(\BFx))^2] \geq  c n^{-1/d}.
\]
\end{proposition}

\subsection{Model Misspecification}\label{sec:misspec}

As discussed in Section~\ref{sec:non-diff}, sample paths of a TMGP are nowhere differentiable.
However, most of the response surfaces of practical interest do possess a certain level of smoothness.
To postulate that the true surface is a TMGP sample path is
restrictive.
We study the model misspecification issue in this section.
We show that if the true surface
is realized from a GP with a tensor product kernel having a higher degree of differentiability than TM kernels, thereby inducing smoother sample paths than TMGPs, then kriging with TM kernels and TSGs still breaks the curse of dimensionality, which is obviously reassuring for practitioners.

To facilitate presentation, we will use the notion of reproducing kernel Hilbert space (RKHS); see \citet[Chapter 6]{RasmussenWilliams06} for an introduction.
For a kernel $k$, we use $\ScrH_k$ to denote its associated RKHS and $\norm{\cdot}_{\ScrH_k}$ to denote the norm of the space.

\begin{assumption}\label{assump:smooth-kernel}
Let $k(\BFx,\BFx') = \prod_{j=1}^d k_j(x_j,x'_j)$ be a TM kernel that satisfies Assumption~\ref{assump:SL}.
Let $k^*(\BFx,\BFx')=\prod_{j=1}^d k_j^*(x_j, x'_j)$ be a tensor product kernel for $\BFx,\BFx' \in (0,1)^d$.
For each $j=1,\ldots,d$, $k^*_j$ satisfies $\sup_{x\in(0,1)}\|\CalL_j k^*_j(x,\cdot)\|_{\ScrH_{k_j}} <\infty$, where
the differential operator $\CalL_j$ is defined in \eqref{eq:diff-op-L}.
\end{assumption}

\begin{remark}
It can be shown that Assumption~\ref{assump:smooth-kernel} holds if for each $j$, $k^*_j(x,x')$ has a second-order weak derivative with respect to $x$ and $\partial_{x}^2 k^*_j(x,x')$ has a first-order weak derivative with respect to $x'$ for all $x,x'\in(0,1)$. For example, $k^*_j$ can be the Mat\'ern($\nu$) kernel with $\nu\geq \sfrac{3}{2}$ or the Gaussian kernel.
For another example,
 $k^*$ can be the GIBF($\BFell$) kernel where $\BFell =(\ell_1,\ldots,\ell_d)$ with $\ell_j\geq 2$ for each $j$; see Example \ref{example:GIBF}.
\end{remark}

\begin{theorem}\label{theo:misspec-determ}
Let $k$ be a TM kernel that satisfies Assumption~\ref{assump:SL} and $k^*$ be a tensor product kernel that satisfies Assumption~\ref{assump:smooth-kernel}.
Let $\{\SFY^*(\BFx):\BFx \in (0,1)^d\}$ be a zero mean GP with kernel $k^*$.
Suppose that the true response surface is a realization of $\SFY^*$, the simulation has no noise, and the design points $\{\BFx_1,\ldots,\BFx_n\}$  form a TSG.
Let $\widehat \SFY_n^{\mathsf{mis}}(\BFx)=\BFk^\intercal(\BFx)\BFK^{-1}\BFy^*$ be the misspecified kriging predictor with kernel $k$ and the TSG design, where $\BFy^*=(\SFY^*(\BFx_1),\ldots,\SFY^*(\BFx_n))^\intercal$.
Then, as $n\to\infty$,
\begin{equation}\label{eq:upper-bound-misspec}
   \sup_{\BFx\in(0,1)^d}\E[(\widehat \SFY_n^{\mathsf{mis}}(\BFx) - \SFY^*(\BFx))^2] = \CalO\left(n^{-2}(\log n)^{3(d-1)}\right).
\end{equation}
\end{theorem}

Note that the upper bound \eqref{eq:upper-bound-misspec} decays to zero at a rate that is substantially faster than the upper bound \eqref{eq:upper-bound-determ-TSG} in Theorem~\ref{theo:determ-TSG}.
This is somewhat surprising; it appears as if the convergence rate of kriging with TM kernels and TSG designs were benefiting positively from the model misspecification.
The seemingly counterintuitive result stems from the fact that, in general, optimal convergence rates for estimating unknown functions depend on the properties of the true function rather than on those of the postulated model.
For instance, the lower bound \eqref{eq:lower-bound-Matern}  does not depend on the smoothness parameter of the Mat\'ern kernel used for kriging but instead depends on the smoothness parameter of the GP that generates the true surface.

Clearly, the tensor product kernel $k^*$ in Theorem~\ref{theo:misspec-determ} has a higher degree of differentiability than the TM kernel $k$,
as
each component of the latter, having the form $p_j(x\wedge \cdot )q_j(x\vee \cdot)$, is non-differentiable.
Therefore, the true surface is significantly smoother than TMGP sample paths.
Conceivably, if one uses $k^*$ for kriging so that the model is correctly specified, then one might be able to prove an even faster rate than \eqref{eq:upper-bound-misspec}, possibly under a different experimental design than TSGs.
However, this is beyond the scope of the present paper, and we leave it to future research.

\subsection{Stochastic Simulation and Budget Allocation}

The rate results for deterministic simulation can be generalized to the stochastic setting as follows.

\begin{theorem}\label{theo:stoch-TSG}
Let $\{\SFY(\BFx):\BFx\in(0,1)^d\}$ be a zero mean TMGP with kernel $k$  that satisfies Assumption~\ref{assump:SL}.
Suppose that the true response surface is a realization of $\SFY$,
the simulation noise has variance $\sigma^2(\BFx)$, the design points $\{\BFx_1,\ldots,\BFx_n\}$  form a TSG, and the number of replications taken at each $\BFx_i$ is $m_i$.
Let $\widehat \SFY_n(\BFx)$ be the stochastic kriging predictor \eqref{eq:SK-predictor} with kernel $k$ and the TSG design.
If
$
(\log n)^d \max_{1\leq i\leq n} m_i^{-1/2}\sigma(\BFx_i) \to 0$
as $n\to\infty$, then
\begin{equation}\label{eq:upper-bound-stoch-TSG}
    \sup_{\BFx\in(0,1)^d}\E[(\widehat \SFY_n(\BFx) - \SFY(\BFx))^2] = \CalO\left(n^{-1}(\log n)^{2(d-1)} + (\log n)^d \max_{1\leq i\leq n} m_i^{-1/2}\sigma(\BFx_i)\right).
\end{equation}
\end{theorem}

\begin{theorem}\label{theo:misspec-stoch}
Let $k$ be a TM kernel that satisfies Assumption~\ref{assump:SL} and $k^*$ be a tensor product kernel that satisfies Assumption~\ref{assump:smooth-kernel}.
Let $\{\SFY^*(\BFx):\BFx \in (0,1)^d\}$ be a zero mean GP with kernel $k^*$.
Suppose that the true response surface is a realization of $\SFY^*$, the simulation noise has variance $\sigma^2(\BFx)$,  the design points $\{\BFx_1,\ldots,\BFx_n\}$  form a TSG, and the number of replications taken at each $\BFx_i$ is $m_i$.
Let $\widehat \SFY_n^{\mathsf{mis}}(\BFx)=\BFk^\intercal(\BFx)(\BFK+\BFSigma)^{-1}\bar\BFY^*$ be the misspecified stochastic kriging predictor with kernel $k$ and the TSG design, where $\bar\BFY^*$ denotes the vector composed of entries  $\SFY^*(\BFx_i)+m_i^{-1}\sum_{r=1}^{m_i}\epsilon_r(\BFx_i)$ for $i=1,\ldots,n$.
If $(\log n)^d \max_{1\leq i\leq n} m_i^{-1/2}\sigma(\BFx_i) \to 0$
as $n\to\infty$, then
\begin{equation}\label{eq:upper-bound-misspec-stoch}
\sup_{\BFx\in(0,1)^d}\E[(\widehat \SFY_n^{\mathsf{mis}}(\BFx) - \SFY^*(\BFx))^2] = \CalO\left(n^{-2}(\log n)^{3(d-1)}  + (\log n)^d\max_{1\leq i\leq n} m_i^{-1/2}\sigma(\BFx_i) \right).
\end{equation}
\end{theorem}

Comparing the above results with their deterministic counterparts  (Theorems~\ref{theo:determ-TSG} and \ref{theo:misspec-determ}), we can easily see that the MSE upper bounds \eqref{eq:upper-bound-stoch-TSG} and \eqref{eq:upper-bound-misspec-stoch} can both be decomposed into two parts, the function approximation error and the simulation error, thereby reflecting the bias--variance trade-off.
Intuitively, in order for the MSE to diminish, the number of replications at each design point needs to grow to infinity to reduce the simulation error to zero.
The upper bounds \eqref{eq:upper-bound-stoch-TSG} and \eqref{eq:upper-bound-misspec-stoch} further suggest that each $m_i$ needs to grow fast enough, as $n\to\infty$.

In a stochastic simulation experiment, one is usually constrained by a sampling budget $B$.
For simplicity, we assume that the sample cost is the same for all $\BFx$ and
write $B=\sum_{i=1}^n m_i$.
To gain insight into the implications of budget allocation,
we further assume homoscedasticity (i.e.,  $\sigma(\BFx)\approx \sigma$ for all $\BFx$) and an equal number of replications at the design points  (i.e., $m_i\approx m$ for all $i=1,\ldots,m$).
Then, $B=mn$
and the upper bound \eqref{eq:upper-bound-stoch-TSG} becomes
$n^{-1}(\log n)^{2(d-1)} +  \sigma m^{-1/2} (\log n)^d$,
which is minimized when the two summands are approximately equal, resulting in $m \approx (n/(\log n)^{d-2})^2$.
Further, for a fixed $d$,
the relation is approximately reduced to $m \approx n^2$ for large $n$.
Therefore, the budget allocation minimizing the upper convergence rate is roughly $m\approx B^{2/3}$ and $n\approx B^{1/3}$, leading to an MSE upper bound approximately of order $B^{-1/3}$.

The above heuristic analysis can be applied to the model misspecification case as well.
In this case, the asymptotically optimal budget allocation is such that $m\approx B^{4/5}$ and $n\approx B^{1/5}$, and the corresponding MSE upper bound is approximately of order  $ B^{-2/5}$.

\section{Fast Computation}\label{sec:computation}

\cite{Plumlee14}  develops fast algorithms for kriging of tensor product kernels and classical SG designs.
However, they are only valid if
(i) the design points must form a (complete) SG, and
(ii) the simulation model is deterministic, so that the matrix to invert is $\BFK$ rather than $(\BFK+\BFSigma)$.
In this section, we relax the above two restrictions
to allow design points to form a TSG and cover the case of stochastic simulation.

\subsection{Kernel Matrix Inversion}\label{sec:kernel-inv}

We develop a fast algorithm for exact computation of $\BFK^{-1}$ for TSGs in four steps, each analyzing a case that is more general than the last.
The four cases are: (i) one-dimensional grids, (ii) lattices, (iii) classical SGs, and (iv) TSGs,
resulting in Algorithms~\ref{alg:TM-1D}--\ref{alg:TM-TSG}, respectively.
The algorithm developed in an earlier step becomes a subroutine for the more general cases in later steps.

The computation of $\BFK^{-1}\BFk(\BFx)$ for classical SGs is an important building block for the inversion of the kernel matrix for TSGs.
It turns out that by taking advantage of special structures related to TM kernels, we may directly compute this vector
without going through the two-step approach that computes $\BFK^{-1}$ first and does matrix--vector multiplication later.
We therefore present Algorithms~\ref{alg:TM-1D}--\ref{alg:TM-SG} to include
both the methods for computing $\BFK^{-1}$ and that for $\BFK^{-1}\BFk(\BFx)$.

For the ease of presentation henceforth, given two sets of design points $\CalX'$ and $\CalX''$,  we use $k(\CalX', \CalX'')$ to denote the matrix consisting of entries $k(\BFx',\BFx'')$ for all $\BFx'\in\CalX'$ and $\BFx''\in\CalX''$.
With this notation, for example, we have $\BFK = k(\CalX,\CalX)$ and $\BFk(\BFx) = k(\CalX, \{\BFx\})$.

\subsubsection{Computation in One Dimension}

With $d=1$, kernel matrices for TM kernels are particularly tractable, with the inverse being specified explicitly in Proposition~\ref{prop:inverse},
which immediately suggests Algorithm~\ref{alg:TM-1D} for  computing $\BFK^{-1}$ and $\BFK^{-1}\BFk(x)$.

\begin{algorithm}[ht]
\SingleSpacedXI
\SetAlgoLined
\DontPrintSemicolon
\SetKwInOut{Input}{Input}
\SetKwInOut{Output}{Output}
\Input{TM kernel $k$ in one dimension, design points $x_1<\ldots<x_n$, and prediction point $x$ }
\Output{$\BFK^{-1}$ and $\BFK^{-1}\BFk(x)$}
\BlankLine
Initialize $\BFA \leftarrow \BFzero\in\Real^{n\times n}$ and $\BFb \leftarrow \BFzero \in \Real^{n \times 1}$ \;
Update the tridiagonal entries of $\BFA$ using \eqref{eq:K-inverse}\;
Search for $i^*\in\{0,1,\ldots,n\}$ such that $x\in [x_{i^*}, x_{i^*+1})$, where $x_0 = -\infty$ and $x_{n+1}=\infty$ \;
Update $\BFb_{i^*}$ and $\BFb_{i^*+1}$ using \eqref{eq:K-inverse-times-k}\;
Return $\BFK^{-1}\leftarrow \BFA$ and $\BFK^{-1}\BFk(x)\leftarrow \BFb$\;
\caption{Computing $\BFK^{-1}$ and $\BFK^{-1}\BFk(x)$ for One-Dimensional Grids}\label{alg:TM-1D}
\end{algorithm}

\subsubsection{Computation on Lattices}

In light of the tensor structure of TM kernels, it is straightforward to generalize the computation of $\BFK^{-1}$ and $\BFK^{-1}\BFk(\BFx)$ from one-dimensional grids to multidimensional lattices \citep{OHagan91}. In Algorithm~\ref{alg:TM-lattice},
$\BFK_j$ denotes the matrix composed of entries $k_j(x,  x')$ for all $x, x'\in \CalX_j$, $\BFk_j(x)$ denotes the vector composed of entries $k_j(x,  x')$ for all $  x' \in \CalX_j $, and
$\mathrm{vec}(\cdot)$ denotes the vectorization of a matrix.

\begin{algorithm}[ht]
\SingleSpacedXI
\SetAlgoLined
\DontPrintSemicolon
\SetKwInOut{Input}{Input}
\SetKwInOut{Output}{Output}
\Input{TM kernel $k$ in $d$ dimensions,  lattice design  $\CalX=\bigtimes_{j=1}^d \CalX_j$ with $\abs{\CalX_j}=n_j$, and prediction point $\BFx=(x_1,\ldots,x_d)$ }
\Output{$\BFK^{-1}$ and $\BFK^{-1}\BFk(\BFx)$}
\BlankLine
\For{$j\leftarrow 1$ \KwTo $d$}{
Compute $\BFK_j^{-1} \in \Real^{n_j\times n_j}$ and $\BFK_j^{-1}\BFk_j(x_j)\in \Real^{n_j\times 1}$ via Algorithm~\ref{alg:TM-1D} with inputs $(k_j,\CalX_j, x_j)$ \;
}
Return $\BFK^{-1} = \bigotimes_{j=1}^d \BFK_j^{-1}$ and
$\BFK^{-1}\BFk(\BFx) = \mathrm{vec}\left(\bigotimes_{j=1}^d \BFK_j^{-1}\BFk_j(x_j) \right)$ \;
\caption{Computing $\BFK^{-1}$ and $\BFK^{-1}\BFk(\BFx)$ for Lattices}\label{alg:TM-lattice}
\end{algorithm}

\subsubsection{Computation on Classical Sparse Grids}

Algorithms for computing quantities involved in kriging with SG designs have been developed in \cite{Plumlee14} for general tensor product kernels.
We now tailor them to our context, resulting in Algorithm~\ref{alg:TM-SG}, which we briefly explain below;  see Appendix B of \cite{Plumlee14} for the proof of its validity.

Let $\CalX_{j,l} = \{\SFc_{l,i}:1\leq i\leq 2^l-1\}$ with $\SFc_{l,i} = i\cdot 2^{-l}$ being a component design defined in \eqref{eq:dyadic}.
For any multi-index $\BFl\in\NatInt^d$, $\CalX_{\BFl}^{\mathsf{FG}} \coloneqq \bigtimes_{j=1}^d \CalX_{j, l_j} $ defines a full grid (i.e., lattice).
By its definition in \eqref{eq:SG}, the classical SG of level $\tau$ is expressed as $\CalX^{\mathsf{SG}}_\tau = \bigcup_{\abs{\BFl}\leq \tau+d-1} \CalX_{\BFl}^{\mathsf{FG}}$,
which drives the updating schemes in Algorithm~\ref{alg:TM-SG}.
Let
$\BFK_{\BFl} \coloneqq k(\CalX_{\BFl}^{\mathsf{FG}},\CalX_{\BFl}^{\mathsf{FG}})$
and $\BFk_\BFl(\BFx) \coloneqq k(\CalX_{\BFl}^{\mathsf{FG}}, \{\BFx\}) $.
In general, for any matrix $\BFA\in\Real^{n\times n}$ with $n=\abs{\CalX^{\mathsf{SG}}_\tau}$ whose entries are indexed by $(\BFx,\BFx')$ for $\BFx,\BFx'\in\CalX_\tau^{\mathsf{SG}}$, we may write $\BFA_\BFl$ to denote the part of $\BFA$ having entries indexed by $(\BFx,\BFx')$ for all $\BFx,\BFx'\in \CalX_\BFl^{\mathsf{FG}}$.
Likewise, we may define $\BFb_\BFl$ for any vector $\BFb\in\Real^n$.
In  each iteration $\BFl$ of Algorithm~\ref{alg:TM-SG}, the part of $\BFK^{-1}$ that corresponds to $\CalX_\BFl^{\mathsf{FG}}$ is updated; so is $\BFK^{-1}\BFk(\BFx)$.
Note that both $\BFK_{\BFl}$ and $\BFk_\BFl(\BFx)$ are defined on a lattice.
Therefore, they can be computed via Algorithm~\ref{alg:TM-lattice}.

\begin{algorithm}[ht]
\SingleSpacedXI
\SetAlgoLined
\DontPrintSemicolon
\SetKwInOut{Input}{Input}
\SetKwInOut{Output}{Output}
\Input{TM kernel $k$ in $d$ dimensions,  classical SG design  $\CalX_\tau^{\mathsf{SG}}$ with size $n$, and prediction point $\BFx$ }
\Output{$\BFK^{-1}$ and $\BFK^{-1}\BFk(\BFx)$}
\BlankLine
Initialize $\BFA\leftarrow \BFzero \in\Real^{n\times n}$ and $\BFb\leftarrow \BFzero\in\Real^{n\times 1}$ \;
\For{all $\BFl\in\NatInt^d$ with $ \tau\leq \abs{\BFl} \leq \tau+d-1$ }{
Compute $\BFK_\BFl^{-1}$ and $\BFK_\BFl^{-1}\BFk_\BFl(\BFx)$ via Algorithm~\ref{alg:TM-lattice} with inputs   $(k,\bigtimes_{j=1}^d \CalX_{j, l_j}, \BFx) $ \;
Update $\BFA_\BFl$ and $\BFb_\BFl$ via
\vspace{-2ex}
\begin{align}
\BFA_\BFl \leftarrow{}& \BFA_\BFl + (-1)^{\tau+d-1-\abs{\BFl}} \binom{d-1}{\tau+d-1-\abs{\BFl}} \BFK_\BFl^{-1} \label{eq:update-A}\\[0.5ex]
\BFb_\BFl \leftarrow{}& \BFb_\BFl + (-1)^{\tau+d-1-\abs{\BFl}} \binom{d-1}{\tau+d-1-\abs{\BFl}} \BFK_\BFl^{-1} \BFk_\BFl(\BFx)\label{eq:update-b}
\end{align}
\vspace{-2ex}
}
Return $\BFK^{-1} \leftarrow \BFA$ and
$\BFK^{-1}\BFk(\BFx) \leftarrow \BFb$ \;
\caption{Computing $\BFK^{-1}$ and $\BFK^{-1}\BFk(\BFx)$ for Classical Sparse Grids}\label{alg:TM-SG}
\end{algorithm}

\subsubsection{Computation on Truncated Sparse Grids}

Being part of a larger classical SG by definition, a TSG can be viewed as an ``incomplete'' SG.
The lack of completeness breaks the tensor structure of SGs, meaning that a TSG cannot be expressed as a union of tensor products of one-dimensional designs in the same way as SGs can in \eqref{eq:SG}.
This is the key reason why Algorithm~\ref{alg:TM-SG} and the algorithms proposed in \cite{Plumlee14} cease to work for TSGs.

To address the issue, we take advantage of the fact that a TSG is the union of a classical SG and a set of design points belonging to the classical SG of the next level, that is, $\CalX_n^{\mathsf{TSG}} = \CalX_\tau^{\mathsf{SG}}\cup \CalA_{\tilde n}$ with $\CalA_{\tilde n}\subseteq \CalD_{\tau+1} = \CalX_{\tau+1}^{\mathsf{SG}}\setminus \CalX_\tau^{\mathsf{SG}}$.
In other words, the ``incompleteness'' only occurs for the highest level of the SG involved.
It can be shown that TMGPs possess a nice conditional independence structure that specifies that
the samples taken at design points belonging to $\CalD_{\tau+1}$ are conditionally independent given the observations made at all the design points of $\CalX_\tau^{\mathsf{SG}}$.
This conditional independence structure then induces the sparseness of $\BFK^{-1}$.
In particular, part (i) of Theorem~\ref{theo:K-inverse} asserts that $\BFK^{-1}$ can be expressed as a $2\times 2$ block matrix in which the bottom-right block is a diagonal submatrix and the other blocks are all sparse as well.
We characterize the sparse structure explicitly in part (ii) of Theorem~\ref{theo:K-inverse}, providing sufficient conditions for entries of $\BFK^{-1}$ to be zero.
Knowing which entries are guaranteed to be zero, we only need to compute and store those that are likely to be nonzero, which yields substantial savings in both computational time and matrix storage.
As the computational efficiency for computing $\BFK^{-1}$ is essentially  determined by its sparsity,
we give an estimate of the proportion of nonzero entries in $\BFK^{-1}$ in part (iii) of Theorem~\ref{theo:K-inverse}.

\begin{theorem}
\label{theo:K-inverse}
Let $k(\BFx,\BFx')=\prod_{j=1}^d p_j(x_j\wedge x_j') q_j(x_j\vee x_j')$ be a TM kernel that satisfies Assumption~\ref{assump:SL} and $n\geq 3$. Suppose that the design points $\{\BFx_1,\ldots,\BFx_n\}$ form a TSG
$\CalX_n^{\mathsf{TSG}}=\CalX_\tau^{\mathsf{SG}}\cup \CalA_{\Tilde{n}}$.
\begin{enumerate}[label=(\roman*)]
\item
The inverse of the kernel matrix on $\CalX_n^{\mathsf{TSG}}$ can be expressed as the following block matrix,
\begin{equation}\label{eq:K-inverse-TSG}
\BFK^{-1} =
\begin{pNiceMatrix}[first-row, columns-width=auto]
\scriptstyle{|\CalX_\tau^{\mathsf{SG}}|\;\,\mathrm{dim.}} & \scriptstyle{\Tilde{n} \;\,\mathrm{dim.}} \\[2pt]
\BFE & - \BFB \BFD \\[3pt]
- \BFD \BFB^\intercal &  \BFD
\end{pNiceMatrix},
\end{equation}
where
$\BFE = \BFA^{-1} + \BFB \BFD \BFB^\intercal$,
$\BFA = k(\CalX_\tau^{\mathsf{SG}},\CalX_\tau^{\mathsf{SG}})$,
$\BFB = \BFA^{-1} k(\CalX_\tau^{\mathsf{SG}},\CalA_{\tilde n})$,
and $\BFD$ is a $\tilde n\times \tilde n$ diagonal matrix the diagonal entries of which are given by
\begin{equation}\label{eq:cond-var}
\BFD_{(\BFl,\BFi),(\BFl,\BFi)} =
\prod_{j=1}^d \left(\frac{\SFp_{j, l_j,i_j+1} \SFq_{j,l_j, i_j-1} - \SFp_{j,l_j,i_j-1} \SFq_{j,l_j,i_j+1}}{\left(\SFp_{j,l_j,i_j} \SFq_{j,l_j,i_j-1} - \SFp_{j,l_j,i_j-1}  \SFq_{j,l_j,i_j} \right) \left(\SFp_{j,l_j,i_j+1}  \SFq_{j,l_j,i_j} - \SFp_{j,l_j,i_j} \SFq_{j,l_j,i_j+1}\right) }\right)
\end{equation}
for all $(\BFl,\BFi)$ such that $\BFc_{\BFl,\BFi} \in \CalA_{\Tilde{n}}$,
where $\SFp_{j,l,i}= p_j(\SFc_{l, i})$ and
$\SFq_{j,l,i}= q_j(\SFc_{l, i})$ for all $j$, $l$, and $i$.
\item
For each $(\BFl,\BFi)$, define a hyperrectangle
$
\CalR_{(\BFl,\BFi)} \coloneqq \bigtimes_{j=1}^d (\SFc_{l_j, i_j-1}, \SFc_{l_j, i_j+1})
$.
Suppose that one of the following conditions is satisfied:
\begin{enumerate}[label=(\alph*)]
    \item   $\BFc_{\BFl',\BFi'}, \BFc_{\BFl'',\BFi''}\in\CalX_{\tau+1}^{\mathsf{SG}}$, and $\BFc_{\BFl',\BFi'} \neq \BFc_{\BFl'',\BFi''}$;
    \item $\BFc_{\BFl',\BFi'}\in\CalX_\tau^{\mathsf{SG}}$, $\BFc_{\BFl'',\BFi''}\in\CalA_{\tilde n}$, and  $\BFc_{\BFl'',\BFi''}\notin\CalR_{\BFl',\BFi'}$;
    \item
    $\BFc_{\BFl',\BFi'}, \BFc_{\BFl'',\BFi''}\in\CalX_\tau^{\mathsf{SG}}$,
    \begin{equation}\label{eq:no-overlap-1}
    \Bigl|i_j'\cdot 2^{l_j'\vee l_j''-l_j'} - i_j''\cdot 2^{l_j'\vee l_j'' - l_j''}\Bigr| \geq 2,\quad\mbox{for some } j=1,\ldots,d,
    \end{equation}
    and
    \begin{equation}\label{eq:no-overlap-2}
    \BFc_{\BFl,\BFi}\notin  \CalR_{\BFl',\BFi'}\cap\CalR_{\BFl'',\BFi''},\quad \mbox{for all }\BFc_{\BFi,\BFl}\in\CalA_{\tilde n}.
    \end{equation}
\end{enumerate}
Then,
$(\BFK^{-1})_{(\BFl',\BFi'),(\BFl'',\BFi'')} = (\BFK^{-1})_{(\BFl'',\BFi''),(\BFl',\BFi')} = 0$.
\item
Both $\BFB$ and $\BFB\BFD$ have $\CalO(\tilde n \log(|\CalX_\tau^{\mathsf{SG}}|)^{d-1})$ nonzero entries,
and $\BFE$ has $\CalO(|\CalX_\tau^{\mathsf{SG}}| + \tilde n\tau^{2d})$ nonzero entries.
The density of $\BFK^{-1}$ (i.e., the proportion of nonzero entries therein) is $\CalO(n^{-1}(\log n)^{2d})$.
\end{enumerate}
\end{theorem}

The three conditions (a), (b), and (c) in part (ii) of Theorem~\ref{theo:K-inverse} correspond to different blocks of $\BFK^{-1}$ in \eqref{eq:K-inverse-TSG}, namely, the bottom-right, off-diagonal, and top-left blocks,
respectively.
Each block can be computed efficiently.
First, the diagonal entries of $\BFD$ can be computed via \eqref{eq:cond-var}.
Second, $\BFA$ is simply the kernel matrix on the classical SG $\CalX_{\tau}^{\mathsf{SG}}$,  and each column of $\BFB$ is
of the form $\BFA^{-1} k(\CalX_{\tau}^{\mathsf{SG}}, \{\BFx\})$ for $\BFx\in \CalA_{\Tilde{n}}$,
so both $\BFA^{-1}$ and $\BFB$ can be computed via Algorithm~\ref{alg:TM-SG}.
It is also easy to compute $\BFB\BFD\BFB^\intercal$ because $\BFD$ is diagonal and $\BFB$ is a sparse matrix by part (iii) of Theorem~\ref{theo:K-inverse}.
This gives rise to  Algorithm~\ref{alg:TM-TSG}.

\begin{algorithm}[ht]
\SingleSpacedXI
\SetAlgoLined
\DontPrintSemicolon
\SetKwInOut{Input}{Input}
\SetKwInOut{Output}{Output}
\Input{TM kernel $k$ in $d$ dimensions,  TSG design  $\CalX_n^{\mathsf{TSG}} = \CalX_\tau^{\mathsf{SG}}\cup \CalA_{\Tilde{n}}$, and prediction point $\BFx$ }
\Output{$\BFK^{-1}$}
\BlankLine
With $\BFA\coloneqq k(\CalX_\tau^{\mathsf{SG}}, \CalX_\tau^{\mathsf{SG}})$, compute $\BFA^{-1}$ via Algorithm~\ref{alg:TM-SG} with inputs $(k, \CalX_\tau^{\mathsf{SG}})$\;
Initialize $\BFB \leftarrow \BFzero \in\Real^{|\CalX_{\tau}^{\mathsf{SG}}| \times \Tilde{n} }$ and $\BFD\leftarrow \BFzero \in\Real^{\Tilde{n}\times \Tilde{n}} $\;
\For{all $\BFx\in \CalA_{\Tilde{n}}$}{
Compute $\BFb \leftarrow \BFA^{-1}k(\CalX_{\tau}^{\mathsf{SG}}, \{\BFx\})$ via Algorithm~\ref{alg:TM-SG} with inputs $(k, \CalX_\tau^{\mathsf{SG}}, \BFx)$ \;
Update the $\BFx$-th column of $\BFB$ to $\BFb$\;
Update the $\BFx$-th diagonal entry of $\BFD$ via  \eqref{eq:cond-var} \;
}
Compute $\BFK^{-1}$ via \eqref{eq:K-inverse-TSG} \;
\caption{Computing $\BFK^{-1}$ for Truncated Sparse Grids}\label{alg:TM-TSG}
\end{algorithm}

\begin{remark}
The validity of Theorem~\ref{theo:K-inverse} relies on two critical conditions.
First, the kernel is a TM kernel, so Algorithm~\ref{alg:TM-TSG} does not work for general tensor product kernels.
Second, the experimental design is a TSG, so the algorithm does not work for ``incomplete'' SGs in general.
\end{remark}

\subsection{Computing Stochastic Kriging with TM Kernels and TSGs}

As fast computation of $\BFK^{-1}$ is made available by Algorithm~\ref{alg:TM-TSG} and $\BFSigma$ is a diagonal matrix,
we can apply the Woodbury matrix identity, a standard trick in machine learning \citep{BinoisGramacyLudkovski18,LuRudiBorgonovoRosasco20}, to obtain fast computation of $(\BFK+\BFSigma)^{-1}$.

Specifically, the Woodbury matrix identity asserts that
\[(\BFK+\BFSigma)^{-1} = \BFSigma^{-1} - \BFSigma^{-1}(\BFK^{-1} + \BFSigma^{-1})^{-1}\BFSigma^{-1}.\]
Note that $(\BFK^{-1}+\BFSigma^{-1})$ is a sparse matrix, so computing its inverse can benefit from sparse linear algebra.
Indeed, the specific sparse structure of $\BFK^{-1}$ in \eqref{eq:K-inverse-TSG} permits fast Cholesky decomposition $(\BFK^{-1} + \BFSigma^{-1}) = \BFL\BFL^\intercal$, where $\BFL$ is a lower triangular matrix.
Then, the stochastic kriging predictor \eqref{eq:SK-predictor} and its MSE \eqref{eq:SK-MSE} can be expressed as
\begin{align}
\widehat \SFY(\BFx) ={}& \BFk^\intercal(\BFx)\BFSigma^{-1} \bar\BFY -\BFb_1^\intercal\BFb_2, \label{eq:SK-pred-Woodbury} \\[0.5ex]
\MSE[\widehat \SFY(\BFx)]={}& k(\BFx,\BFx) - \BFk^\intercal(\BFx) \BFSigma^{-1} \BFk(\BFx) + \BFb_1^\intercal\BFb_1, \label{eq:SK-MSE-Woodbury}
\end{align}
where $\BFb_1$ and $\BFb_2$ are the solutions of the following systems of linear equations,
\begin{equation}\label{eq:linear-sys}
\BFL^\intercal \BFb_1 = \BFSigma^{-1}\BFk(\BFx) \qq{and}
\BFL^\intercal \BFb_2 = \BFSigma^{-1}\bar{\BFY}.
\end{equation}
We summarize the above discussion in Algorithm~\ref{alg:SK}.

\begin{algorithm}[ht]
\SingleSpacedXI
\SetAlgoLined
\DontPrintSemicolon
\SetKwInOut{Input}{Input}
\SetKwInOut{Output}{Output}
\Input{TM kernel $k$ in $d$ dimensions,  TSG design  $\CalX_n^{\mathsf{TSG}}$, observations $\bar{\BFY}$,  covariance matrix $\BFSigma$, and prediction point $\BFx$ }
\Output{$\widehat \SFY(\BFx)$ and $\MSE[\widehat \SFY(\BFx)]$}
\BlankLine
Compute $\BFK^{-1}$ via Algorithm~\ref{alg:TM-TSG} \;
Perform the Cholesky decomposition $\BFL \BFL^\intercal = \BFK^{-1}+\BFSigma^{-1}$\;
Solve the equations \eqref{eq:linear-sys} for $\BFb_1$ and $\BFb_2$\;
Compute $\widehat \SFY(\BFx)$ and $\MSE[\widehat \SFY(\BFx)]$ via \eqref{eq:SK-pred-Woodbury} and \eqref{eq:SK-MSE-Woodbury}, respectively
\caption{Computing Stochastic Kriging with TM Kernels and TSGs}\label{alg:SK}
\end{algorithm}

\begin{remark}
In addition to the benefit of fast Cholesky decomposition,
the use of the Woodbury matrix identity yields substantial savings in matrix storage in the computation of stochastic kriging.
Instead of storing the dense matrix $\BFK$, it suffices to store the sparse matrix $\BFK^{-1}$ (the locations of its possible nonzero entries are given by part (ii) of Theorem~\ref{theo:K-inverse}).
Moreover, the savings in matrix storage increase as $n$ increases, because the density of $\BFK^{-1}$ is $\CalO(n^{-1}(\log n)^{2d})$ by part~(iii) of Theorem~\ref{theo:K-inverse}.
\end{remark}

\section{Numerical Experiments}\label{sec:numerical}
We conduct extensive numerical experiments to assess both the prediction capability and the computational efficiency of our methodology, with a focus on high-dimensional problems.
We present three sets of experiments---artificial test functions in up to 500 dimensions in Section~\ref{sec:test-func},
the expected revenue of a production line problem in up to 100 dimensions in Section~\ref{sec:prod-line},  and the optimal value of a linear program as a function of 16675 relevant parameters in Section~\ref{sec:stocfor}.

\subsection{General Setup}\label{sec:setup}

In each experiment, the TM kernel we use is the tensor product of the Laplace kernel, that is,
\begin{equation}\label{eq:TM-Laplace}
k(\BFx,\BFx') = \prod_{j=1}^d \exp\left(- |x_j- x'_j|\right) = \exp\biggl(-\sum_{j=1}^d |x_j- x'_j|\biggr).
\end{equation}
We have also tried other TM kernels, for example, replacing the Laplace kernel with the Brownian motion kernel, and the numerical results are similar.
Moreover, we use RTSG designs for simplicity.

We assess the prediction accuracy via the
root mean squared error (RMSE).
First, we randomly select a set of prediction points of size $n_{\text{pred}}$ from a $d$-dimensional lattice with four equally spaced points in each dimension, that is, a lattice of size $4^{d}$. (We do not perform random selection from the entire design space because the points would mostly be sampled from areas close to the boundary of the design space, making it difficult to differentiate the prediction performance of the competing methods.)
Then, we calculate the RMSE with respect to the prediction points $\{\BFp_j\}_{j=1}^{n_{\text{pred}}}$:
\[
\widehat{\text{RMSE}}=\biggl(\frac{1}{n_{\text{pred}}}\sum_{i=1}^{n_{\text{pred}}}[\widehat{\SFM}(\BFp_i)-y(\BFp_i)]^2\biggr)^{1/2},
\]
where $\widehat{\SFM}(\BFp_i)$ is the value of a simulation metamodel predictor $\widehat{\SFM}$  at $\BFp_i$.

We repeat each experiment $R$ times and call each a macro-replication from which we obtain  $\widehat{\text{RMSE}}_r$, $r=1,\ldots,R$.
To assess the error in estimating the RMSE, we calculate 95\% confidence intervals  (CIs) with a
half-length
\[\frac{1.96}{\sqrt{R}}\biggl[\frac{1}{R-1}\sum_{r=1}^R \biggl(\widehat{\text{RMSE}}_r - \frac{1}{R}\sum_{s=1}^R\widehat{\text{RMSE}}_s \biggr)^2\biggr]^{1/2} .
\]
In each macro-replication $r$, we also record the elapsed time to complete the prediction over all the prediction points. Then, we report the average time over $R$ macro-replications.

In each experiment, we compare our approach (the TM kernel \eqref{eq:TM-Laplace} plus the RTSG design), denoted by \texttt{TM-RTSG} in the sequel, against the following methods having different combinations of kernels and designs.

\begin{enumerate}[label=(\roman*)]
    \item \texttt{M-LHD}: the Mat\'ern($\sfrac{5}{2}$) kernel plus the maximin LHD that maximizes the minimum distance between points \citep{van2007maximin}.
    \cite{TuoWang20} show that the maximin LHD design is asymptotically optimal with respect to the $L_p$ norm for kriging with Mat\'ern kernels.
    \item \texttt{M-RTSG}: the Mat\'ern($\sfrac{5}{2}$) kernel plus the RTSG design.
     \item \texttt{TM-LHD}: the TM kernel plus the maximin LHD.
     \item \texttt{FK}: Fast Kriging \citep{LuRudiBorgonovoRosasco20} is a recent development of the Nystr\"om method. The number of Nystr\"om centers is set to be $0.1n$. We use the Gaussian kernel, and the design points are uniformly drawn from  the input domain.
      \item \texttt{RFF}: Random Fourier Feature \citep{rahimi2007random} is a classical random features method; see \cite{LiuHuangChenSuykens22} for a recent survey. We use random Fourier components generated from the Gaussian kernel. The number of random Fourier components is set to be $0.1n$. The experimental design is set to be the RTSG design.
\end{enumerate}

Both the sample efficiency analysis in Section~\ref{sec:sample} and the algorithms in Section~\ref{sec:computation} require the \emph{joint} use of TM kernels and TSG designs.
Thus, Algorithm~\ref{alg:SK} does not apply to \texttt{M-LHD}, \texttt{M-RTSG}, or \texttt{TM-LHD}.
Instead, generic matrix inversion algorithms are used in their  implementations.

\texttt{FK} and \texttt{RFF} represent two popular classes of approximation methods---the Nystr\"om method and the random feature method, respectively---for handling large datasets  in machine-learning literature (see Section~\ref{sec:apprx}).
The performances of these methods depend on the choice of the rank of the approximating matrix. (For \texttt{FK}, this rank equals the number of Nystr\"om centers, while for \texttt{RFF} it equals the number of random Fourier components.)
Lowering the rank improves the computational efficiency;
meanwhile, doing so generally reduces the prediction accuracy of these approximation methods, although it is not always the case possibly due to the potential benefit of  \emph{implicit regularization} \citep{JacotSimsekSpadaroHonglerGabriel20,FanuelSchreursSuykens21}.
Having tried different proportions of the sample size as the rank, we find that overall, $0.1n$ yields a good trade-off between computational efficiency and prediction accuracy
for both \texttt{FK} and \texttt{RFF} in our numerical experiments.
(Setting the rank to be a higher value such as $0.3n$ would lead to similar prediction accuracy but significantly reduce the computational efficiency.)

All the experiments are implemented in \textsc{Matlab} (version 2018a) on a laptop computer with macOS, 3.3 GHz Intel Core i5 CPU, and 8 GB of RAM (2133 Mhz).
The \textsc{Matlab} codes of the competing approaches are all publicly available;
the code of
\texttt{M-LHD}, \texttt{M-RTSG}, and \texttt{TM-LHD} is from \cite{AnkenmanNelsonStaum10}  (available at \url{users.iems.northwestern.edu/~nelsonb/SK}), the code of \texttt{FK} is from  \cite{LuRudiBorgonovoRosasco20} (available at \url{github.com/LuXuefei/FastKriging}), and the code of \texttt{RFF} is  from \cite{LiuHuangChenSuykens22} (available at \url{www.lfhsgre.org}).

\subsection{Test Functions} \label{sec:test-func}
\begin{figure}[ht]
\FIGURE{
\includegraphics[width=0.8\textwidth]{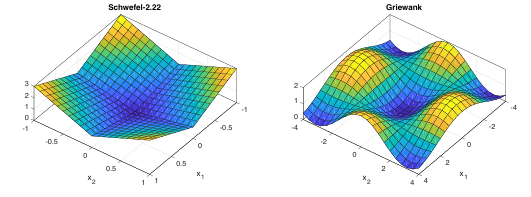}
}
{The Schwefel-2.22 and Griewank Functions in Two Dimensions. \label{fig:test-func-surfaces}}
{}
\end{figure}

In this section, we use the following two  test functions in varying dimensions ($d=20$, $100$, and $500$) to determine the sample and computational efficiency of our methodology:
\begin{align*}
        &y_{\text{Schwefel-2.22}}(\BFx)=\sum_{j=1}^d|x_j|+\prod_{j=1}^d|x_j|,\quad  \BFx\in(-1 ,1)^d, \\
        &y_{\text{Griewank}}(\BFx)=\sum_{j=1}^d\frac{x_j^2}{4000}-\prod_{j=1}^d\cos(\frac{x_j}{\sqrt{t}})+1,\quad \BFx\in(-4 ,4)^d.
\end{align*}
These two functions have different smoothness.
The two-dimensional versions of these functions are shown in Figure~\ref{fig:test-func-surfaces}.
The {Schwefel-2.22} function is non-differentiable, having ``wedges'' on the surface, due to the presence of the absolute value functions in its definition, whereas
the Griewank function is infinitely differentiable.

We investigate the prediction performance and the computational cost of each method as the sampling budget and dimension increase.
We construct a set of prediction points of size $n_{\text{pred}}=1000$, and
compute $\widehat{\text{RMSE}}$ of the predictor of each method, its associated 95\% CI, and the average time to complete the $n_{\text{pred}}$ predictions based on $R=100$ macro-replications, as detailed in Section~\ref{sec:setup}.

\begin{figure}[ht]
\FIGURE
{
\includegraphics[width=\textwidth]{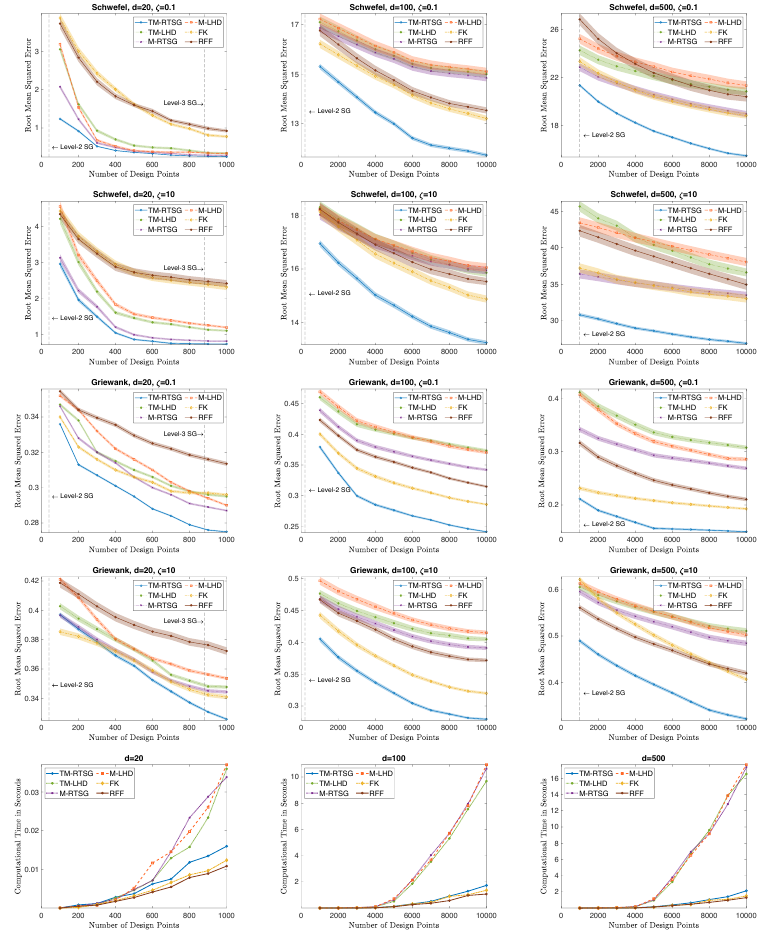}
}
{Prediction Errors and Computational Time for the Schewefel-2.22 and Griewank  Functions. \label{fig:test_func}}
{The three columns correspond to $d=20$, $100$, and $500$, respectively. The shaded areas represent 95\% CIs. The black dashed vertical lines indicate the number of design points for which RTSG is
identical to a classical SG.}
\end{figure}

We consider the case of noisy observations by adding to the test function value $y(\BFx)$ a zero mean Gaussian random variable with standard deviation that is proportional to $y(\BFx)$. That is, the observations are heteroscedastic Gaussian random variables,
$Y_r(\BFx_i)\sim \mathcal{N}\big(y(\BFx_i), \zeta y^2(\BFx_i)\big)$, for all $r=1,\ldots,m_i$ and $i=1,\ldots,n$.
The experiments are run under two scenarios with different noise levels---a low-noise scenario with $\zeta=0.1$ and a high-noise scenario with $\zeta=10$.
We set $m_i\equiv 10$ for all $i$ and use the estimated variance in stochastic kriging.
The sampling budget is set as follows. We set $n=100, 200, \ldots,1000$ for $d=20$ and
set $n=1000, 2000, \ldots,10000$ for $d=100,500$.
The experiment results are shown in  Figures~\ref{fig:test_func}.

First, our approach \texttt{TM-RTSG}, in general, outperforms the others significantly in prediction accuracy.
Not only does
\texttt{TM-RTSG} yield the lowest $\widehat{\text{RMSE}}$ in almost all the cases---regardless of the test function, noise level, dimensionality, or sample size---that are considered,
but the corresponding CIs of $\widehat{\text{RMSE}}$ also have the  smallest widths.
Because the CIs are calculated via multiple macro-replications in each of which a random set of prediction points are sampled,
this suggests that the predictions given by \texttt{TM-RTSG} are more stable than the competing approaches.
In particular, the fact that \texttt{TM-RTSG} has a significantly higher prediction accuracy than \texttt{TM-LHD}, \texttt{M-RTSG}, and \texttt{M-LHD}---which results from different combinations of kernels and experimental designs---demonstrates that TM kernels and TSG designs complement each other and that both contribute substantially to the superior performance of \texttt{TM-RTSG}.

Second, the advantage  of \texttt{TM-RTSG} in prediction accuracy increases as the dimensionality $d$ grows.
In relatively low dimensions ($d=20$), the RMSEs for \texttt{TM-RTSG} may be comparable to those of some of the other methods.
For example, when the test function is Schwefel-2.22, the performances of \texttt{TM-RTSG} and \texttt{M-RTSG} are similar (see the first two subfigures of the first column in Figure~\ref{fig:test_func}).
However, the prediction accuracy of \texttt{TM-RTSG} is dominant in high dimensions ($d=100, 500$).
This is consistent with our theoretical analysis of the prediction errors of \texttt{TM-RTSG} in Section~\ref{sec:sample},
which asserts that its sample complexity grows very slightly in the dimensionality.

Third, other things being equal, the difference in $\widehat{\text{RMSE}}$ between \texttt{TM-RTSG} and the other approaches is generally greater for the Schwefel-2.22 function than for the Griewank function.
For example, consider the setup with $d=20$ and $\zeta=10$ (see the second and fourth subfigures of the first column in Figure~\ref{fig:test_func}).
In this setup,
\texttt{TM-RTSG} performs the best when predicting the Schwefel-2.22 function for all sample sizes,
whereas it is outperformed by other approaches when predicting the Griewank function using a small sample size $n\leq 300$.
This difference in prediction accuracy may be attributed to the different levels of smoothness of the two test functions; the former has non-differentiable regions, whereas the latter is infinitely differentiable.

Last, the bottom row in Figure~\ref{fig:test_func}
shows the average computational time of each approach in predicting the two test functions in different dimensions.
Compared with the three approaches that are not based on matrix approximations (i.e., \texttt{TM-LHD}, \texttt{M-RTSG}, and \texttt{M-LHD}), the advantage of \texttt{TM-RTSG} is clear, especially when the dimensionality is high and the sample size is large.
The stark difference stems from the careful design of the algorithm that leverages the sparseness of the inverse kernel matrix associated with TM kernels and RTSG designs. We can see from Theorem~\ref{theo:K-inverse} that the sparseness is increasing as the sample size grows, with the proportion of nonzero entries decaying to zero at the rate $n^{-1}(\log n)^{2d}$. In contrast, \texttt{TM-LHD}, \texttt{M-RTSG}, and \texttt{M-LHD} all suffer heavily from the numerical inversion of dense matrices.

However, compared with the two approximation approaches (i.e., \texttt{FK} and \texttt{RFF}), the computational efficiency of \texttt{TM-RTSG} is lower but only by a small margin, especially when the sample size is large.
This is not surprising because both \texttt{FK} and \texttt{RFF} exploit low-rank approximations to accelerate matrix inversion.
However, the acceleration in computation is achieved at the cost of prediction accuracy.
Indeed, the $\widehat{\text{RMSE}}$ associated with \texttt{TM-RTSG} is markedly lower than that  associated with both \texttt{FK} and \texttt{RFF} in almost all cases.
In a nutshell, \texttt{TM-RTSG} achieves a much higher prediction accuracy with a slightly lower computational efficiency than the two approximation approaches.

\subsection{A Production Line Problem}\label{sec:prod-line}
In this section, we consider a production line problem from \cite{BuchholzThummler05}.
We consider production line queueing systems with $d=20$, $50$ and $100$ tandem servers, respectively.
Each server has exponentially distributed service times.
The response surface of interest is the expected revenue of the production line as a function of the vector of service rates $\BFx\in(0,2)^d$.

Parts arrive to the first server of the production line following a Poisson process with rate $\lambda$.
The parts are processed at each queue on a first-come-first-serve basis.
All servers have a finite capacity $K$.
Upon completion of the service at server $i$, a part will be transferred to server $i+1$ if there is room at the next server.
Otherwise, if there are already $K$ parts at server $i+1$,
server $i$ is said to be blocked, and the part being served at server $i$ cannot leave even if its service is completed.
The throughput of the production line, denoted by $\mathsf{ Th}(\BFx)$, is defined as the number of parts leaving the last server within a given time horizon $T$.
The expected revenue is defined as
\[y(\BFx)=\E \left[\frac{r\cdot\mathsf{ Th}(\BFx)}{c_0+\BFc^\intercal \BFx}\right]. \]
The parameters are specified as follows:
\begin{enumerate}[label=(\roman*)]
    \item The design space is $\ScrX = (0, 2)^d$.
    \item Capacity $K=10$, unit $r=2\times 10^5$, and time horizon $T=10d$.
    \item The vector $\BFc=(c_1,c_2,\ldots,c_d)$ with $c_i=i$ denoting the cost to run server $i$ and the initial cost $c_0=1$.
\end{enumerate}

In this example, the simulation noise has an unknown variance.
We run $m=10$ replications at each design point during the training process to
estimate the variance $\sigma^2(\BFx)$ and use the sample estimate in stochastic kriging.
We construct the set of prediction points of size $n_{\text{pred}}=1000$ as instructed in Section~\ref{sec:setup}.
Because $y(\BFx)$ has no closed form,
for each prediction point we run 100 replications to compute the ``true'' value of the response there.
Moreover, we run the experiments for two values of the arrival rate, $\lambda=0.5$ and $\lambda=2$.
The variances of the simulation outputs  for the former case are lower than those for the latter.
Therefore, similar to Section~\ref{sec:test-func},
we call them the low-noise scenario and the high-noise scenario, respectively.
For $d=20$, $50$, and $100$, we set the sampling budget to be $n=100, 200, \ldots, 1000$, $n=500, 1000, \ldots, 5000$, and $n=1000, 2000, \ldots, 10000$, respectively.
We set the number of macro-replications to be $R=30$.

\begin{figure}[ht]
\FIGURE
{
\includegraphics[width=\textwidth]{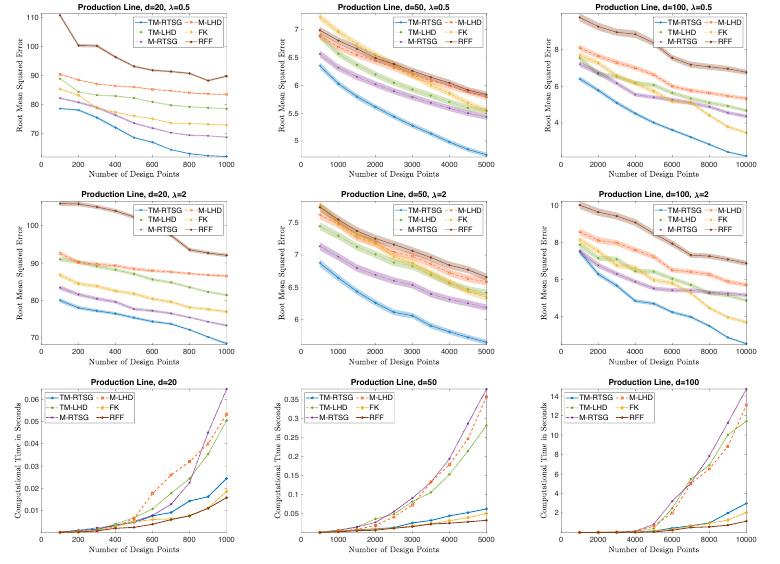}
}
{Prediction Errors and Computational Time for the Production Line Problem. \label{fig:Product_Line} }
{The three columns correspond to $d=20$, $50$, and $100$, respectively. The shaded areas represent 95\% CIs.}
\end{figure}

\begin{figure}[ht]
\FIGURE
{
\includegraphics[width=\textwidth]{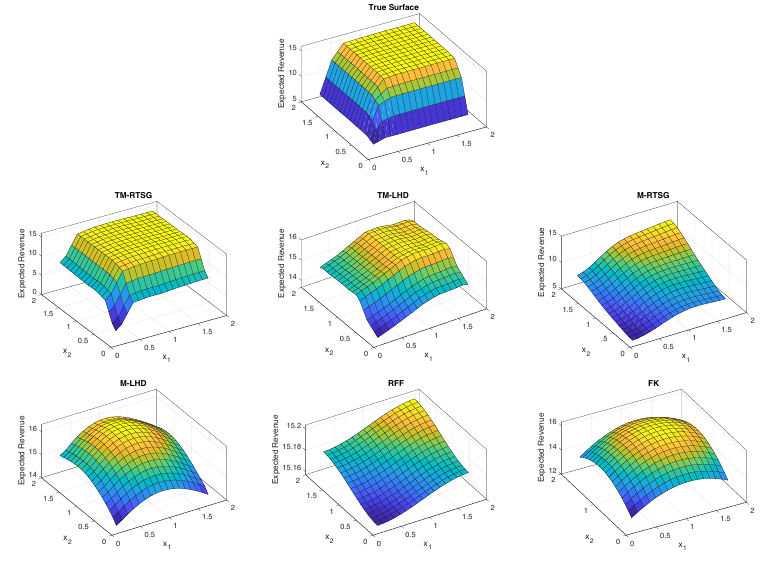}
}
{Projections of the True and Predicted Surfaces of the Production Line Problem ($\lambda=0.5$ and $d=100$). \label{fig:prod_line_surface}}
{The predicted surfaces are constructed with a sample size $n=10000$.}
\end{figure}

Figure~\ref{fig:Product_Line} shows $\widehat{\text{RMSE}}$ (with the associated 95\% CI) and the computational time of different approaches  for $d=20$, $50$, and $100$ and for both the low-noise and high-noise scenarios.
The results are consistent with the experiments in Section~\ref{sec:test-func}.
\texttt{TM-RTSG} significantly outperforms the other approaches in terms of prediction accuracy in all the cases considered,
and its advantage tends to be more prominent as the dimensionality and sample size increase.
Meanwhile, in terms of the computational efficiency, \texttt{TM-RTSG} is significantly faster than the three approaches involving no approximations (i.e., \texttt{TM-LHD}, \texttt{M-RTSG}, and \texttt{M-LHD})
but slightly slower than the two approximation approaches \texttt{FK} and \texttt{RFF}.

In Figure~\ref{fig:prod_line_surface}, we visualize  the true surface (for $\lambda=0.5$ and $d=100$) and the predicted surfaces using different approaches that are projected onto the dimensions spanned by $x_{1}$ and $x_{2}$ of the design space.
Specifically, the projections are formed by fixing $x_j=1$ for $j\neq 1, 2$ with a sample size $n=10000$.
We can see that \texttt{TM-RTSG} successfully captures the shape of the true surface, whereas the projections formed by other approaches fail to do so.
We have also examined projections onto other dimensions and the results are similar.

\subsection{A Large Linear Program}\label{sec:stocfor}
In this section, we challenge the prediction capability of \texttt{TM-RTSG}  with a very high-dimensional example with $d=16675$.
The response surface of interest is the optimal value of STOCFOR3, a linear programming (LP) model---the largest one available in the NETLIB Library (\url{netlib.org/lp})---to decide what parts of a forest
should be harvested to prevent wildfires.
The design variable $\BFx\in(0,1)^d$ is taken to be
the right-hand-side term of the constraints of the LP model.
Specifically, the LP model is of the form
\begin{align*}
y(\BFx)\coloneqq  \mbox{Minimize}_{\BFw}\quad &\BFc^\intercal \BFw \\
\mbox{subject to}\quad & \BFA \BFw = \BFx \\
& \BFl \leq \BFw \leq \BFu,
\end{align*}
where the values of $\BFc$, $\BFA$, $\BFl$, and $\BFu$ are publicly available.
The value of $y(\BFx)$ can be computed with the LP solver in \textsc{Matlab}.
Similar to Section~\ref{sec:test-func},
we add artificial Gaussian noise to $y(\BFx)$ but with equal variance $\sigma^2(\BFx)\equiv 1$.
For simplicity, we assume the variance is known to the competing methods and take one replication at each design point.
The sampling budget is set to be  $n=3000, 6000, \ldots, 30000$, and  $n_{\text{pred}}=1000$ prediction points are selected in the way detailed in Section~\ref{sec:setup}.
We set the number of macro-replications to be $R=30$.

We exclude \texttt{TM-LHD}, \texttt{M-RTSG}, and \texttt{TM-LHD} here, however, because they involve both storing training data of size $n\times d$ and doing computations on $n\times n$ dense matrices, both of which exceed the capacity of our computer when $n$ is larger than 10000.
In contrast, $\texttt{TM-RTSG}$, $\texttt{RFF}$, and $\texttt{FK}$ are not subject to this limitation.

\begin{figure}[ht]
\FIGURE
{
\includegraphics[width=0.8\textwidth]{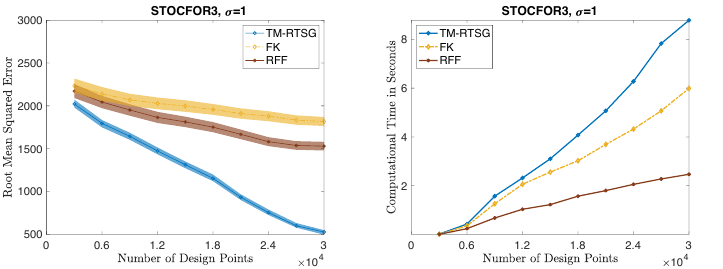}
}
{Prediction Errors (left) and Computational Time (right) for the STOCFOR3 Problem. \label{fig:LP_STOCFOR3}}
{The shaded areas represent 95\% CIs.}
\end{figure}

\begin{figure}[ht]
\FIGURE
{
\includegraphics[width=0.8\textwidth]{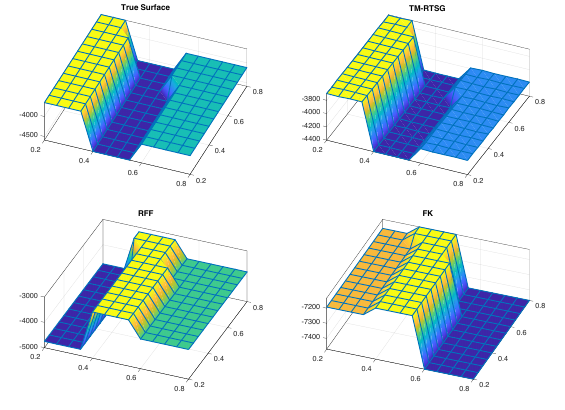}
}
{Projections of the True and Predicted Surfaces of the STOCFOR3 Problem.\label{fig:STOCFOR3_2Dslice}}
{
The predicted surfaces are constructed with a sample size $n=30000$.
}
\end{figure}

We can see from Figure~\ref{fig:LP_STOCFOR3} that \texttt{TM-RTSG} outperforms \texttt{RFF} and \texttt{FK} in prediction accuracy by a large margin, while its computational efficiency is lower but still at an acceptable level.
Further, the advantage of \texttt{TM-RTSG} appears to be increasing as the number of design points increases.
Such an advantage in prediction accuracy is also reflected clearly in Figure~\ref{fig:STOCFOR3_2Dslice}.
It shows projections of the true response surface and the predicted surfaces onto
two arbitrarily chosen dimensions, with the values of the design variable in the other dimensions being randomly chosen from $\sfrac{1}{4}$, $\sfrac{1}{2}$, and $\sfrac{3}{4}$.

\section{Concluding Remarks}\label{sec:conclusions}
In this paper, we develop a new methodology for simulation metamodeling in high dimensions.
We show that stochastic kriging with TM kernels and TSG designs is both sample-efficient and computationally efficient.
First, we establish an upper bound on the convergence rate  under both
correct model specification and model misspecification.
The upper bounds imply that for both cases,
the curse of dimensionality on sample complexity is significantly alleviated.
The convergence analysis under model misspecification is particularly reassuring for practitioners,
as it proves the strong robustness of our methodology.

Second, we develop algorithms that perform fast, exact computation of stochastic kriging with TM kernels and TSG designs.
They are fundamentally driven by our explicit characterization of the sparse structure of the inverse kernel matrix.
This characterization holds only for TM kernels.
Compared to existing algorithms for kriging with SG designs, our  algorithms allow simulation noise and are not restricted to ``complete'' SG designs.
However, our algorithms do not apply to general tensor product kernels, which may not possess the sparse structure.

The present paper can potentially be extended in several ways.
First, we only establish upper bounds on the convergence rate, and whether they are tight is unknown.
It is of great interest to establish minimax lower bounds, although new theoretical tools may be needed for this purpose.

Second, kriging metamodels often suffer from numerical instability due to the ill-conditionedness of the kernel matrix.
The kernel matrices associated with TM kernels and TSGs appear to be well-conditioned in our numerical experiments, but we do not analyze them theoretically.
Further studies on the eigenvalues of these kernel matrices might provide insightful guidance.

Third, this paper focuses on prediction of the response surface.
Another main use of simulation metamodeling is simulation optimization.
Given a sampling budget, a simple algorithm for computing the global optimal solution of an unknown response surface would be to take samples following the TSG design, construct the predicted surface using stochastic kriging with TM kernels, and then select from the TSG the design point that has the optimal predicted value as the solution.
As the upper convergence rates that we establish are uniform in the design variable,
it is conceivable that this algorithm may substantially mitigate the curse of dimensionality.
One might devise a more efficient algorithm that adaptively selects sampling locations from an SG design, which is made possible by the theory developed  for the TSG designs here.

\ACKNOWLEDGMENT{We thank the area editor (L. Jeff Hong), the associate editor and anonymous referees for their thoughtful and constructive comments that have significantly improved this paper. We also thank Jin Li (University of Hong Kong) for helpful discussions which led to the term “truncated sparse grids”. We gratefully acknowledge financial support from the Hong Kong Research Grants Council (GRF 17206821).
}

\bibliographystyle{informs2014} %
\bibliography{ScalableSK.bib}

\ECSwitch

\EquationsNumberedBySection
\ECHead{Supplemental Material}

\section{An Orthogonal Expansion of TM Kernels}

We first introduce a technical result that is fundamental for both the convergence rate analysis and the kernel matrix  inversion.
It summarizes Theorem~2 and Corollary~1 in \cite{Ding2020_ec}.

\begin{lemma}\label{lemma:TMK_expansion}
Let $k(\BFx,\BFx') = \prod_{j=1}^d p_j(x_j\wedge x'_j) q_j(x_j\vee  x'_j)$ be a TM kernel that  satisfies Assumption~\ref{assump:SL}.
Let $\SFc_{l, i}\coloneqq i\cdot 2^{-l}$ for $l\geq 1$ and $i=1,\ldots,2^l -1$.
For each $j=1,\ldots,d$, define the following continuous function with support $(\SFc_{l, i-1}, \SFc_{l, i+1}) = ((i-1)\cdot 2^{-l}, (i+1)\cdot 2^{-l})$:
\begin{equation}\label{eq:basis-func}
\phi_{j, l, i}(x) \coloneqq \left\{
\begin{array}{ll}
\displaystyle \frac{p_j(x)\SFq_{j,l,i-1} - \SFp_{j,l,i-1}q_j(x)}{\SFp_{j,l,i}\SFq_{j,l,i-1} - \SFp_{j,l,i-1}\SFq_{j,l,i}},     & \quad\mbox{if } x \in (\SFc_{l, i-1}, \SFc_{l, i}], \\[2ex]
\displaystyle \frac{\SFp_{j,l,i+1} q_j(x) - p_j(x) \SFq_{j,l,i+1} }{\SFp_{j,l,i+1}\SFq_{j,l,i} - \SFp_{j,l,i}\SFq_{j,l,i+1}},
     & \quad\mbox{if } x \in (\SFc_{l, i}, \SFc_{l, i+1}), \\[2ex]
0, & \quad\mbox{otherwise },
\end{array}
\right.
\end{equation}
where
$\SFp_{j,l,i}=p_j(\SFc_{l, i})$ and $\SFq_{j,l,i}=q_j(\SFc_{l, i})$.
For any multi-indices $\BFl,\BFi\in\NatInt^d$, define
\begin{equation}\label{eq:basis-func-tensor}
\phi_{\BFl,\BFi}(\BFx) \coloneqq \prod_{j=1}^d \phi_{j,l_j,i_j}(x_j).
\end{equation}
Let $\ScrH_k$ denote the reproducing kernel Hilbert space (RKHS) generated by the kernel $k$, with inner product $\langle\cdot, \cdot\rangle_{\ScrH_k}$ and norm $\norm{\cdot}_{\ScrH_k}$.
Then, $\{\phi_{\BFl,\BFi}: \BFl\in\NatInt^d,\,\BFi\in\rho(\BFl)\}$ is an orthogonal basis for $\ScrH_k$ with respect to
$\langle\cdot, \cdot\rangle_{\ScrH_k}$,
where $\rho(\BFl) = \bigtimes_{j=1}^d  \{i_j:1\leq i_j\leq  2^{l_j}-1,\;i_j \mbox{ odd} \}$.
In particular,
we have the following expansion for $k$:
\begin{equation}
    \label{eq:TMK_expansion}
    k(\BFx,\BFx')= \sum_{\BFl\in\NatInt^d}\sum_{\BFi\in\rho(\BFl)}\frac{\phi_{\BFl,\BFi}(\BFx)\phi_{\BFl,\BFi}(\BFx')}{\|\phi_{\BFl,\BFi}\|_{\ScrH_k}^2},
\end{equation}
for all $\BFx,\BFx'\in(0,1)^d$.
\end{lemma}

The following result calculates the RKHS norm of $\phi_{\BFl,\BFi}$ and establishes its decay rate as $\abs{\BFl}\to\infty$.
This result will be used in the analysis of convergence rates of stochastic kriging with TM kernels and TSGs.

\begin{lemma}\label{lemma:RKHS_norm_estimate}
Let $k(\BFx,\BFx') = \prod_{j=1}^d p_j(x_j\wedge x'_j) q_j(x_j\vee  x'_j)$ be a TM kernel that satisfies Assumption~\ref{assump:SL}.
Let $\{\phi_{\BFl,\BFi}:\BFl\in\NatInt^d,\BFi\in\rho(\BFl)\}$ be the orthogonal basis defined by \eqref{eq:basis-func-tensor} for $\ScrH_k$. Then,
\begin{equation}\label{eq:phi-norm}
\begin{aligned}
    \|\phi_{\BFl,\BFi}\|^2_{\ScrH_k} = \prod_{j=1}^d \left(\frac{\SFp_{j, l_j,i_j+1} \SFq_{j,l_j, i_j-1} - \SFp_{j,l_j,i_j-1} \SFq_{j,l_j,i_j+1}}{\left(\SFp_{j,l_j,i_j} \SFq_{j,l_j,i_j-1} - \SFp_{j,l_j,i_j-1}  \SFq_{j,l_j,i_j} \right) \left(\SFp_{j,l_j,i_j+1}  \SFq_{j,l_j,i_j} - \SFp_{j,l_j,i_j} \SFq_{j,l_j,i_j+1}\right) }\right),
\end{aligned}
\end{equation}
where $\SFp_{j,l,i}= p_j(\SFc_{l, i})$ and
$\SFq_{j,l,i}= q_j(\SFc_{l, i})$ for all $j$, $l$, and $i$.
Moreover, as $\abs{\BFl}\to\infty$,
\begin{equation}
\label{eq:basisRKHSestimate}
    \|\phi_{\BFl,\BFi}\|^2_{\ScrH_k}\asymp 2^{|\BFl|}.
\end{equation}
\end{lemma}
\proof{Proof.}
If $d=1$, we suppress the dependence on $j$ and write $k(x,x')=p(x\wedge x')q(x\vee x')$ for all $x,x'\in[0,1]$. Then,
\eqref{eq:phi-norm} is reduced to
\begin{equation}\label{eq:phi-norm-1d}
\|\phi_{l,i}\|_{\ScrH_{k}}^2 = \frac{\SFp_{l,i+1} \SFq_{l, i-1} - \SFp_{j,l,i-1} \SFq_{l,i+1}}{\left(\SFp_{l,i} \SFq_{l,i-1} - \SFp_{l,i-1}  \SFq_{l,i} \right) \left(\SFp_{l,i+1}  \SFq_{l,i} - \SFp_{l,i} \SFq_{l,i+1}\right) },
\end{equation}
where $\phi_{l,i}$ is defined as \eqref{eq:basis-func}, and  $\SFp_{l,i}= p(\SFc_{l, i})$ and
$\SFq_{l,i}= q(\SFc_{l, i})$ for all $l$ and $i$.
Moreover, \eqref{eq:basisRKHSestimate} is reduced to
\begin{equation}\label{eq:basisRKHSestimate-1d}
     \|\phi_{l,i}\|^2_{\ScrH_k}\asymp 2^{l},
\end{equation}
as $l\to\infty$.
To prove \eqref{eq:phi-norm-1d},  we show that $\phi_{l,i}(x)$ can be expressed as a linear combination of $k(x, \SFc_{l,i-1})$, $k(x, \SFc_{l,i})$, and $k(x, \SFc_{l,i+})$, and then apply the definition of the RKHS norm.
Specifically, it can be shown by direct, albeit tedious, calculation that
\begin{equation}\label{eq:phi_li}
\phi_{l,i}(x) =
\eta_{l,i-1} k(x, \SFc_{l,i-1}) + \eta_{l,i} k(x, \SFc_{l,i}) + \eta_{l,i+1} k(x, \SFc_{l,i+1}),
\end{equation}
for all $x\in[0,1]$, $l\geq 1$, and $i\in\rho(l)$, where
\begin{equation}\label{eq:eta_li}
\begin{aligned}
    \eta_{l,i} ={}& \frac{\SFp_{l,i+1} \SFq_{l, i-1} - \SFp_{l,i-1} \SFq_{l,i+1}}{\left(\SFp_{l,i} \SFq_{l,i-1} - \SFp_{l,i-1}  \SFq_{l,i} \right) \left(\SFp_{l,i+1}  \SFq_{l,i} - \SFp_{l,i} \SFq_{l,i+1}\right) }, \\
    \eta_{l,i-1} ={}& \frac{-1}{\SFp_{l,i}\SFq_{l,i-1}-\SFp_{l,i}\SFq_{l,i-1}}, \\
    \eta_{l,i+1} ={} & \frac{-1}{\SFp_{l,i+1}\SFq_{l,i} - \SFp_{l,i}\SFq_{l,i+1} }.
\end{aligned}
\end{equation}
Hence, by the definition of the RKHS norm,
\begin{align*}
\|\phi_{l,i}\|_{\ScrH_{k}}^2 = \sum_{i'=i-1}^{i+1}\sum_{i''=i-1}^{i+1} \eta_{l,i'}\eta_{l,i''} k(\SFc_{l,i'},\SFc_{l,i''}) = \eta_{l, i},
\end{align*}
where the second equality, again, follows direct calculation. This proves \eqref{eq:phi-norm-1d}.

To prove \eqref{eq:basisRKHSestimate-1d}, we notice that
\begin{align*}
\SFp_{l,i+1}  \SFq_{l,i} - \SFp_{l,i} \SFq_{l,i+1} ={}& (\SFp_{l,i+1} - \SFp_{l,i}) \SFq_{l,i} - \SFp_{l,i} (\SFq_{l,i+1}-\SFq_{l,i})  \\
={}& [p(\SFc_{l,i+1})-p(\SFc_{l,i}) ]q(\SFc_{l,i}) - p(\SFc_{l,i})[q(\SFc_{l,i+1}) - q(\SFc_{l,i})].
\end{align*}
Hence,
\begin{align*}
2^{-l}(\SFp_{l,i+1}  \SFq_{l,i} - \SFp_{l,i} \SFq_{l,i+1}) ={}& \frac{p((i+1)\cdot 2^{-l}) - p(i\cdot 2^{-l})}{2^l} q(\SFc_{l,i}) - p(\SFc_{l,i})\frac{q((i+1)\cdot 2^{-l}) - q(i\cdot 2^{-l})}{2^l} \\
\asymp{}& p'(\SFc_{l,i})q(\SFc_{l,i}) - p(\SFc_{l,i})q'(\SFc_{l,i}),
\end{align*}
as $l\to\infty$,
where $p'$ and $q'$ denote the derivatives of $p$ and $q$, respectively.
Similarly, we have
\begin{equation}\label{eq:eta-estimate}
\begin{aligned}
2^{-l}(\SFp_{l,i}  \SFq_{l,i-1} - \SFp_{l,i-1} \SFq_{l,i}) \asymp {}& p'(\SFc_{l,i-1})q(\SFc_{l,i-1}) - p(\SFc_{l,i-1})q'(\SFc_{l,i-1}), \\
2^{-(l-1)}(\SFp_{l,i+1}  \SFq_{l,i-1} - \SFp_{l,i-1} \SFq_{l,i+1}) \asymp {}& p'(\SFc_{l,i-1})q(\SFc_{l,i-1}) - p(\SFc_{l,i-1})q'(\SFc_{l,i-1}),
\end{aligned}
\end{equation}
as $l\to\infty$.
It follows that, by \eqref{eq:phi-norm-1d},
\begin{equation}\label{eq:phi-norm-magnitude}
\|\phi_{l,i}\|_{\ScrH_{k}}^2 \asymp \frac{2^{l+1}}{p'(\SFc_{l,i})q(\SFc_{l,i}) - p(\SFc_{l,i})q'(\SFc_{l,i})},
\end{equation}
as $l\to\infty$.
Furthermore, by Lemma~\ref{lemma:PD}, $p(x)/q(x)$ is strictly increasing in $x$ on $(0,1)$.
Thus,
\[\left(\frac{p(x)}{q(x)}\right)' = \frac{p'(x)q(x)-p(x)q'(x)}{q^2(x)} > 0\]
for all $x\in[0,1]$.
Therefore, $|p'(x)q(x)-p(x)q'(x)|\in [a, b]$ for some positive constants $a$ and $b$, since $p$ and $q$ are continuously differentiable by Assumption~\ref{assump:SL}.
We thus conclude from \eqref{eq:phi-norm-magnitude} that $\|\phi_{l,i}\|_{\ScrH_{k}}^2 \asymp 2^l$, proving \eqref{eq:phi-norm-1d}.

For $d>1$, in light of the tensor structure of both $k$ and $\phi_{\BFl,\BFi}$, the RKHS norm of $\phi_{\BFl,\BFi}$ possesses a product form as follows,
\[\|\phi_{\BFl,\BFi}\|_{\ScrH_k}^2=\prod_{j=1}^d\|\phi_{l_j,i_j}\|_{\ScrH_{k_j}}^2,\]
where  $\ScrH_{k_j}$ denotes the RKHS induced by the component kernel $k_j$.
Then, \eqref{eq:phi-norm} and \eqref{eq:basisRKHSestimate} follow immediately from \eqref{eq:phi-norm-1d} and \eqref{eq:basisRKHSestimate-1d}, respectively, to each dimension $j=1,\ldots,d$.
\Halmos \endproof

Next, we establish an important connection between the functions $\{\phi_{\BFl,\BFi}:\BFl\in\NatInt^d, \BFi\in \rho(\BFl)\}$ and classical SGs.
In particular, given a design point $\BFc_{\BFl', \BFi'}$ on a classical SG, $\phi_{ \BFl,  \BFi}(\BFc_{\BFl',\BFi'})=0$ for most of the pairs $( \BFl, \BFi)$ with $|\BFl|\geq |\BFl'|$.
Since the notation is a bit involved, we use a one-dimensional example for illustration before presenting the formal statement.

\begin{figure}[ht]
\FIGURE{
\includegraphics[width=\textwidth]{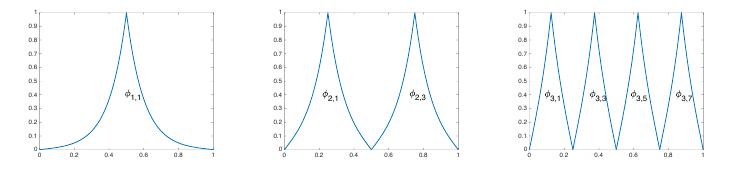}
}
{The one-dimensional orthogonal basis $\{\phi_{l,i}:l\geq 1,\; 1\leq i\leq 2^l-1,\;i\mbox{ odd}\}$ associated with the TM kernel $k(x, x') = \exp(-5\abs{x- x'})$ on $(0,1)$.\label{fig:waveletExpansion}}
{Left: $l=1$; Middle: $l=2$; Right: $l=3$.}
\end{figure}

In Figure~\ref{fig:waveletExpansion}, we plot the first few functions of the orthogonal basis $\{\phi_{l,i}(x):l\geq 1,\; 1\leq i\leq 2^l-1,\;i\mbox{ odd}\}$ associated with the TM kernel $k(x, x')=\exp(-5\abs{x- x'})$ on $[0,1]$.
Recall that when $d=1$, the classical SG of level $\tau$ is simply the one-dimensional dyadic grid $\{i\cdot 2^{-\tau}: 1\leq i\leq 2^{-\tau}-1\}$.
For $\tau=1$, $\CalX_1^{\mathsf{SG}}=\{\sfrac{1}{2}\}$.
It is straightforward to see that
$\phi_{2, 1}(\sfrac{1}{2}) = \phi_{2, 3}(\sfrac{1}{2}) = 0$, and $\phi_{3, 1}(\sfrac{1}{2}) = \phi_{3, 3}(\sfrac{1}{2}) = \phi_{3, 5}(\sfrac{1}{2}) = \phi_{3, 7}(\sfrac{1}{2}) = 0$.
Likewise, for $\tau=2$, consider the design point $\sfrac{1}{4}\in\CalX_2^{\mathsf{SG}}$.
Clearly, $\phi_{3, i}(\sfrac{1}{4})= 0$ for all $i=1,3, 5, 7$. In general, we have $\phi_{l, i}(\SFc_{\tau,  i'})=0$ if $l>\tau$.
This is because in this case, $\SFc_{\tau,  i'}$ happens to be outside the support of $\phi_{l,i}$ by definition.
We now present formally the general result regarding the function $\phi_{\BFl,\BFi}$ evaluated at design points on a classical SG.

\begin{lemma}\label{lemma:basis-eval}
Consider the functions $\{\phi_{\BFl,\BFi}:\BFl\in\NatInt^d, \BFi\in\rho(\BFl)\}$ defined by \eqref{eq:basis-func-tensor}, and design points on a classical SG that are of the form $\BFc_{\BFl,\BFi} = (\SFc_{l_1,i_1}, \ldots, \SFc_{l_d, i_j})$.
Then,
\begin{enumerate}[label=(\roman*)]
\item $\phi_{\BFl,\BFi}(\BFc_{\BFl,\BFi}) = 1$ for all $\BFl\in\NatInt^d$ and $\BFi\in\rho(\BFl)$;
\item $\phi_{\BFl,\BFi}(\BFc_{\BFl', \BFi'}) = 0$ for all $\BFl,\BFl'\in\NatInt^d$, $\BFi\in\rho(\BFl)$, and $\BFi'\in\rho(\BFl')$ with
$l_j>l_j'$ for some $j=1,\ldots,d$.
\item $\phi_{\BFl,\BFi}(\BFc_{\BFl', \BFi'}) = 0$ for all $\BFl,\BFl'\in\NatInt^d$, $\BFi\in\rho(\BFl)$, and $\BFi'\in\rho(\BFl')$ with $\BFl = \BFl'$ and $\BFi \neq \BFi'$.
\end{enumerate}
\end{lemma}

\proof{Proof of Lemma~\ref{lemma:basis-eval}.}
These properties are essentially a result of the tensor structure and the nested structure of $\phi_{\BFl,\BFi}$'s and classical SGs.
\paragraph{Part (i).}
By the definition \eqref{eq:basis-func}, for each tuple $(j,l,i)$,
\[
\phi_{j,l,i}(\SFc_{l,i}) =
\frac{p_j(\SFc_{l,i})\SFq_{j,l,i-1} - \SFp_{j,l,i-1}q_j(\SFc_{l,i})}{\SFp_{j,l,i}\SFq_{j,l,i-1} - \SFp_{j,l,i-1}\SFq_{j,l,i}} = 1,
\]
since $p_j(\SFc_{l,i}) = \SFp_{j,l,i}$ and $q_j(\SFc_{l,i}) = \SFq_{j,l,i}$.
Hence, $\phi_{\BFl,\BFi}(\BFc_{\BFl,\BFi})=\prod_{j=1}^d \phi_{j,l_j,i_j}(\SFc_{l_j,i_j}) = 1$.

\paragraph{Part (ii).}
Without loss of generality, assume $j=1$, i.e., $l_1 > l_1'$.
Note that $\SFc_{l_1', i_1'}=i_1'\cdot 2^{-l_1'} = \bigl(i_1'\cdot 2^{l_1-l_1'}\bigr)\cdot 2^{-l_1}$.
Since $i_1'$ is an odd number between 1 and $2^{l_1'}-1$, we know that $i_1'\cdot 2^{l_1-l_1'}$ is an even number between $2^{l_1-l_1'}$ and $2^{l_1} -  2^{l_1-l_1'}$.
Hence, there does not exist an odd number $i_1$ between 1 and $2^{l_1}-1$ such that
$i_1-1 < i_1'\cdot 2^{l_1-l_1'} < i_1+1 $.
This implies that
\[\SFc_{l_1',i_1'}= \bigl(i_1'\cdot 2^{l_1-l_1'}\bigr)\cdot 2^{-l_1} \notin \bigl((i_1-1)\cdot 2^{-l_1}, \; (i_1+1)\cdot 2^{-l_1} \bigr), \]
for any odd number $i_1$ between 1 and $2^{l_1}-1$.
It follows that
$\phi_{1, l_1, i_1}(\SFc_{l_1, i_1}) = 0$, since the support of $\phi_{1, l_1, i_1}$ is $\bigl((i_1-1)\cdot 2^{-l_1}, \; (i_1+1)\cdot 2^{-l_1} \bigr)$.
Therefore, $\phi_{\BFl, \BFi}(\BFc_{\BFl',\BFi'}) = \phi_{1,l_1,i_1}(\SFc_{l_1',i_1'})\cdot\prod_{j=2}^d \phi_{j, l_j, i_j}(\SFc_{l_j',i_j'})=0$.

\paragraph{Part (iii).}
Since $\BFi\neq \BFi'$, there must exist $j$ such that $i_j\neq i_j'$.
Without loss of generality, assume $j=1$.
Note that $l_1=l_1'$, so
$i_1'$ is an odd number between 1 and $2^{l_1}-1$.
It follows that $i_1'\notin (i_1 -1, i_1+1)$, the only odd number in the open interval is $i_1$, which is not $i_1'$.
Thus, $\SFc_{l_1',i_1'}=i_1'\cdot 2^{-l_1'} = i_1'\cdot 2^{-l_1}$ does not belong to $\bigl((i_1-1)\cdot 2^{-l_1}, \; (i_1+1)\cdot 2^{-l_1} \bigr)$, the support of $\phi_{j, l_j, i_j}$.
Hence,  $\phi_{\BFl, \BFi}(\BFc_{\BFl',\BFi'}) = \phi_{1,l_1,i_1}(\SFc_{l_1',i_1'})\cdot\prod_{j=2}^d \phi_{j, l_j, i_j}(\SFc_{l_j',i_j'})=0$.
\Halmos
\endproof

Last, we show that the orthogonal basis $\{\phi_{\BFl,\BFi}:\BFl\in\NatInt^d,\BFi\in\rho(\BFl)\}$ associated with the TM kernel $k$ induces an expansion of the TMGP with the same kernel in terms of i.i.d. standard normal random variables.
This is an immediate consequence of Theorems~3.1.1 and 3.1.2 in \cite{AdlerRF&G09_ec}.

\begin{lemma}\label{lemma:TMGP-expansion}
Let $\{\SFY(\BFx):\BFx\in(0,1)^d\}$ be a mean zero TMGP with kernel $k$ that satisfies Assumption~\ref{assump:SL}.
Let $\{\phi_{\BFl,\BFi}:\BFl\in\NatInt^d,\BFi\in\rho(\BFl)\}$ be the orthogonal basis defined by \eqref{eq:basis-func-tensor} for $\ScrH_k$. Then,
\begin{equation}\label{eq:GP-expansion}
\SFY(\BFx) = \sum_{\BFl\in\NatInt^d}\sum_{\BFi\in\rho(\BFl)}
\frac{\phi_{\BFl,\BFi}(\BFx)}{\|\phi_{\BFl,\BFi}\|_{\ScrH_k}}Z_{\BFl,\BFi},
\end{equation}
where $Z_{\BFl,\BFi}$'s are i.i.d. standard normal variables, and the sum converges almost surely and the convergence is uniform over $x\in(0,1)^d$.
\end{lemma}

We can apply Lemmas~\ref{lemma:basis-eval} and \ref{lemma:TMGP-expansion} to express $\SFY(\BFc_{\BFl,\BFi})$ as a linear combination of
a \emph{finite} set of i.i.d normal random variables.
\begin{lemma}\label{lemma:finite-expansion}
Let $\{\SFY(\BFx):\BFx\in(0,1)^d\}$ be a mean zero TMGP with kernel $k$ that satisfies Assumption~\ref{assump:SL}.
Let $\{\phi_{\BFl,\BFi}:\BFl\in\NatInt^d,\BFi\in\rho(\BFl)\}$ be the orthogonal basis defined by \eqref{eq:basis-func-tensor} for $\ScrH_k$.
Then,
\begin{equation}\label{eq:finite-expansion}
\SFY(\BFc_{\BFl',\BFi'}) = \sum_{\abs{\BFl}\leq \tau + d - 1} \sum_{\BFi\in\rho(\BFl)}  \frac{\phi_{\BFl,\BFi}(\BFc_{\BFl',\BFi'})}{\|\phi_{\BFl,\BFi}\|_{\ScrH_k}}Z_{\BFl,\BFi} + \frac{1}{\|\phi_{\BFl',\BFi'}\|_{\ScrH_k}}Z_{\BFl',\BFi'},
\end{equation}
for all $\BFc_{\BFl',\BFi'}\in \CalD_{\tau+1} = \CalX_{\tau+1}^{\mathsf{SG}}\setminus \CalX_{\tau}^{\mathsf{SG}}$, where $Z_{\BFl,\BFi}$'s are i.i.d. standard normal variables.
\end{lemma}

\proof{Proof of Lemma~\ref{lemma:finite-expansion}.}
By the definition of classical SGs, particularly the formula \eqref{eq:disjoint-rep}, we know that
$\abs{\BFl'} = \tau +d$ if $\BFc_{\BFl',\BFi'}\in\CalD_{\tau+1}$.
Hence, if $\abs{\BFl}\geq \tau +d + 1$, then there must exist $j$ such that $l_j>l_j'$, and thus $\phi_{\BFl,\BFi}(\BFc_{\BFl',\BFi'})=0$ by part (ii) of Lemma~\ref{lemma:basis-eval}.
In addition, it follows from part (ii) and part (iii) of Lemma~\ref{lemma:basis-eval} that for all $\BFl$ such that $\abs{\BFl}=\abs{\BFl'}$, $\phi_{\BFl,\BFi}(\BFc_{\BFl',\BFi'})=0$ unless $(\BFl,\BFi)=(\BFl',\BFi')$.
Consequently, all the terms in the expansion \eqref{eq:GP-expansion} with $\abs{\BFl}\geq \abs{\BFl'}$ are zero except for $(\BFl,\BFi)=(\BFl',\BFi')$, yielding that
\[\SFY(\BFc_{\BFl',\BFi'}) = \sum_{\abs{\BFl}\leq \tau + d - 1} \sum_{\BFi\in\rho(\BFl)}  \frac{\phi_{\BFl,\BFi}(\BFc_{\BFl',\BFi'})}{\|\phi_{\BFl,\BFi}\|_{\ScrH_k}}Z_{\BFl,\BFi} + \frac{\phi_{\BFl',\BFi'}(\BFc_{\BFl',\BFi'})}{\|\phi_{\BFl',\BFi'}\|_{\ScrH_k}}Z_{\BFl',\BFi'}.\]
Then, \eqref{eq:finite-expansion} follows immediately from part (i) of Lemma~\ref{lemma:basis-eval}.
\Halmos\endproof

Following Lemma~\ref{lemma:finite-expansion}, we show that
given a TSG $\CalX_{n}^{\mathsf{TSG}}$,
observing $\{\SFY(\BFc_{\BFl,\BFi}): \BFc_{\BFl,\BFi}\in\CalX_n^{\mathsf{TSG}}\}$ is equivalent to observing
$\{Z_{\BFl,\BFi}: \BFc_{\BFl,\BFi}\in\CalX_n^{\mathsf{TSG}}\}$.

\begin{lemma}\label{lemma:sigma-algebra}
Let $\{\SFY(\BFx):\BFx\in(0,1)^d\}$ be a mean zero TMGP with kernel $k$ that satisfies Assumption~\ref{assump:SL}.
Consider the expansion \eqref{eq:GP-expansion}.
Then,
\begin{equation}\label{eq:sigma-algebra-equiv}
\varsigma\bigl(\SFY(\BFc_{\BFl,\BFi}): \BFc_{\BFl,\BFi}\in\CalX_n^{\mathsf{TSG}}\bigr) =  \varsigma\bigl(Z_{\BFl,\BFi}:\BFc_{\BFl,\BFi}\in\CalX_n^{\mathsf{TSG}}\bigr),
\end{equation}
for all $n\geq 1$,
where  $\varsigma(\mathcal S)$ denotes the sigma-algebra generated by $\mathcal S$.
\end{lemma}
\proof{Proof of Lemma~\ref{lemma:sigma-algebra}.}
We first prove a weaker result as follows,
\begin{equation}\label{eq:sigma-algebra-equiv-SG}
\varsigma\bigl(\SFY(\BFc_{\BFl,\BFi}): \BFc_{\BFl,\BFi}\in\CalX_\tau^{\mathsf{SG}}\bigr) =  \varsigma\bigl(Z_{\BFl,\BFi}:\BFc_{\BFl,\BFi}\in\CalX_\tau^{\mathsf{SG}}\bigr),
\end{equation}
for all $\tau\geq 1$. We do so by induction on $\tau$.
If $\tau=1$,  then $\CalX_\tau^{\mathsf{SG}}=\{\SFc_{\vec{\BFone}, \vec{\BFone}}\}$, where $\vec{\BFone}$ is the $d$-dimensional vector of ones.
Thus, by \eqref{eq:finite-expansion},
$
Z_{\vec{\BFone},\vec{\BFone}} = \|\phi_{\vec{\BFone},\vec{\BFone}}\|_{\ScrH_k} \SFY(\SFc_{\vec{\BFone}, \vec{\BFone}})
$,
so \eqref{eq:sigma-algebra-equiv-SG} follows immediately.

Assume that \eqref{eq:sigma-algebra-equiv-SG} holds for some $\tau\geq 1$.
That is,
$\{Z_{\BFl,\BFi}: \BFc_{\BFl,\BFi}\in\CalX_\tau^{\mathsf{SG}}\}$
are uniquely determined by
$\{\SFY(\BFc_{\BFl,\BFi}):\BFc_{\BFl,\BFi}\in\CalX_\tau^{\mathsf{SG}}\}$, and vice versa.
By \eqref{eq:finite-expansion},
\begin{equation}\label{eq:calculate-Z-by-Y}
Z_{\BFl',\BFi'} =\|\phi_{\BFl',\BFi'}\|_{\ScrH_k} \Biggl[\SFY(\BFc_{\BFl',\BFi'}) - \sum_{(\BFl,\BFi): \BFc_{\BFl,\BFi}\in\CalX_\tau^{\mathsf{SG}}}  \frac{\phi_{\BFl,\BFi}(\BFc_{\BFl',\BFi'})}{\|\phi_{\BFl,\BFi}\|_{\ScrH_k}}Z_{\BFl,\BFi}\Biggr],
\end{equation}
for all $\BFc_{\BFl',\BFi'}\in\CalX_{\tau+1}^{\mathsf{SG}} \setminus \CalX_{\tau}^{\mathsf{SG}} $.
Hence,
$\{Z_{\BFl,\BFi}: \BFc_{\BFl,\BFi}\in\CalX_{\tau+1}^{\mathsf{SG}}\}$
are uniquely determined by
$\{\SFY(\BFc_{\BFl,\BFi}):\BFc_{\BFl,\BFi}\in\CalX_{\tau+1}^{\mathsf{SG}}\}$, and vice versa, completing the induction.
So, \eqref{eq:sigma-algebra-equiv-SG} holds for all $\tau\geq 1$.

To prove \eqref{eq:sigma-algebra-equiv},
let $\tau$ be such that $\CalX_\tau^{\mathsf{SG}} \subseteq \CalX_n^{\mathsf{TSG}}\subset \CalX_{\tau+1}^{\mathsf{SG}}$.
It follows from \eqref{eq:finite-expansion} and \eqref{eq:calculate-Z-by-Y} that
for any $\CalS \subseteq \CalX_{\tau+1}^{\mathsf{SG}}\setminus \CalX_\tau^{\mathsf{SG}}$,
$\{Z_{\BFl,\BFi}: \BFc_{\BFl,\BFi}\in\CalX_\tau^{\mathsf{SG}} \cup \CalS  \}$
are uniquely determined by
$\{\SFY(\BFc_{\BFl,\BFi}):\BFc_{\BFl,\BFi}\in\CalX_\tau^{\mathsf{SG}} \cup \CalS\}$, and vice versa.
Since $\CalX_n^{\mathsf{TSG}} = \CalX_\tau^{\mathsf{SG}}\cup \CalA_{\tilde n}$ with $\CalA_{\tilde n}\subset \CalX_{\tau+1}^{\mathsf{SG}}$, we conclude that
\eqref{eq:sigma-algebra-equiv} holds.
\Halmos
\endproof

We now prove a general result regarding the conditional distribution of $\SFY(\cdot)$ given $\{\SFY(\BFx):\BFx\in\CalX_n^{\mathsf{TSG}}\}$.
This result will be used later in convergence rate analysis and the kernel matrix inversion.

\begin{proposition}\label{prop:cond_dist}
Let $\{\SFY(\BFx):\BFx\in(0,1)^d\}$ be a zero mean TMGP with kernel $k$ that satisfies Assumption~\ref{assump:SL}.
Let $\{\SFz_{\BFl,\BFi}:\BFc_{\BFl,\BFi}\in\CalX_n^{\mathsf{TSG}}\}$ denote the realization of $Z_{\BFl,\BFi}$'s that are uniquely determined by observations of $\{\SFY(\BFc_{\BFl,\BFi}):\BFc_{\BFl,\BFi}\in\CalX_n^{\mathsf{TSG}}\}$ via \eqref{eq:calculate-Z-by-Y}.
Then,
conditioning on $\{\SFY(\BFc_{\BFl,\BFi}):\BFc_{\BFl,\BFi}\in\CalX_n^{\mathsf{TSG}}\}$,  $\SFY(\cdot)$ is a GP with mean function $\mu_n^{\mathsf{TSG}}$ and kernel function $k_n^{\mathsf{TSG}}$ as follows,
\begin{align}
\mu_n^{\mathsf{TSG}}(\BFx') ={}& \sum_{(\BFl,\BFi):\BFc_{\BFl,\BFi}\in\CalX_n^{\mathsf{TSG}}}\frac{\phi_{\BFl,\BFi}(\BFx')}{\|\phi_{\BFl,\BFi}\|_{\ScrH_k}} \SFz_{\BFl,\BFi}, \label{eq:posterior_mean}\\
k_n^{\mathsf{TSG}}(\BFx',\BFx'') ={}& \sum_{(\BFl,\BFi):\BFl\in\NatInt^d,\BFi\in\rho(\BFl),\BFc_{\BFl,\BFi}\notin\CalX_n^{\mathsf{TSG}}}
\frac{\phi_{\BFl,\BFi}(\BFx')\phi_{\BFl,\BFi}(\BFx'')}{\|\phi_{\BFl,\BFi}\|_{\ScrH_k}^2}, \label{eq:posterior_kernel}
\end{align}
for all $\BFx',\BFx''\in(0,1)^d$.
\end{proposition}

\proof{Proof of Proposition~\ref{prop:cond_dist}.}
Note that conditioning on $\{\SFY(\BFx):\BFx\in\CalX_n^{\mathsf{TSG}}\}$,
\begin{equation}\label{eq:cond_expression}
\SFY(\BFx') = \sum_{(\BFl,\BFi):\BFc_{\BFl,\BFi}\in\CalX_n^{\mathsf{TSG}}}  \frac{\phi_{\BFl,\BFi}(\BFx')}{\|\phi_{\BFl,\BFi}\|_{\ScrH_k}}\SFz_{\BFl,\BFi} +
\sum_{(\BFl,\BFi):\BFl\in\NatInt^d,\BFi\in\rho(\BFl),\BFc_{\BFl,\BFi}\notin\CalX_n^{\mathsf{TSG}}}
\frac{\phi_{\BFl,\BFi}(\BFx')}{\|\phi_{\BFl,\BFi}\|_{\ScrH_k}}Z_{\BFl,\BFi},
\end{equation}
by  \eqref{eq:GP-expansion} and \eqref{eq:sigma-algebra-equiv}.
It is straightforward to see that $\mu_n^{\mathsf{TSG}}(\BFx') = \E[\SFY(\BFx')\,|\, \SFY(\BFx):\BFx\in\CalX_n^{\mathsf{TSG}} ]$ is given by \eqref{eq:posterior_mean}.
Moreover,  $k_n^{\mathsf{TSG}}(\BFx',\BFx'')=\Cov[\SFY(\BFx'), \SFY(\BFx'')\,|\, \SFY(\BFx):\BFx\in\CalX_n^{\mathsf{TSG}} ] $ equals
\begin{align*}
& \E\Biggl[ \Biggl(\sum_{(\BFl,\BFi):\BFl\in\NatInt^d,\BFi\in\rho(\BFl),\BFc_{\BFl,\BFi}\notin\CalX_n^{\mathsf{TSG}}}\frac{\phi_{\BFl,\BFi}(\BFx')Z_{\BFl,\BFi}}{\|\phi_{\BFl,\BFi}\|_{\ScrH_k}}\Biggr) \Biggl(\sum_{(\BFl,\BFi):\BFl\in\NatInt^d,\BFi\in\rho(\BFl),\BFc_{\BFl,\BFi}\notin\CalX_n^{\mathsf{TSG}}}\frac{\phi_{\BFl,\BFi}(\BFx'')Z_{\BFl,\BFi}}{\|\phi_{\BFl,\BFi}\|_{\ScrH_k}}\Biggr)\Biggr] \\
={}& \sum_{(\BFl,\BFi):\BFl\in\NatInt^d,\BFi\in\rho(\BFl),\BFc_{\BFl,\BFi}\notin\CalX_n^{\mathsf{TSG}}}
\frac{\phi_{\BFl,\BFi}(\BFx')\phi_{\BFl,\BFi}(\BFx'')}{\|\phi_{\BFl,\BFi}\|_{\ScrH_k}^2},
\end{align*}
proving \eqref{eq:posterior_kernel}.
\Halmos\endproof

We further show that TMGPs possess a nice structure of conditional independence between observations at design points on $\CalD_{\tau+1}$.

\begin{proposition}
\label{prop:cond_var}
Let $\{\SFY(\BFx):\BFx\in(0,1)^d\}$ be a zero mean TMGP with kernel $k$  that satisfies Assumption~\ref{assump:SL}. Consider a TSG $\CalX_n^{\mathsf{TSG}} = \CalX_\tau^{\mathsf{SG}}\cup \CalA_{\tilde n}$.
For all $\BFx',\BFx''\in \CalA_{\tilde n}$ with $\BFx'\neq \BFx''$,
$\SFY(\BFx')$ and $\SFY(\BFx'')$ are conditionally independent
given $\{\SFY(\BFx):\BFx \in \CalX_n^{\mathsf{TSG}}\setminus\{\BFx',\BFx''\}\}$.
Moreover, for all $\BFc_{\BFl,\BFi}\in\CalA_{\tilde n}$,
\begin{equation}\label{eq:cond-var-ec}
\Var\left[\SFY(\BFc_{\BFl,\BFi})\,|\, \SFY(\BFx):\BFx \in \CalX_n^{\mathsf{TSG}}\setminus \{\BFc_{\BFl,\BFi}\}\right] = \norm{\phi_{\BFl,\BFi}}_{\ScrH_k}^{-2}.
\end{equation}
\end{proposition}

\proof{Proof of Proposition~\ref{prop:cond_var}.}
Fix arbitrary $\BFc_{\BFl',\BFi'},\BFc_{\BFl'',\BFi''}\in \CalA_{\tilde n}$ and assume that $(\BFl',\BFi')\neq (\BFl'', \BFi'')$.
Let $\CalX_{n-2}^{\mathsf{TSG}} \coloneqq \CalX_n^{\mathsf{TSG}}\setminus\{\BFc_{\BFl',\BFi'},\BFc_{\BFl'',\BFi''}\}$.
Since  $\CalX_{n-2}^{\mathsf{TSG}}$ is a TSG,
it follows from Proposition~\ref{prop:cond_dist}, particularly \eqref{eq:posterior_kernel} therein, that
\[
\Cov[\SFY(\BFc_{\BFl',\BFi'}),\SFY(\BFc_{\BFl'',\BFi''})\,|\,\SFY(\BFx),\BFx\in\CalX_{n-2}^{\mathsf{TSG}}] =
\sum_{(\BFl,\BFi):\BFl\in\NatInt^d,\BFi\in\rho(\BFl),\BFc_{\BFl,\BFi}\notin\CalX_{n-2}^{\mathsf{TSG}}}
\frac{\phi_{\BFl,\BFi}(\BFc_{\BFl',\BFi'})\phi_{\BFl,\BFi}(\BFc_{\BFl'',\BFi''})}{\|\phi_{\BFl,\BFi}\|_{\ScrH_k}^2}. \]

Further, note that
for any $(\BFl,\BFi)$ such that  $\BFl\in\NatInt^d$, $\BFi\in\rho(\BFl)$, and $(\BFl,\BFi)\notin \CalX_{n-2}^{\mathsf{TSG}}$,
it must satisfy one of the following conditions:
\begin{enumerate}[label=(\roman*)]
    \item $(\BFl,\BFi) \in \CalX_{\tau+1}^{\mathsf{SG}}\setminus \CalX_\tau^{\mathsf{SG}}$, in which case $\abs{\BFl} = \abs{\BFl'} = \abs{\BFl''}=\tau+d$;
    \item $(\BFl,\BFi)\in \CalX_{\tilde\tau}^{\mathsf{SG}}\setminus \CalX_{\tau+1}^{\mathsf{SG}}$ for some $\tilde\tau\geq \tau+2 $, in which case $\abs{\BFl} \geq \tau+d+1$.
\end{enumerate}
In either case, we can invoke Lemma~\ref{lemma:basis-eval} to deduce that
$\phi_{\BFl,\BFi}(\BFl',\BFi') = \phi_{\BFl,\BFi}(\BFl'',\BFi'') = 0$ unless $(\BFl,\BFi)\in\{(\BFl',\BFi'),(\BFl'',\BFi'')\}$.
Therefore,
\[
\Cov[\SFY(\BFc_{\BFl',\BFi'}),\SFY(\BFc_{\BFl'',\BFi''})\,|\,\SFY(\BFx),\BFx\in\CalX_{n-2}^{\mathsf{TSG}}]
=\sum_{(\BFl,\BFi) \in\{ (\BFl',\BFi'), (\BFl'',\BFi'')\}}\frac{\phi_{\BFl,\BFi}(\BFc_{\BFl',\BFi'})\phi_{\BFl,\BFi}(\BFc_{\BFl'',\BFi''})}{\|\phi_{\BFl,\BFi}\|_{\ScrH_k}^2}
= 0,
\]
where the second equality can be easily seen from part (iii) of Lemma~\ref{lemma:basis-eval}.

Likewise, we can calculate the conditional variance of $\SFY(\BFc_{\BFl',\BFi'})$.
Let
$\CalX_{n-1}^{\mathsf{TSG}} \coloneqq \CalX_n^{\mathsf{TSG}}\setminus\{\BFc_{\BFl',\BFi'}\}$.
Then,
\[
\Var\left[\SFY(\BFc_{\BFl',\BFi'})\,|\, \SFY(\BFx):\BFx \in \CalX_{n-1}^{\mathsf{TSG}}\right]
=\sum_{(\BFl,\BFi):\BFl\in\NatInt^d,\BFi\in\rho(\BFl),\BFc_{\BFl,\BFi}\notin\CalX_{n-1}^{\mathsf{TSG}}}
\frac{\phi_{\BFl,\BFi}^2(\BFc_{\BFl',\BFi'})}{\|\phi_{\BFl,\BFi}\|_{\ScrH_k}^2}
=\frac{\phi_{\BFl',\BFi'}^2(\BFc_{\BFl',\BFi'})}{\|\phi_{\BFl',\BFi'}\|_{\ScrH_k}^2}=
\frac{1}{\|\phi_{\BFl',\BFi'}\|_{\ScrH_k}^2}. \Halmos\]
\endproof

\section{Convergence Rate Analysis for Deterministic Simulation}

In this section, we analyze the convergence rates for the (deterministic) kriging predictor with TM kernels and TSG designs.
Specifically, we prove  Theorem~\ref{theo:determ-TSG}, Proposition \ref{prop:FG-convergence}, and Theorem~\ref{theo:misspec-determ}.

\subsection{Proof of Theorem~\ref{theo:determ-TSG}}

The proof consists of two main steps.
First, we establish the convergence rate if the design points happen to form a classical SG $\CalX_\tau^{\mathsf{SG}}$.
Second, we generalize the result to the TSG $\CalX_{n}^{\mathsf{TSG}}$, by studying the asymptotic difference between the MSE conditioned on $\CalX_\tau^{\mathsf{SG}}$ and that on $\CalX_{\tau+1}^{\mathsf{SG}}$, the two classical SGs of consecutive levels that enclose $\CalX_{n}^{\mathsf{TSG}}$.

\begin{proposition}\label{prop:determ-SG}
Let $\{\SFY(\BFx):\BFx\in(0,1)^d\}$ be a zero mean TMGP with kernel $k$  that satisfies Assumption~\ref{assump:SL}.
Suppose that the true response surface is a realization of $\SFY$, the simulation has no noise, and the design points $\{\BFx_1,\ldots,\BFx_n\}$  form a classical SG.
Let $\widehat \SFY_n(\BFx)$ be the kriging predictor with kernel $k$ and the classical SG design.
Then, as $n\to\infty$,
\begin{equation}\label{eq:upper-bound-determ-SG}
    \sup_{\BFx\in(0,1)^d}\E[(\widehat \SFY_n(\BFx) - \SFY(\BFx))^2] = \CalO\left(n^{-1}(\log n)^{2(d-1)}\right).
\end{equation}
\end{proposition}

\proof{Proof of Proposition~\ref{prop:determ-SG}.}
In the absence of simulation noise, the MSE of $\widehat \SFY_n(\BFx)$ is simply the posterior variance of $\SFY(\BFx)$ given the observations of $\SFY$ on the classical SG $\CalX_\tau^{\mathsf{SG}}$.
It then follows from Proposition~\ref{prop:cond_dist} that
\begin{equation}\label{eq:MSE-bound-1}
\E[(\widehat \SFY_n(\BFx)-\SFY(\BFx))^2] =
\sum_{|\BFl|>\tau+d-1}\sum_{\BFi\in\rho(\BFl)}\frac{\phi^2_{\BFl,\BFi}(\BFx)}{\|\phi_{\BFl,\BFi}\|_{\ScrH_k}^2}.
\end{equation}
Note that for a fixed $\BFl$, $\phi_{\BFl,\BFi}$'s have disjoint supports by definition, and they form a partition of the design space $(0,1)^d$.
Thus, there exists one and only one $\BFi^*(\BFl)$ such that $\BFx\in \supp(\phi_{\BFl,\BFi^*(\BFl)})$, implying that
$\phi_{\BFl,\BFi^*(\BFl)}(\BFx)=0$ for all $\BFi\neq \BFi^*(\BFl)$.
Hence, by \eqref{eq:MSE-bound-1},
\begin{equation}\label{eq:MSE-bound-2}
\sup_{\BFx\in(0,1)^d}\E[(\widehat \SFY_n(\BFx)-\SFY(\BFx))^2]
=
\sup_{\BFx\in(0,1)^d}
\sum_{|\BFl|>\tau+d-1}\frac{\phi^2_{\BFl,\BFi^*(\BFl)}(\BFx)}{\|\phi_{\BFl,\BFi^*(\BFl)}\|_{\ScrH_k}^2}
\leq \sum_{|\BFl|>\tau+d-1} \frac{1}{\|\phi_{\BFl,\BFi^*(\BFl)}\|_{\ScrH_k}^2},
\end{equation}
where the inequality holds because
$\phi_{\BFl,\BFi}(\BFx)\in[0,1]$ for all $(\BFl,\BFi)$, which can be easily seen from its definition.
We now apply Lemma~\ref{lemma:RKHS_norm_estimate} to obtain the following estimate,
\begin{equation}\label{eq:MSE-bound-3}
\sum_{|\BFl|>\tau+d-1} \frac{1}{\|\phi_{\BFl,\BFi^*(\BFl)}\|_{\ScrH_k}^2}
\asymp \sum_{|\BFl|>\tau+d-1} 2^{-\abs{\BFl}}
=\CalO( 2^{-\tau} \tau^{d-1}),
\end{equation}
where the last equality follows from Lemma~3.7 in  \cite{BungartzGriebel04_ec}.
Last, note that $n=\abs{\CalX_\tau^{\mathsf{ SG}}}\asymp 2^\tau \tau^{d-1}$ by \eqref{eq:SSGNumpt}, so
\begin{equation}\label{eq:bigO-tau-to-n}
    \CalO( 2^{-\tau} \tau^{d-1}) = \CalO(n^{-1}\tau^{2(d-1)}) = \CalO\left(n^{-1}(\log n)^{2(d-1)}\right).
\end{equation}
We
combine \eqref{eq:MSE-bound-1}--\eqref{eq:bigO-tau-to-n} to conclude that
\[
\sup_{\BFx\in(0,1)^d}\E[(\widehat \SFY_n(\BFx)-\SFY(\BFx))^2] =
\CalO\left(n^{-1}(\log n)^{2(d-1)}\right). \Halmos
\]
\endproof

Recall that classical SGs are not specified directly via the sample size $n$ but via the level.
We now relax this restriction and generalize Proposition~\ref{prop:determ-SG} to Theorem~\ref{theo:determ-TSG}.
The key is to analyze the change in MSE between two classical SGs of consecutive levels.

\proof{Proof of Theorem \ref{theo:determ-TSG}.}
Fix an arbitrary $\BFx'\in(0,1)^d$.
By definition, $\CalX_\tau^{\mathsf{SG}} \subseteq \CalX_{n}^{\mathsf{TSG}} \subset \CalX_{\tau+1}^{\mathsf{SG}}$.
Hence,
\[
\underbrace{\Var[\SFY(\BFx')\,|\, \SFY(\BFx),\BFx\in\CalX_{\tau+1}^{\mathsf{SG}}]}_{V(\BFx'; \CalX_{\tau+1}^{\mathsf{SG}})}
\leq
\underbrace{
\Var[\SFY(\BFx')\,|\, \SFY(\BFx),\BFx\in\CalX_{n}^{\mathsf{TSG}}]}_{V(\BFx';\CalX_{n}^{\mathsf{TSG}})}
\leq
\underbrace{
\Var[\SFY(\BFx')\,|\, \SFY(\BFx),\BFx\in\CalX_\tau^{\mathsf{SG}}]}_{
V(\BFx';\CalX_\tau^{\mathsf{SG}})}.
\]
Now, we provide an upper bound on the difference between the conditional variances.
Specifically, following Proposition~\ref{prop:cond_dist},
\begin{equation}\label{eq:MSE-bound-6}
V(\BFx';\CalX_\tau^{\mathsf{SG}})
-
V(\BFx';\CalX_{\tau+1}^{\mathsf{SG}})
=\sum_{\abs{\BFl}=\tau+d} \sum_{\BFi\in\rho(\BFl)} \frac{\phi^2_{\BFl,\BFi}(\BFx')}{\|\phi_{\BFl,\BFi}\|_{\ScrH_k}^2}
\leq
\sum_{\abs{\BFl}=\tau+d} \frac{1}{\|\phi_{\BFl,\BFi}\|_{\ScrH_k}^2},
\end{equation}
where the inequality can be shown  with an argument similar to that for \eqref{eq:MSE-bound-2}.
By Lemma~\ref{lemma:RKHS_norm_estimate},
\begin{equation}\label{eq:MSE-bound-7}
\sum_{\abs{\BFl}=\tau+d} \frac{1}{\|\phi_{\BFl,\BFi}\|_{\ScrH_k}^2}
\asymp
\sum_{\abs{\BFl}=\tau+d} 2^{-\abs{\BFl}} = 2^{-(\tau+d)}  \binom{\tau+d-1}{d-1} =\CalO(2^{-\tau}\tau^{d-1}),
\end{equation}
where the last step can be shown via Stirling's approximation for factorials.
Therefore, combining \eqref{eq:MSE-bound-5} and  \eqref{eq:MSE-bound-6} yields
\begin{align*}
V(\BFx';\CalX_n^{\mathsf{TSG}})
- V(\BFx';\CalX_{\tau+1}^{\mathsf{SG}})
\leq
V(\BFx';\CalX_\tau^{\mathsf{SG}})
-
V(\BFx';\CalX_{\tau+1}^{\mathsf{SG}})
=\CalO(2^{-\tau}\tau^{d-1}).
\end{align*}
Moreover, we know from
the proof of Proposition~\ref{prop:determ-SG}, particularly
\eqref{eq:MSE-bound-1}--\eqref{eq:MSE-bound-3},
that
$
V(\BFx';\CalX_{\tau+1}^{\mathsf{SG}}) = \CalO(2^{-(\tau+1)}(\tau+1)^{d-1})
$.
Hence,
\[V(\BFx';\CalX_n^{\mathsf{TSG}})
= \CalO(2^{-\tau}\tau^{d-1}) + \CalO(2^{-(\tau+1)}(\tau+1)^{d-1}) = \CalO(2^{-\tau}\tau^{d-1})
= \CalO\left(n^{-1}(\log n)^{2(d-1)}\right)
.
\]
Note that by \eqref{eq:MSE-bound-6},
the upper bound is independent of $\BFx'$.
Hence,
\[\sup_{\BFx\in (0,1)^d}\E[(\widehat \SFY_n(\BFx)- \SFY(\BFx))^2] = \sup_{\BFx\in (0,1)^d} V(\BFx; \CalX_n^{\mathsf{TSG}})
= \CalO\left(n^{-1}(\log n)^{2(d-1)}\right).\Halmos
\]
\endproof

\subsection{Proof of Proposition \ref{prop:FG-convergence}}

\proof{Proof of Proposition~\ref{prop:FG-convergence}.}

Let $\CalX_{\tau}^{\mathsf{FG}} = \bigtimes_{j=1}^d\CalX_{j,\tau}$ denote the full grid (i.e., lattice) of level $\tau$.
Note that $\CalX_{\tau}^{\mathsf{FG}}$ can be expressed as
\[
\CalX_{\tau}^{\mathsf{FG}} = \bigcup_{\norm{\BFl}_\infty\leq \tau} \{\BFc_{\BFl,\BFi}:\BFi\in\rho(\BFl)\},
\]
where $\|\BFl\|_\infty=\max(l_1,\ldots,l_d)$.
Then, following the expansion \eqref{eq:GP-expansion} and a proof similar to that of Proposition~\ref{prop:cond_var}, it can be shown that
the conditional variance of $\SFY(\BFx)$ given the observations of $\SFY$ on the full grid $\CalX_{\tau}^{\mathsf{FG}}$  is given by
\begin{equation}\label{eq:MSE-bound-4}
\E[(\widehat \SFY_n(\BFx)-\SFY(\BFx))^2] =  \sum_{\|\BFl\|_\infty>\tau}\sum_{\BFi\in\rho(\BFl)}\frac{\phi^2_{\BFl,\BFi}(\BFx)}{\|\phi_{\BFl,\BFi}\|_{\ScrH_k}^2},
\end{equation}
for all $\BFx\in(0,1)^d$.

We fix an $\tilde\BFx\notin \CalX_\tau^{\mathsf{FG}}$.
Let $\tilde\BFl = (\tau+1,1,\ldots,1)$ and $\tilde\BFi$ be the index such that $\tilde \BFx\in\supp(\phi_{\tilde\BFl,\tilde\BFi})$.
Then, $\|\tilde\BFl\|_\infty = \tau+1$, $\abs{\BFl}=\tau+d$,   and
$\phi_{\tilde\BFl,\tilde\BFi}(\tilde\BFx) > 0$.
Hence, by\eqref{eq:MSE-bound-4},
\begin{equation}\label{eq:MSE-bound-5}
\sup_{\BFx\in(0,1)^d} \E[(\widehat \SFY_n(\BFx)-\SFY(\BFx))^2]
\geq
\E[(\widehat \SFY_n(\tilde\BFx)-\SFY(\tilde\BFx))^2]
\geq \frac{\phi^2_{\tilde\BFl,\tilde \BFi}(\tilde\BFx)}{\|\phi_{\tilde\BFl,\tilde\BFi}\|_{\ScrH_k}^2} \asymp
\phi^2_{\tilde\BFl,\tilde \BFi}(\tilde\BFx) \cdot 2^{-\abs{\BFl}} =
\phi^2_{\tilde\BFl,\tilde \BFi}(\tilde\BFx) \cdot 2^{-(\tau+d)},
\end{equation}
where the asymptotic equivalence follows from Lemma~\ref{lemma:RKHS_norm_estimate}
Last, note that  $n=\abs{\CalX_\tau^{\mathsf{FG}}}=(2^\tau-1)^d\asymp 2^{\tau d}$, so $2^\tau \asymp n^{1/d}$ as $n\to\infty$.
Hence, the proof is completed by using \eqref{eq:MSE-bound-5}.
\Halmos\endproof

\subsection{Proof of Theorem \ref{theo:misspec-determ}}
To simplify the notation, for each $n$ we define an operator $\CalP_n$ that takes effect on a function $f:(0,1)^d\mapsto \Real$ as follows,
\begin{equation}
(\CalP_n f)(\BFx)\coloneqq \BFk^\intercal (\BFx) \BFK^{-1} \BFf,\label{eq:projection-P},
\end{equation}
where $\BFk(\cdot)=(k(\BFx_1,\cdot),\ldots, k(\BFx_n,\cdot))^\intercal $, $\BFf=(f(\BFx_1),\ldots,f(\BFx_n))^\intercal$.

We also define an operator $\CalL_{\BFx}$ that take effect on a function $f(\BFx,\tilde\BFx):(0,1)^d\times (0,1)^d\to\Real$ as follows,
\begin{align}
(\CalL_{\BFx} f(\cdot,\tilde\BFx))(\BFx)\coloneqq{} & (\CalL_{1,x_1}\cdots \CalL_{d,x_d} f(\cdot,\tilde\BFx))(\BFx), \label{eq:operator-L}
\end{align}
where $\CalL_{j,x_j}$ is the differential operator $\CalL_j$, defined in \eqref{eq:diff-op-L}, acting on the $j$-th variable $x_j$ of $\BFx$, i.e.,
\[
\bigl(\CalL_{j,x_j} f(\cdot,\tilde \BFx)\bigr)(\BFx)= \biggl(\frac{\dd }{\dd x_j}u_j(x_j) \frac{\dd }{\dd x_j} + v_j(x_j)  f(\cdot,\tilde\BFx)\biggr)(\BFx).
\]

\proof{Proof of Theorem \ref{theo:misspec-determ}.}
Note that the misspecified kriging predictor $\widehat \SFY_n^{\mathsf{mis}}(\BFx) = (\CalP_n \SFY^*)(\BFx)$, so
\begin{equation}\label{eq:MSE-misspec}
\begin{aligned}
&\E[(\SFY^*(\BFx) - (\CalP_n \SFY^*)(\BFx))^2] \\
={}& \underbrace{\E[(\SFY^*(\BFx))^2 - \SFY^*(\BFx) (\CalP_n \SFY^*)(\BFx)]}_{I_1(\BFx)}
- \underbrace{\E[\SFY^*(\BFx) (\CalP_n \SFY^*)(\BFx) - ((\CalP_n \SFY^*)(\BFx))^2]}_{I_2(\BFx)}.
\end{aligned}
\end{equation}
We first consider $I_1(\BFx)$.
Note that $\E[(\SFY^*(\BFx))^2] = \Var[\SFY^*(\BFx)] = k^*(\BFx,\BFx)$,
since  $\SFY^*$ is a zero mean GP with kernel $k^*$; moreover,
\begin{align}
\E[\SFY^*(\BFx)(\CalP_n \SFY^*)(\BFx)] ={}& \E[\BFk^\intercal(\BFx)\BFK^{-1} (\SFY^*(\BFx_1),\ldots,\SFY^*(\BFx_n))^\intercal \SFY^*(\BFx)] \nonumber\\
={}& \BFk^\intercal(\BFx)\BFK^{-1} (k^*(\BFx_1,\BFx),\ldots,k^*(\BFx_n,\BFx))^\intercal \nonumber\\
={}& (\CalP_n k^*(\cdot,\BFx))(\BFx). \label{eq:cross-term}
\end{align}
Since we assume that $k^*$ has a product form with twice differentiable factor, we have that $k^*\in\ScrH_k$.

For all $f\in\ScrH_k$, note that $(\CalP_n f)(\cdot)$ can be expressed as a linear combination of $\{k(\BFx_i,\cdot):i=1,\ldots,n\}$, so it lies in $\ScrF_n$, the span of $\{k(\BFx_i,\cdot):i=1,\ldots,n\}$.
Since the design points $\{\BFx_1,\ldots,\BFx_n\}$ form a TSG $\CalX_n^{\mathsf{TSG}}$,
by the orthogonal expansion \eqref{eq:TMK_expansion},
$\ScrF_n$ is identical to $\ScrG_n$, the span of $\{\phi_{\BFl,\BFi}: \BFc_{\BFl,\BFi}\in\CalX_n^{\mathsf{TSG}}\}$.
Thus,
$(\CalP_n f)(\cdot)\in \ScrG_n$, and it can decomposed as follows.
\begin{equation}\label{eq:proj-expansion}
(\CalP_n f)(\cdot) =
\sum_{(\BFl,\BFi):\BFc_{\BFl,\BFi}\in\CalX_n^{\mathsf{TSG}}} \frac{\langle \CalP_n f, \phi_{\BFl,\BFi} \rangle_{\ScrH_k}}{ \norm{\phi_{\BFl,\BFi}}^2_{\ScrH_k}} \phi_{\BFl,\BFi}(\cdot)
= \sum_{(\BFl,\BFi):\BFc_{\BFl,\BFi}\in\CalX_n^{\mathsf{TSG}}} \frac{\langle f, \phi_{\BFl,\BFi} \rangle_{\ScrH_k}}{ \norm{\phi_{\BFl,\BFi}}^2_{\ScrH_k}} \phi_{\BFl,\BFi}(\cdot).
\end{equation}
It follows from Lemma~\ref{lemma:TMK_expansion} and \eqref{eq:proj-expansion} that for all $\BFx,\BFx'\in (0,1)^d$,
\begin{align}
& |k^*(\BFx,\BFx') -  (\CalP_n k^*(\cdot,\BFx'))(\BFx)| \nonumber \\
={}& \biggl|\sum_{(\BFl,\BFi):\BFl\in\NatInt^d,\BFi\in\rho(\BFl)}\frac{\langle k^*(\cdot,\BFx'), \phi_{\BFl,\BFi} \rangle_{\ScrH_k}}{ \norm{\phi_{\BFl,\BFi}}^2_{\ScrH_k}} \phi_{\BFl,\BFi}(\BFx)
 -
\sum_{(\BFl,\BFi):\BFc_{\BFl,\BFi}\in\CalX_n^{\mathsf{TSG}}} \frac{\langle k^*(\cdot,\BFx'), \phi_{\BFl,\BFi} \rangle_{\ScrH_k}}{ \norm{\phi_{\BFl,\BFi}}^2_{\ScrH_k}} \phi_{\BFl,\BFi}(\BFx) \biggr| \nonumber \\
\leq{}&\sum_{(\BFl,\BFi):\BFl\in\NatInt^d,\BFi\in\rho(\BFl), \BFc_{\BFl,\BFi}\notin\CalX_n^{\mathsf{TSG}}} \frac{|\langle k^*(\cdot,\BFx'), \phi_{\BFl,\BFi} \rangle_{\ScrH_k}|}{ \norm{\phi_{\BFl,\BFi}}^2_{\ScrH_k}} \phi_{\BFl,\BFi}(\BFx) \nonumber \\
\leq{}&  \sum_{|\BFl|> \tau+d-1}\sum_{\BFi\in\rho(\BFl)}  \frac{|\langle k^*(\cdot,\BFx'), \phi_{\BFl,\BFi} \rangle_{\ScrH_k}|}{ \norm{\phi_{\BFl,\BFi}}^2_{\ScrH_k}} \phi_{\BFl,\BFi}(\BFx)  \nonumber \\
={}&  \sum_{|\BFl|> \tau+d-1}  \frac{|\langle k^*(\cdot,\BFx'), \phi_{\BFl,\BFi^*(\BFl)} \rangle_{\ScrH_k}|}{ \norm{\phi_{\BFl,\BFi^*(\BFl)}}^2_{\ScrH_k}} \phi_{\BFl,\BFi^*(\BFl)}(\BFx) \nonumber \\
\leq{}& \sum_{|\BFl|> \tau+d-1}  \frac{|\langle k^*(\cdot,\BFx'), \phi_{\BFl,\BFi^*(\BFl)} \rangle_{\ScrH_k}|}{ \norm{\phi_{\BFl,\BFi^*(\BFl)}}^2_{\ScrH_k}},
\label{eq:residual-expansion}
\end{align}
where the second inequality holds because $\CalX_n^{\mathsf{TSG}}\supseteq \CalX_\tau^{\mathsf{SG}} = \{\BFc_{\BFl,\BFi}:|\BFl|\leq \tau+d-1,\BFi\in\rho(\BFl)\}$; the last equality holds because given $\BFl$, there exists a unique $\BFi^*(\BFl)$ such that $\BFx'\in \supp(\phi_{\BFl,\BFi^*(\BFl)})$, implying that
$\phi_{\BFl,\BFi}(\BFx')=0$ for all $\BFi\neq \BFi^*(\BFl)$, since $\phi_{\BFl,\BFi}$'s have disjoint supports by definition, and they form a partition of the design space $(0,1)^d$; the last inequality holds because $\phi_{\BFl,\BFi}\in[0,1]$ by definition.
In the following, we give an estimate of the inner product $\langle k^*(\cdot,\BFx'), \phi_{\BFl,\BFi} \rangle_{\ScrH_k}$.

Since both $k^*$ and $\phi_{\BFl,\BFi}$ are of product form,
for all $(\BFl,\BFi)$ and $\BFx'=( x_1',\ldots, x_d')\in(0,1)^d$,
\begin{align}
|\langle k^*(\cdot,\BFx'),\phi_{\BFl,\BFi}\rangle_{\ScrH_k}|
={}&\biggl|\int_{(0,1)^d} \bigl(\CalL_{\BFx} k^*(\cdot,\BFx')\bigr)(\BFx)\phi_{\BFl,\BFi}(\BFx)\dd{\BFx}\biggr|\nonumber \\
={}& \biggl|\prod_{j=1}^d \int_0^1\phi_{l_j,i_j}(x_j) \bigl(\CalL_{j,x_j} k_j^*(\cdot, x_j')\bigr)(x_j)\dd{x_j} \biggr| \nonumber \\
\leq{}& \prod_{j=1}^d  \int_0^1\phi_{l_j,i_j}(x)\dd{x} \cdot \sup_{x_j\in(0,1)}\biggl| \bigl(\CalL_{j,x_j} k_j^*(\cdot,x_j')\bigr)(x_j) \biggr|. \label{eq:inner-prod-estimate}
\end{align}
By the Cauchy–Schwarz inequality, for all $x'_j\in [0,1]$,
\begin{align}
\sup_{x_j\in(0,1)}\bigl|\bigl(\CalL_{j,x_j} k_j^*(\cdot,x'_j)\bigr)(x_j)\bigr|
={}&  \sup_{x_j\in(0,1)}\bigl|\bigl\langle \bigl(\CalL_{j,x_j} k_j^*(\cdot,x_j)\bigr),k(\cdot,x'_j)\bigr\rangle_{\ScrH_k}\bigr| \nonumber \\
\leq{} &\sup_{x'_j\in(0,1)}\sqrt{k_j(x'_j,x'_j)} \cdot \sup_{x_j\in(0,1)}\bigl\|\bigl(\CalL_{j,x_j} k_j^*(x_j,\cdot)\bigr)\bigr\|_{\ScrH_{k_j}} \nonumber\\
\coloneqq{}& C_1 <\infty, \label{eq:inner-prod-estimate-2}
\end{align}
where the last step holds because of Assumption~\ref{assump:smooth-kernel}.

Further, note that
\begin{equation}\label{eq:volume}
\prod_{j=1}^d\int_0^1\phi_{l_j,i_j}(x) \dd{x} \leq  \prod_{j=1}^d \int_{(i-1)\cdot 2^{-l_j}}^{(i+1)\cdot 2^{-l_j}} \dd{x} = \prod_{j=1}^d 2^{-(l_j-1)} = 2^{d-\abs{\BFl}},
\end{equation}
since $\phi_{l_j,i_j}\in[0,1]$ with support  $((i-1)\cdot 2^{-l_j}, (i+1)\cdot 2^{-l_j})$.

Hence, by \eqref{eq:inner-prod-estimate}, \eqref{eq:inner-prod-estimate-2}, and \eqref{eq:volume},
\begin{equation}\label{eq:k*projectEstimate}
    |\langle k^*(\cdot,\BFx'),\phi_{\BFl,\BFi}\rangle_{\ScrH_k}| \leq C_1 2^{d-\abs{\BFl}},
\end{equation}
for all $\BFx'\in(0,1)^d$.
Applying \eqref{eq:residual-expansion} and Lemma~\ref{lemma:RKHS_norm_estimate},  \begin{equation}\label{eq:I_1-estimate}
|k^*(\BFx,\BFx') -  (\CalP_n k^*(\cdot,\BFx'))(\BFx)|
\leq \sum_{|\BFl|> \tau+d-1} C_2 2^{d-2\abs{\BFl}}
\leq C_3 2^{-2\tau}\tau^{d-1} \leq C_4 n^{-2} (\log n)^{3(d-1)},
\end{equation}
for all $\BFx,\BFx'\in(0,1)^d$ and some constants $C_i>0$, $i=2,3,4$,
where the second equality follows from Lemma~3.7 in \cite{BungartzGriebel04_ec}, while the last from
an argument similar to \eqref{eq:bigO-tau-to-n}.
By \eqref{eq:I_1-estimate},
\begin{equation}\label{eq:I_1-estimate-2}
|I_1(\BFx)| = |k^*(\BFx,\BFx) -  (\CalP_n k^*(\cdot,\BFx))(\BFx)| \leq C_4 n^{-2} (\log n)^{3(d-1)}.
\end{equation}

We now consider $I_2(\BFx)$ in \eqref{eq:MSE-misspec}.
Let $h(\BFx,\BFx')\coloneqq (\CalP_n k^*(\cdot,\BFx'))(\BFx)$.
Then,
\begin{equation}\label{eq:h-Ph}
h(\BFx,\BFx') - (\CalP_n h(\BFx', \cdot))(\BFx)
= \Bigl( \CalP_n \bigl(k^*(\cdot, \BFx') - h(\BFx',\cdot) \bigr)\Bigr)(\BFx).
\end{equation}
Applying \eqref{eq:proj-expansion} to expand   $k^*(\BFx, \BFx') - h(\BFx',\BFx)$ as a function in the variable $\BFx$, and switching the order of summation and $\CalP_n $,  \eqref{eq:h-Ph} then reads:
\begin{equation}\label{eq:h-Ph-2}
h(\BFx,\BFx') - (\CalP_n h(\BFx', \cdot))(\BFx) = \sum_{(\BFl,\BFi):\BFl\in\NatInt^d,\BFi\in\rho(\BFl), \BFc_{\BFl,\BFi}\notin\CalX_n^{\mathsf{TSG}}} \frac{\CalP_n\bigl( \langle k^*(\cdot,\cdot), \phi_{\BFl,\BFi} \rangle_{\ScrH_k}\bigr)(\BFx)}{ \norm{\phi_{\BFl,\BFi}}^2_{\ScrH_k}} \phi_{\BFl,\BFi}(\BFx'),
\end{equation}
where
\begin{align*}
    \quad & \CalP_n\bigl( \langle k^*(\cdot,\cdot), \phi_{\BFl,\BFi} \rangle_{\ScrH_k}\bigr)(\BFx)\nonumber \\
    = & \CalP_n\biggl( \int_{(0,1)^d}\phi_{\BFl,\BFi}(\tilde\BFx) \bigl(\CalL_{\tilde\BFx} k^*(\cdot, \tilde\BFx)\bigr)\dd{\tilde\BFx} \biggr)(\BFx)\nonumber\\
    ={} & \int_{(0,1)^d}\phi_{\BFl,\BFi}(\tilde\BFx) \CalL_{\tilde\BFx} \Bigl(\CalP_n\bigl(k^*(\cdot, \tilde\BFx)\bigr)(\BFx)\Bigr)\dd{\tilde\BFx}\nonumber\\
    ={}& \int_{(0,1)^d}\phi_{\BFl,\BFi}(\tilde\BFx) \CalL_{\tilde\BFx} \biggl(\sum_{(\BFl',\BFi'):\BFc_{\BFl',\BFi'}\in\CalX_n^{\mathsf{TSG}}} \frac{\langle k^*(\cdot,\tilde\BFx), \phi_{\BFl',\BFi'} \rangle_{\ScrH_k}}{ \norm{\phi_{\BFl',\BFi'}}^2_{\ScrH_k}} \phi_{\BFl',\BFi'}(\BFx)\biggr)\dd{\tilde\BFx}\nonumber\\
    ={} & \sum_{(\BFl',\BFi'):\BFc_{\BFl',\BFi'}\in\CalX_n^{\mathsf{TSG}}}\frac{\phi_{\BFl',\BFi'}(\BFx)}{ \norm{\phi_{\BFl',\BFi'}}^2_{\ScrH_k}} {\int_{(0,1)^d}\int_{(0,1)^d}\phi_{\BFl,\BFi}(\tilde\BFx)\bigl(\CalL_{\tilde\BFx}k^*(\BFx',\tilde\BFx)\bigr)\bigl(\CalL_{\BFx'}\phi_{\BFl',\BFi'}(\BFx')\bigr)\dd{\tilde\BFx}\dd{\BFx'}}.
\end{align*}
Therefore,
\begin{align}
& \bigl|\CalP_n\bigl( \langle k^*(\cdot,\cdot), \phi_{\BFl,\BFi} \rangle_{\ScrH_k}\bigr)(\BFx)\bigr| \nonumber\\
    \leq{} & \sum_{(\BFl',\BFi'):\BFc_{\BFl',\BFi'}\in\CalX_n^{\mathsf{TSG}}}\frac{\phi_{\BFl',\BFi'}(\BFx)}{ \norm{\phi_{\BFl',\BFi'}}^2_{\ScrH_k}}
    \int_{(0,1)^d} \phi_{\BFl,\BFi}(\tilde\BFx)\dd{\tilde\BFx}
    \cdot \sup_{\tilde\BFx\in(0,1)^d}\biggl|\int_{(0,1)^d}\bigl(\CalL_{\tilde\BFx}k^*(\BFx',\tilde\BFx)\bigr)\bigl(\CalL_{\BFx'} \phi_{\BFl',\BFi'}(\BFx')\bigr)\dd{\BFx'}\biggr|\nonumber\\
    = {}& \int_{(0,1)^d} \phi_{\BFl,\BFi}(\tilde\BFx)\dd{\tilde\BFx} \cdot  \sum_{(\BFl',\BFi'):\BFc_{\BFl',\BFi'}\in\CalX_n^{\mathsf{TSG}}}\frac{\phi_{\BFl',\BFi'}(\BFx)}{ \norm{\phi_{\BFl',\BFi'}}^2_{\ScrH_k}} \cdot \sup_{\tilde\BFx\in(0,1)^d}\bigl| \bigl\langle \CalL_{\tilde\BFx}k^*(\cdot,\tilde\BFx), \phi_{\BFl',\BFi'} \bigr\rangle_{\ScrH_k} \bigr|
    \nonumber \\
    \leq{}& \int_{(0,1)^d} \phi_{\BFl,\BFi}(\tilde\BFx)\dd{\tilde\BFx} \cdot  \sum_{(\BFl',\BFi'):\BFc_{\BFl',\BFi'}\in\CalX_n^{\mathsf{TSG}}}\frac{\phi_{\BFl',\BFi'}(\BFx)}{ \norm{\phi_{\BFl',\BFi'}}_{\ScrH_k}} \cdot \sup_{\tilde\BFx\in(0,1)^d} \bigl\|\CalL_{\tilde\BFx} k^*(\cdot,\tilde\BFx)\bigr\|_{\ScrH_k},
    \label{eq:estimate-Pn-h}
\end{align}
where the last inequality follows from the Cauchy--Schwarz inequality.

By Assumption~\ref{assump:smooth-kernel},
\begin{equation}\label{eq:C_5}
C_5 \coloneqq \sup_{\tilde\BFx\in(0,1)^d} \bigl\|\bigl(\CalL_{\tilde\BFx} k^*(\cdot,\tilde\BFx)\bigr)\bigr\|_{\ScrH_k} = \prod_{j=1}^d \bigl\|\bigl(\CalL_{j,\tilde x_j} k^*_j(\cdot,\tilde x_j)\bigr)\bigr\|_{\ScrH_{k_j}} <\infty.
\end{equation}

Moreover, applying Lemma~\ref{lemma:RKHS_norm_estimate}, we have
\begin{align}
\sum_{(\BFl',\BFi'):\BFc_{\BFl',\BFi'}\in\CalX_n^{\mathsf{TSG}}}
\frac{\phi_{\BFl',\BFi'}(\BFx)}{ \norm{\phi_{\BFl',\BFi'}}_{\ScrH_k}}
\leq{}& \sum_{|\BFl'|\geq d} \sum_{\BFi'\in\rho(\BFl')}
\frac{\phi_{\BFl',\BFi'}(\BFx)}{ \norm{\phi_{\BFl',\BFi'}}_{\ScrH_k}}
\nonumber \\
={}& \sum_{|\BFl'|\geq d} \frac{\phi_{\BFl',\BFi^*(\BFl')}(\BFx)}{ \norm{\phi_{\BFl',\BFi^*(\BFl')}}_{\ScrH_k}}   \nonumber \\
\leq{}& C_6 \sum_{|\BFl'|\geq d}   2^{-|\BFl'|/2}  \coloneqq C_7 <\infty,  \label{eq:sum-HKnorm}
\end{align}
for some constant $C_6>0$,
where $\BFi^*(\BFl')$ is the unique index in $\rho(\BFl')$ for which $\BFx\in \supp(\phi_{\BFl',\BFi^*(\BFl')})$, and the last step follows from Lemma~3.7 in  \cite{BungartzGriebel04_ec}.

Combining \eqref{eq:volume}, \eqref{eq:estimate-Pn-h}, \eqref{eq:C_5}, and \eqref{eq:sum-HKnorm} yields
\[
\bigl|\CalP_n\bigl( \langle k^*(\cdot,\cdot), \phi_{\BFl,\BFi} \rangle_{\ScrH_k}\bigr)(\BFx)\bigr|
\leq C_5 C_7 2^{d -|\BFl|},
\]
for all $\BFx\in(0,1)^d$.
It then follows from \eqref{eq:h-Ph-2} that
\begin{align*}
|h(\BFx,\BFx') - (\CalP_n h(\BFx', \cdot))(\BFx) | \leq {}&
\sum_{(\BFl,\BFi):\BFl\in\NatInt^d,\BFi\in\rho(\BFl), \BFc_{\BFl,\BFi}\notin\CalX_n^{\mathsf{TSG}}}
C_5C_7 2^{d-|\BFl|} \frac{\phi_{\BFl,\BFi}(\BFx')}{\norm{\phi_{\BFl,\BFi}}^2_{\ScrH_k}} \\
\leq{}& \sum_{|\BFl|> \tau+d-1}  \sum_{\BFi\in\rho(\BFl)}
C_5C_7 2^{d-|\BFl|} \frac{\phi_{\BFl,\BFi}(\BFx')}{\norm{\phi_{\BFl,\BFi}}^2_{\ScrH_k}} \\
= {}&   \sum_{|\BFl|> \tau+d-1} C_5C_7 2^{d-|\BFl|} \frac{\phi_{\BFl,\BFi^*(\BFl)}(\BFx')}{\norm{\phi_{\BFl,\BFi^*(\BFl)}}^2_{\ScrH_k}} \\
\leq{}& \sum_{|\BFl|> \tau+d-1} C_8 2^{d-2|\BFl|} \\
\leq{}& C_9 n^{-2} (\log n)^{3(d-1)},
\end{align*}
for some positive constants $C_8$ and $C_9$,
where $\BFi^*(\BFl)$ is the unique index in $\rho(\BFl)$ for which $\BFx'\in \supp(\phi_{\BFl,\BFi^*(\BFl)})$, and the last step follows an argument similar to   \eqref{eq:I_1-estimate}.
Hence,
\begin{equation}\label{eq:h-Ph-3}
\sup_{\BFx\in(0,1)^d}|h(\BFx,\BFx') - (\CalP_n h(\BFx', \cdot))(\BFx) | \leq C_9 n^{-2}(\log n)^{3(d-1)},
\end{equation}
for all $\BFx'\in(0,1)^d$.

It can be shown via a calculation similar to \eqref{eq:cross-term} that $\E[((\CalP_n \SFY^*)(\BFx))^2] = (\CalP_n h(\BFx,\cdot))(\BFx)$.
Therefore, by \eqref{eq:h-Ph-3},
\begin{equation}\label{eq:I_2-estimate}
    |I_2(\BFx)| = |h(\BFx,\BFx) - (\CalP_n h(\cdot, \BFx))(\BFx)| \leq C_9 n^{-2} (\log n)^{3(d-1)}.
\end{equation}
The proof is completed in light of \eqref{eq:MSE-misspec}, \eqref{eq:I_1-estimate-2}, \eqref{eq:I_2-estimate}, and the fact that the constants $C_i$'s are independent of $\BFx$.
\Halmos\endproof

\section{Convergence Rate Analysis for Stochastic Simulation}

In this section, we analyze the convergence rates for the stochastic kriging predictor with TM kernels and TSG designs.
Specifically, we prove Theorem~\ref{theo:stoch-TSG} and Theorem~\ref{theo:misspec-stoch}.

In addition to the operator $\CalP_n$, we define the following operator for each $\lambda>0$,
\begin{equation}\label{eq:projection-P_lambda}
(\CalP_n^\lambda f)(\cdot) \coloneqq \BFk^\intercal(\cdot) (\BFK+\lambda \BFI)^{-1} \BFf,
\end{equation}
where $\BFk(\cdot)=(k(\BFx_1,\cdot),\ldots, k(\BFx_n,\cdot))^\intercal $, $\BFf=(f(\BFx_1),\ldots,f(\BFx_n))^\intercal$, and $\BFI$ is the $n\times n$ identity matrix.

\begin{lemma}\label{lemma:inf-norm-subspace}
Suppose the design points $\{\BFx_1,\ldots,\BFx_n\}$ form a TSG $\CalX_n^{\mathsf{TSG}}$.
Let $\ScrG_n$ be the span of $\{\phi_{\BFl,\BFi}: \BFc_{\BFl,\BFi}\in\CalX_n^{\mathsf{TSG}}\}$.
Then,
\[
    \sup_{\BFx\in(0,1)^d}|g(\BFx)| \leq C (\log n)^d \cdot \sup_{1\leq i \leq n}|g(\BFx_i)|,
\]
for all $g\in\ScrG_n$ and some constant $C>0$ that is independent of $g$ and $n$.
\end{lemma}
\proof{Proof of Lemma~\ref{lemma:inf-norm-subspace}.}

Fix an arbitrary $g\in\ScrG_n$ and decompose it as follows,
\begin{equation}\label{eq:g-decomp}
g(\BFx) = \sum_{(\BFl,\BFi):\BFc_{\BFl,\BFi}\in\CalX_n^{\mathsf{TSG}}}
\biggl\langle g,\frac{\phi_{\BFl,\BFi}}{\|\phi_{\BFl,\BFi}\|^2_{\ScrH_k}}\biggr\rangle_{\ScrH_k}
\phi_{\BFl,\BFi}(\BFx).
\end{equation}
By virtue of Lemma~\ref{lemma:RKHS_norm_estimate} and \eqref{eq:phi_li} , we have
\begin{equation}\label{eq:prod-neighbors}
\frac{\phi_{\BFl,\BFi}(\BFx)}{\|\phi_{\BFl,\BFi}\|^2_{\ScrH_k}}=\prod_{j=1}^d\biggl(k_j(x_j,\SFc_{l_j,i_j})+\frac{\eta_{l_j,i_j-1}}{\eta_{l_j,i_j}}k_j(x_j,\SFc_{l_j,i_j-1})+\frac{\eta_{l_j,i_j+1}}{\eta_{l_j,i_j}}k_j(x_j, \SFc_{l_j,i_j+1})\biggr),
\end{equation}
where  $\eta_{l,i}$ is given by \eqref{eq:eta_li}.
For simplicity, we first consider $d=1$ and suppress the dependence on $j$.
For all $(l,i)$ such that $\SFc_{l,i}\in \CalX_n^{\mathsf{TSG}}$,
\begin{align}
\biggl|\biggl\langle g,\frac{\phi_{l,i}}{\|\phi_{l,i}\|^2_{\ScrH_k}}\biggr\rangle_{\ScrH_k} \biggr|
={}& \biggl|\biggl\langle g,k(\SFc_{l,i},\cdot)+\frac{\eta_{l,i-1}}{\eta_{l,i}}k(\SFc_{l,i-1},\cdot)+\frac{\eta_{l,i+1}}{\eta_{l,i}}k(\SFc_{l,i+1},\cdot)\biggr\rangle_{\ScrH_k}\biggr| \nonumber\\
={}&\biggl|g(\SFc_{l,i})+\frac{\eta_{l,i-1}}{\eta_{l,i}}g(\SFc_{l,i-1})+ \frac{\eta_{l,i+1}}{\eta_{l,i}}g(\SFc_{l,i+1})\biggr| \label{eq:reproducing} \\
\leq{}& \biggl|1+\frac{\eta_{l,i-1}}{\eta_{l,i}}+ \frac{\eta_{l,i+1}}{\eta_{l,i}}\biggr|\cdot \sup_{i-1\leq i'\leq i+1} |g(\SFc_{l,i'})| \nonumber \\
\leq{}& C_2 \sup_{(l,i):\SFc_{l,i}\in\CalX_n^{\mathsf{TSG}}}|g(\SFc_{l,i})|, \label{eq:max-estimate}
\end{align}
for  some constant $C_2>0$ independent of $g,l,i$.
Here, \eqref{eq:reproducing} follows from the reproducing property of the kernel $k$,
and \eqref{eq:max-estimate} follows from the fact that  ${\eta_{l,i'}}/{\eta_{l,i}}$ converges to $-\frac{1}{2}$, $i'=i\pm 1$ as $l\to\infty$, which can be seen from \eqref{eq:eta-estimate}.

For a general $d\geq 2$, we can apply the same argument to show that,
\begin{align}
     \biggl|\biggl\langle g,\frac{\phi_{\BFl,\BFi}}{\|\phi_{\BFl,\BFi}\|^2_{\ScrH_k}}\biggr\rangle_{\ScrH_k}   \biggr|
    ={}&\biggl| \biggl\langle g,\prod_{j=1}^d\biggl(k_j(\SFc_{l_j,i_j},\cdot)+\frac{\eta_{l_j,i_j-1}}{\eta_{l_j,i_j}}k_j(\SFc_{l_j,i_j-1},\cdot)+\frac{\eta_{l_j,i_j+1}}{\eta_{l_j,i_j}}k_j( \SFc_{l_j,i_j+1},\cdot)\biggr)\biggr\rangle_{\ScrH_k} \biggr| \nonumber \\
    \leq{}& C_2^d \sup_{(\BFl,\BFi):\BFc_{\BFl,\BFi}\in\CalX_n^{\mathsf{TSG}}}|g(\SFc_{\BFl,\BFi})|, \label{eq:max-estimate-general}
\end{align}
for all $(\BFl,\BFi)$ such that $\BFc_{\BFl,\BFi}\in \CalX_n^{\mathsf{TSG}}$.
It follows from \eqref{eq:g-decomp}  and \eqref{eq:max-estimate-general} that
\begin{equation}\label{eq:sample-inequality}
|g(\BFx)| \leq C_2^d \sup_{(\BFl,\BFi):\BFc_{\BFl,\BFi}\in\CalX_n^{\mathsf{TSG}}}|g(\SFc_{\BFl,\BFi})|\cdot \sum_{{(\BFl,\BFi):\BFc_{\BFl,\BFi}\in\CalX_n^{\mathsf{TSG}}}} \phi_{\BFl,\BFi}(\BFx).
\end{equation}
Further, note that
\begin{equation}\label{eq:estimate-sum-phi}
\sum_{{(\BFl,\BFi):\BFc_{\BFl,\BFi}\in\CalX_n^{\mathsf{TSG}}}} \phi_{\BFl,\BFi}(\BFx) \leq
\sum_{|\BFl|\leq \tau+d} \sum_{\BFi\in\rho(\BFl)} \phi_{\BFl,\BFi}(\BFx)
= \sum_{|\BFl|\leq \tau+d} \phi_{\BFl,\BFi^*(\BFl)}(\BFx)
\leq \sum_{|\BFl|\leq \tau+d } 1
= \sum_{\ell=d}^{\tau+d}\binom{\ell-1}{d-1},
\end{equation}
where $i^*(\BFl)$ is  the unique index in $\rho(\BFl)$ for which $\BFx\in \supp(\phi_{\BFl,\BFi^*(\BFl)})$,
and that
\begin{equation}\label{eq:estimate-sum-phi-2}
\sum_{\ell=d}^{\tau+d}\binom{\ell-1}{d-1} \leq (\tau + 1) \binom{\tau+d-1}{d-1} \asymp \tau^d \leq C_3 (\log n)^d,
\end{equation}
for some constant $C_3>0$,
where the asymptotic equivalence can be shown via Stirling's approximation for factorials.
Hence, by \eqref{eq:sample-inequality},
\[
|g(\BFx)| \leq C_2^d C_3 (\log n)^d\cdot \sup_{(\BFl,\BFi):\BFc_{\BFl,\BFi}\in\CalX_n^{\mathsf{TSG}}}|g(\SFc_{\BFl,\BFi})|. \Halmos
\]
\endproof

\begin{lemma}\label{lemma:lambda-diff}
Suppose the design points $\{\BFx_1,\ldots,\BFx_n\}$ form a TSG $\CalX_n^{\mathsf{TSG}}$.
Then,
\begin{equation}\label{eq:P-lambda_norm}
\|\CalP_n^\lambda f\|_\infty  \leq \|f\|_\infty +  \lambda^{1/2} C (\log n)^d  \norm{f}_{\ScrH_k},
\end{equation}
and
\begin{equation}\label{eq:lambda-diff}
\|\CalP_n f - \CalP_n^\lambda f\|_\infty \leq C \lambda^{1/2}(\log n)^d\norm{f}_{\ScrH_k},
\end{equation}
for all $\lambda>0$, all $f\in\ScrH_k$, and some constant $C>0$.
\end{lemma}

\proof{Proof of Lemma~\ref{lemma:lambda-diff}.}

Let $g^\lambda(\cdot)\coloneqq f(\cdot) - (\CalP_n^\lambda f)(\cdot)$.
Note that $g^\lambda(\cdot)$ can be expressed as a linear combination of $\{k(\BFx_i,\cdot):i=1,\ldots,n\}$, so it lies in $\ScrF_n$, the span of $\{k(\BFx_i,\cdot):i=1,\ldots,n\}$.
Since the design points $\{\BFx_1,\ldots,\BFx_n\}$ form a TSG $\CalX_n^{\mathsf{TSG}}$,
by the orthogonal expansion \eqref{eq:TMK_expansion},
$\ScrF_n$ is identical to $\ScrG_n$, the span of $\{\phi_{\BFl,\BFi}: \BFc_{\BFl,\BFi}\in\CalX_n^{\mathsf{TSG}}\}$.
Applying Lemma~\ref{lemma:inf-norm-subspace},
\[
\|g^\lambda\|_\infty \leq C (\log n)^d\max_{1\leq i\leq n} |g^\lambda(\BFx_i)|.
\]
for some constant $C>0$.
Hence,
\begin{equation}\label{eq:P-f-inf-norm}
\|f - \CalP_n^\lambda f\|_\infty \leq
C (\log n)^d\max_{1\leq i\leq n} | f(\BFx_i) - (\CalP_n^\lambda f)(\BFx_i)|.
\end{equation}

In the same vein, we can show that
\[
\|\CalP_nf - \CalP_n^\lambda f\|_\infty \leq
C (\log n)^d\max_{1\leq i\leq n} | (\CalP_n f)(\BFx_i) - (\CalP_n^\lambda f)(\BFx_i)|.
\]
Since $\CalP_n f$ is the kriging predictor of $f$ with kernel $k$ and design $\CalX_n^{\mathsf{TSG}}$, it performs interpolation on $\{f(\BFx_i):\BFx_i\in \CalX_n^{\mathsf{TSG}}\}$, that is, $(\CalP_n f)(\BFx_i) = f(\BFx_i)$, for all $i=1,\ldots,n$.
Hence,
\begin{equation}\label{eq:g_lambda_bound}
\|\CalP_nf - \CalP_n^\lambda f\|_\infty \leq
C (\log n)^d\max_{1\leq i\leq n} | f(\BFx_i) - (\CalP_n^\lambda f)(\BFx_i)|.
\end{equation}

The representer theorem \cite[Section 4.2]{ScholkopfSmola02_ec} implies that $\CalP_n^\lambda f$ solves the minimization problem below
\[\min_{h\in\ScrH_k}\sum_{i=1}^n\left(h(\BFx_i)-f(\BFx_i)\right)^2+\lambda\norm{h}_{\ScrH_k}^2.\]
It then follows immediately from Proposition~3.1 in \cite{WendlandRieger05_ec} that
\begin{equation}\label{eq:error-estimate-2}
|f(\BFx_i)-(\CalP_n^\lambda f)(\BFx_i)| \leq \lambda^{1/2} \|f\|_{\ScrH_k},\quad \forall i=1,\ldots,n.
\end{equation}
Hence,
\eqref{eq:P-lambda_norm} follows from \eqref{eq:P-f-inf-norm} and \eqref{eq:error-estimate-2}, whereas
\eqref{eq:lambda-diff} follows from \eqref{eq:g_lambda_bound} and \eqref{eq:error-estimate-2}.
\Halmos\endproof

\subsection{Proof of Theorem~\ref{theo:stoch-TSG}}

\proof{Proof of Theorem~\ref{theo:stoch-TSG}.}
Let $\lambda\coloneqq\max_{1\leq i\leq n} \sigma^2(\BFx_i)/m_i $.
It is easy to see that the MSE of the stochastic kriging predictor \eqref{eq:SK-MSE} satisfies
\begin{equation}\label{eq:MSE-decomp}
\begin{aligned}
\E[(\widehat\SFY_n(\BFx) - \SFY(\BFx))^2] ={}&
k(\BFx,\BFx)-\BFk^\intercal(\BFx)[\BFK+\BFSigma]^{-1}\BFk(\BFx)  \\
\leq{}& k(\BFx,\BFx)-\BFk^\intercal(\BFx)[\BFK+\lambda\BFI]^{-1}\BFk(\BFx)  \\
={}& \underbrace{k(\BFx,\BFx)-\BFk^\intercal(\BFx)\BFK^{-1}\BFk(\BFx)}_{J_1(\BFx)}
+ \underbrace{\BFk^\intercal(\BFx)\BFK^{-1}\BFk(\BFx)- \BFk^\intercal(\BFx)[\BFK+\lambda\BFI]^{-1}\BFk(\BFx) }_{J_2(\BFx)}.
\end{aligned}
\end{equation}
Notice that $J_1(\BFx)$ is the MSE of the kriging predictor in the absence of simulation noise.
Theorem~\ref{theo:determ-TSG} asserts that
\begin{equation}\label{eq:J_1-bound}
\sup_{\BFx\in(0,1)^d} J_1(\BFx) = \CalO\left(n^{-1}(\log n)^{2(d-1)}\right).
\end{equation}

By the definitions of $\CalP_n$ and $\CalP_n^\lambda$ in \eqref{eq:projection-P} and \eqref{eq:projection-P_lambda}, respectively,
\begin{align*}
     |J_2(\BFx)| ={}& |(\CalP_n k(\BFx,\cdot))(\BFx) - (\CalP_n^\lambda k(\BFx,\cdot))(\BFx)| \\
    \leq{} & C_1 \lambda^{1/2} (\log n)^d\norm{k(\BFx,\cdot)}_{\ScrH_k} \\
    ={}&  C_1(\log n)^d\sqrt{\lambda k(\BFx,\BFx)},
\end{align*}
for some constant $C_1>0$, where the inequality follows from \eqref{eq:lambda-diff} in Lemma~\ref{lemma:lambda-diff}.
Therefore,
\begin{equation}\label{eq:J_2-bound}
\sup_{\BFx\in(0,1)^d} |J_2(\BFx)|
\leq C_2 \lambda^{1/2} (\log n)^d,
\end{equation}
where $C_2 = C_1\sup_{\BFx\in(0,1)^d} \sqrt{k(\BFx,\BFx)}<\infty$.
Combining \eqref{eq:MSE-decomp}, \eqref{eq:J_1-bound}, and \eqref{eq:J_2-bound} yields
\begin{equation*}\label{eq:upper-bound-simple}
\sup_{\BFx\in(0,1)^d}\E[(\widehat \SFY_n(\BFx) - \SFY(\BFx))^2] = \CalO\left(n^{-1}(\log n)^{2(d-1)} + (\log n)^d \max_{1\leq i\leq n} m_i^{-1/2}\sigma(\BFx_i)\right). \Halmos
\end{equation*}
\endproof

\subsection{Proof of Theorem~\ref{theo:misspec-stoch}}

The techniques used here are analogous to those in the proofs of Theorems~\ref{theo:misspec-determ} and \ref{theo:stoch-TSG}.

\proof{Proof of Theorem~\ref{theo:misspec-stoch}.}
We first write explicitly the MSE of the  misspecified stochastic kriging predictor $\widehat\SFY_n^{\mathsf{mis}}(\BFx)$  as follows,
\begin{align}
& \E[(\widehat\SFY_n^{\mathsf{mis}}(\BFx) - \SFY^*(\BFx))^2] \nonumber\\
={}& \E[(\SFY^*(\BFx))^2] + \E[(\BFk^\intercal(\BFx)(\BFK+\BFSigma)^{-1}\bar\BFY^* )^2] - 2\E[\SFY^*(\BFx)\BFk^\intercal(\BFx)(\BFK+\BFSigma)^{-1}\bar\BFY^* ] \nonumber\\
={}&
k^*(\BFx,\BFx) + \BFk^\intercal(\BFx)(\BFK+\BFSigma)^{-1}(\BFK^*+\BFSigma)(\BFK+\BFSigma)^{-1}\BFk(\BFx)
-2 \BFk^\intercal(\BFx)(\BFK+\BFSigma)^{-1}\BFk^*(\BFx) \nonumber \\
\leq{}& k^*(\BFx,\BFx) + \BFk^\intercal(\BFx)(\BFK+\lambda \BFI)^{-1}(\BFK^*+\lambda \BFI)(\BFK+\lambda \BFI)^{-1}\BFk(\BFx)
-2 \BFk^\intercal(\BFx)(\BFK+\lambda \BFI)^{-1}\BFk^*(\BFx) \nonumber \\
={}& M_1(\BFx) - M_2(\BFx) + M_3(\BFx), \label{eq:misspec-MSE-decomp}
\end{align}
where $\BFK^*$ denotes the kernel matrix associated with $k^*$, $\lambda\coloneqq\max_{1\leq i\leq n} \sigma^2(\BFx_i)/m_i $, and
\begin{align*}
M_1(\BFx) \coloneqq{}& k^*(\BFx,\BFx) - \BFk^\intercal(\BFx)(\BFK+\lambda \BFI)^{-1}\BFk^*(\BFx), \\
M_2(\BFx) \coloneqq{}&  \BFk^\intercal(\BFx)(\BFK+\lambda \BFI)^{-1}\BFk^*(\BFx)  - \BFk^\intercal(\BFx)(\BFK+\lambda \BFI)^{-1}\BFK^*(\BFK+\lambda \BFI)^{-1}\BFk(\BFx), \\
M_3(\BFx) \coloneqq{}& \lambda \BFk^\intercal(\BFx)(\BFK+\lambda \BFI)^{-2}\BFk(\BFx).
\end{align*}

We express $M_1(\BFx)$ in the way similar to \eqref{eq:MSE-decomp},
\begin{align*}
M_1(\BFx) ={}&
\bigl[k^*(\BFx,\BFx) - \BFk^\intercal(\BFx)\BFK^{-1}k^*(\BFx)\bigr]  +  \bigl[\BFk^\intercal(\BFx)\BFK^{-1}\BFk^*(\BFx) - \BFk^\intercal(\BFx)(\BFK+\lambda \BFI)^{-1}\BFk^*(\BFx)\bigr]\\
={}& \underbrace{k^*(\BFx,\BFx) - (\CalP_n k^*(\BFx,\cdot))(\BFx)}_{M_{1,1}(\BFx)}  +  \underbrace{(\CalP_n k^*(\cdot,\BFx))(\BFx) - (\CalP_n^\lambda k^*(\cdot,\BFx))(\BFx)}_{M_{1,2}(\BFx)}
\end{align*}
By \eqref{eq:I_1-estimate-2}, we immediately have
\[
\sup_{\BFx\in(0,1)^d}|M_{1,1}(\BFx)| \leq C_1 n^{-2} (\log n)^{3(d-1)},
\]
for some constant $C_1>0$.
Moreover, note that for all $\BFx\in(0,1)^d$, $k^*(\BFx,\cdot)\in \ScrH_k$, so
applying \eqref{eq:lambda-diff} in  Lemma~\ref{lemma:lambda-diff} yields
\[
\sup_{\BFx\in(0,1)^d}|M_{1,2}(\BFx)| \leq C_2 \lambda^{1/2}(\log n)^d,
\]
for some constant $C_2>0$.
Hence,
\begin{equation}\label{eq:M_1-estimate}
\sup_{\BFx\in(0,1)^d} |M_1(\BFx)| \leq C_1 n^{-2} (\log n)^{3(d-1)} + C_2 \lambda^{1/2}(\log n)^d.
\end{equation}

For $M_2(\BFx)$, we adopt the approach used to analyze $I_2(\BFx)$ in the proof of Theorem~\ref{theo:misspec-determ}.
Let $h^\lambda(\BFx,\BFx')\coloneqq (\CalP_n^\lambda k^*(\cdot,\BFx'))(\BFx)$.
Then
\begin{equation}\label{eq:h-Ph-lambda}
h^\lambda(\BFx,\BFx') - (\CalP_n^\lambda h^\lambda(\BFx', \cdot))(\BFx)
= \Bigl( \CalP_n^\lambda \bigl(k^*(\cdot, \BFx') - h^\lambda(\BFx',\cdot) \bigr)\Bigr)(\BFx).
\end{equation}
Applying \eqref{eq:P-lambda_norm} in Lemma~\ref{lemma:lambda-diff}, there exists a constant $C_3>0$ such that
\begin{align}
& \bigl\| \CalP_n^\lambda \bigl(k^*(\cdot, \BFx') - h^\lambda(\BFx',\cdot) \bigr)\bigr\|_\infty  \nonumber \\
\leq{}& C_3 \left(\| k^*(\cdot, \BFx') - h^\lambda(\BFx',\cdot)\|_\infty + \lambda^{1/2}(\log n)^d \| k^*(\cdot, \BFx') - h^\lambda(\BFx',\cdot)\|_{\ScrH_k} \right) \nonumber \\
\leq{}& C_3 \Bigl(\| k^*(\cdot, \BFx') - h^\lambda(\BFx',\cdot)\|_\infty + \lambda^{1/2} (\log n)^d \bigl(\| k^*(\cdot, \BFx')\|_{\ScrH_k} + \|h^\lambda(\BFx',\cdot)\|_{\ScrH_k} \bigr) \Bigr).
\label{eq:h-Ph-lambda-2}
\end{align}

Note that
\begin{align}
\| k^*(\cdot, \BFx') - h^\lambda(\BFx',\cdot)\|_\infty
={}& \sup_{\BFx\in(0,1)^d} | k^*(\BFx, \BFx') - h^\lambda(\BFx',\BFx) | \nonumber \\
={}& \sup_{\BFx\in(0,1)^d} |k^*(\BFx', \BFx) - (\CalP_n^\lambda k^*(\cdot, \BFx))(\BFx')| \nonumber \\
={}& \sup_{\BFx\in(0,1)^d}  |M_1(\BFx)|  \nonumber \\
\leq{}&
C_1 n^{-2} (\log n)^{3(d-1)} + C_2 \lambda^{1/2}(\log n)^d.
\label{eq:h-Ph-lambda-3}
\end{align}

We now estimate $\| k^*(\cdot, \BFx')\|_{\ScrH_k}$.
Applying the orthogonal expansion,
\[
\| k^*(\cdot, \BFx')\|_{\ScrH_k}^2 =
\sum_{\BFl\in\NatInt^d}\sum_{\BFi\in\rho(\BFl)}\frac{\langle k^*(\BFx',\cdot), \phi_{\BFl,\BFi} \rangle_{\ScrH_k}^2}{ \norm{\phi_{\BFl,\BFi}}^2_{\ScrH_k}}
\leq  C_4 \sum_{\BFl\in\NatInt^d}\sum_{\BFi\in\rho(\BFl)} \frac{2^{-2{\abs{\BFl}}}}{2^{\abs{\BFl}}}
= C_4 \sum_{\BFl\in\NatInt^d} \sum_{\BFi\in\rho(\BFl)}  2^{-3{\abs{\BFl}}},
\]
where the inequality follows from \eqref{eq:basisRKHSestimate}
and \eqref{eq:k*projectEstimate}.
Further, note that
\begin{equation}\label{eq:rho-l-cardinality}
|\rho(\BFl)| = \prod_{j=1}^d 2^{-l_j/2} = 2^{-\abs{\BFl}/2}.
\end{equation}
Hence,
\begin{equation}\label{eq:k^*-RKHS-norm}
\| k^*(\cdot, \BFx')\|_{\ScrH_k}^2 \leq C_4 \sum_{\BFl\in\NatInt^d} 2^{-5\abs{\BFl}/2} \coloneqq C_5< \infty,
\end{equation}
by Lemma~3.7 in \cite{BungartzGriebel04_ec}.

We next estimate $\|h^\lambda(\BFx',\cdot)\|_{\ScrH_k}$.
To this end,  we rewrite $ h^\lambda(\BFx',\cdot)$ as
\[
h^\lambda(\BFx',\cdot)=\sum_{i=1}^n\alpha^\lambda_i(\BFx')k^*(\BFx_i , \cdot),
\]
where $\alpha^\lambda_i(\BFx')$ is the $i^{\rm th}$ entry of $\BFk^\intercal(\BFx') (\BFK+\lambda \BFI)^{-1}$. Applying the orthogonal expansion again,
\begin{align}
\| h^\lambda(\BFx',\cdot)\|_{\ScrH_k}^2
={}& \sum_{\BFl\in\NatInt^d}\sum_{\BFi\in\rho(\BFl)}\frac{\langle h^\lambda(\BFx',\cdot), \phi_{\BFl,\BFi} \rangle_{\ScrH_k}^2}{ \norm{\phi_{\BFl,\BFi}}^2_{\ScrH_k}}\nonumber\\
={}&  \sum_{\BFl\in\NatInt^d}\sum_{\BFi\in\rho(\BFl)}\frac{1}{ \norm{\phi_{\BFl,\BFi}}^2_{\ScrH_k}} \biggl|\sum_{i=1}^n\alpha^\lambda_i(\BFx')\langle k^*(\BFx_i , \cdot), \phi_{\BFl,\BFi} \rangle_{\ScrH_k}\biggr|^2 \nonumber \\
={}&  \sum_{\BFl\in\NatInt^d}\sum_{\BFi\in\rho(\BFl)}\frac{1}{ \norm{\phi_{\BFl,\BFi}}^2_{\ScrH_k}}
\bigl|\CalP_n^\lambda\langle k^*(\BFx', \cdot), \phi_{\BFl,\BFi} \rangle_{\ScrH_k}\bigr|^2 \nonumber \\
\leq{}&
\sum_{\BFl\in\NatInt^d}\sum_{\BFi\in\rho(\BFl)}\frac{1}{ \norm{\phi_{\BFl,\BFi}}^2_{\ScrH_k}}
\bigl\|\CalP_n^\lambda\langle k^*(\cdot, \cdot), \phi_{\BFl,\BFi} \rangle_{\ScrH_k}\bigr\|_\infty^2.
\label{eq:k*RKHSnorm}
\end{align}
Applying \eqref{eq:P-lambda_norm} in Lemma~\ref{lemma:lambda-diff},
\begin{equation}\label{eq:k*RKHSnorm-2}
\bigl\|\CalP_n^\lambda\langle k^*(\cdot, \cdot), \phi_{\BFl,\BFi} \rangle_{\ScrH_k}\bigr\|_\infty
\leq
\bigl\|\langle k^*(\cdot, \cdot), \phi_{\BFl,\BFi} \rangle_{\ScrH_k} \bigr\|_\infty  + C_6 \lambda^{1/2} (\log n)^d \bigl\| \langle k^*(\cdot, \cdot), \phi_{\BFl,\BFi} \rangle_{\ScrH_k} \bigr\|_{\ScrH_k},
\end{equation}
for some constant $C_6>0$.
It follows from \eqref{eq:k*RKHSnorm}  and \eqref{eq:k*RKHSnorm-2}  that
\begin{align}
& \| h^\lambda(\BFx',\cdot)\|_{\ScrH_k}^2 \nonumber \\
\leq{}& \sum_{\BFl\in\NatInt^d}\sum_{\BFi\in\rho(\BFl)}\frac{1}{ \norm{\phi_{\BFl,\BFi}}^2_{\ScrH_k}}
\Bigl( \bigl\|\langle k^*(\cdot, \cdot), \phi_{\BFl,\BFi} \rangle_{\ScrH_k} \bigr\|_\infty
+C_6  \lambda^{1/2} (\log n)^d \bigl\| \langle k^*(\cdot, \cdot), \phi_{\BFl,\BFi} \rangle_{\ScrH_k} \bigr\|_{\ScrH_k}\Bigr)^2 \nonumber \\
\asymp{}&
\sum_{\BFl\in\NatInt^d}\sum_{\BFi\in\rho(\BFl)} 2^{-\abs{\BFl}} \Bigl( \bigl\|\langle k^*(\cdot, \cdot), \phi_{\BFl,\BFi} \rangle_{\ScrH_k} \bigr\|_\infty
+ \lambda^{1/2}(\log n)^d \bigl\| \langle k^*(\cdot, \cdot), \phi_{\BFl,\BFi} \rangle_{\ScrH_k} \bigr\|_{\ScrH_k}\Bigr)^2.
\label{eq:k*RKHSnorm-3}
\end{align}

For notational simplicity, let $w(\BFx, \tilde \BFx)\coloneqq\bigr(\CalL_{\tilde\BFx}k^*(\cdot,\BFx)\bigl)(\tilde \BFx)$. Then,
\begin{align}
& \norm{\langle k^*(\cdot , \cdot), \phi_{\BFl,\BFi} \rangle_{\ScrH_k}}^2_{\ScrH_k}
\nonumber \\
={}&\int_{(0,1)^d}\int_{(0,1)^d}\int_{(0,1)^d}\Bigl(\phi_{\BFl,\BFi}(\BFs)\bigl(\CalL_{\BFs} k^*(\BFx,\cdot)\bigr)(\BFs)\Bigr)\Bigl(\CalL_{\BFx}\bigl(\CalL_{\BFt}\phi_{\BFl,\BFi}\bigr)(\BFt)k^*(\cdot,\BFt)\Bigr)(\BFx)\dd{\BFx} \dd{\BFt} \dd{\BFs}\nonumber\\
={} &\int_{(0,1)^d}\int_{(0,1)^d}\int_{(0,1)^d}\bigl(\phi_{\BFl,\BFi}(\BFs)w(\BFx,\BFs)\bigr)\bigl(\CalL_{\BFt}\phi_{\BFl,\BFi}\bigr)(\BFt)w(\BFt,\BFx)\dd{\BFx} \dd{\BFt} \dd{\BFs}\nonumber\\
\leq{} & \biggl(\sup_{\BFx\in(0,1)^{d},\BFs\in(0,1)^d} | w(\BFx,\BFs)|\biggr)\cdot \int_{(0,1)^{d}}\phi_{\BFl,\BFi}(\BFs)\dd{\BFs}\cdot\int_{(0,1)^{d}}\biggl|\int_{(0,1)^{d}}\bigl(\CalL_{\BFt}\phi_{\BFl,\BFi}\bigr)(\BFt)w(\BFt,\BFx) \dd{\BFt}\biggr|\dd{\BFx}\nonumber\\
= {}& \biggl(\sup_{\BFs\in(0,1)^d} \| w(\cdot,\BFs)\|_\infty\biggr)\cdot \int_{(0,1)^{d}}\phi_{\BFl,\BFi}(\BFs)\dd{\BFs}\cdot
\int_{(0,1)^d} |\langle w(\cdot,\BFx),\phi_{\BFl,\BFi} \rangle_{\ScrH_k}| \dd{\BFx} \nonumber  \\
\leq{}&
\biggl(\sup_{\BFs\in(0,1)^d} \| w(\cdot,\BFs)\|_\infty\biggr)\cdot \int_{(0,1)^{d}}\phi_{\BFl,\BFi}(\BFs)\dd{\BFs}\cdot
\biggl(
\sup_{\BFx\in(0,1)^d} |\langle w(\cdot,\BFx),\phi_{\BFl,\BFi} \rangle_{\ScrH_k}|. \label{eq:k*dualnorm}
\biggr)
\end{align}
Applying the Cauchy--Schwarz inequality,
\begin{equation}\label{eq:CS}
\sup_{\BFx\in(0,1)^d}|\langle w(\cdot,\BFx),\phi_{\BFl,\BFi} \rangle_{\ScrH_k}| \leq
\|\phi_{\BFl,\BFi}\|_{\ScrH_k}  \cdot \sup_{\BFx\in(0,1)^d}\|w(\cdot, \BFx) \|_{\ScrH_k} =  C_7 \|\phi_{\BFl,\BFi}\|_{\ScrH_k},
\end{equation}
where $C_7 = \sup_{\BFx\in(0,1)^d}\|w(\cdot, \BFx) \|_{\ScrH_k}  <\infty$ by Assumption~\ref{assump:smooth-kernel}.

Moreover,
note that for all $f\in\ScrH_k$,
\[
\|f\|_\infty = \sup_{\BFx\in(0,1)^d} |\langle f, k(\cdot, x) \rangle_{\ScrH_k} | \leq \norm{f}_{\ScrH_k} \cdot \sup_{\BFx\in(0,1)^d} \norm{k(\cdot,x)}_{\ScrH_k} \coloneqq  C_8 \norm{f}_{\ScrH_k}.
\]
Hence,
\begin{equation}\label{eq:w-inf-norm}
\sup_{\BFs\in(0,1)^d} \| w(\cdot,\BFs)\|_\infty
\leq C_8 \sup_{\BFs\in(0,1)^d} \| w(\cdot,\BFs)\|_{\ScrH_k} = C_7C_8.
\end{equation}
Plugging \eqref{eq:volume}, \eqref{eq:CS} and \eqref{eq:w-inf-norm} into \eqref{eq:k*dualnorm} leads to
\begin{equation}\label{eq:k*dualnorm-2}
\norm{\langle k^*(\cdot , \cdot), \phi_{\BFl,\BFi} \rangle_{\ScrH_k}}^2_{\ScrH_k}
\leq
C_9 2^{-\abs{\BFl}} \|\phi_{\BFl,\BFi}\|_{\ScrH_k} \asymp 2^{-\abs{\BFl}/2},
\end{equation}
for some constant $C_9>0$,  where the last step follows from \eqref{eq:basisRKHSestimate}.

In addition, it follows from \eqref{eq:k*projectEstimate} that
\begin{equation}\label{eq:k*-inner-prod-norm}
\bigl\|\langle k^*(\cdot, \cdot), \phi_{\BFl,\BFi} \rangle_{\ScrH_k} \bigr\|_\infty \leq C_{10} 2^{-\abs{\BFl}},
\end{equation}
for some constant $C_{10}>0$.

We now combine \eqref{eq:k*RKHSnorm-3},  \eqref{eq:k*dualnorm-2}, and
\eqref{eq:k*-inner-prod-norm} to derive the following,
\begin{align}
\|h^\lambda(\BFx',\cdot)\|^2_{\ScrH_k}
\leq{}&
C_{11} \sum_{\BFl\in\NatInt^d}\sum_{\BFi\in\rho(\BFl)} 2^{-\abs{\BFl}}
\left(
2^{-\abs{\BFl}} + \lambda^{1/2}(\log n)^d\cdot  2^{-\abs{\BFl}/4} \right)^2 \nonumber\\
={}&
C_{11} \sum_{\BFl\in\NatInt^d} 2^{\abs{\BFl}/2}\cdot 2^{-\abs{\BFl}}
\left(
2^{-\abs{\BFl}} + \lambda^{1/2}(\log n)^d\cdot  2^{-\abs{\BFl}/4} \right)^2  \label{eq:rho-l-setsize} \\
= {}& C_{11}   \sum_{\BFl\in\NatInt^d} \left(
2^{-5\abs{\BFl}/2} + 2 \lambda^{1/2} (\log n)^d \cdot 2^{-7\abs{\BFl}/4} +  \lambda(\log n)^{2d)}\cdot 2^{-\abs{\BFl}} \right)  \nonumber\\
\leq {}& C_{12} \left(1 + \lambda^{1/2}(\log n)^d +\lambda (\log n)^{2d)}\right)  , \label{eq:h-lambda-norm}
\end{align}
for some positive constants $C_{11}$ and $C_{12}$,
where \eqref{eq:rho-l-setsize} follows from
\eqref{eq:rho-l-cardinality},
while \eqref{eq:h-lambda-norm} holds because
$
\sum_{\BFl\in\NatInt^d} 2^{-s \abs{\BFl}} < \infty,
$
for all $s>0$ by Lemma~3.7 in \cite{BungartzGriebel04_ec}.

Therefore, plugging  \eqref{eq:h-Ph-lambda-3}, \eqref{eq:k^*-RKHS-norm}, and \eqref{eq:h-lambda-norm} into
\eqref{eq:h-Ph-lambda-2} yields
\begin{align}
& \sup_{\BFx\in(0,1)^d} |M_2(\BFx)| \nonumber \\
={}& \sup_{\BFx\in(0,1)^d} |  h^\lambda(\BFx,\BFx) - (\CalP_n^\lambda h^\lambda(\BFx, \cdot))(\BFx) | \nonumber\\
\leq{}&
\bigl\| \CalP_n^\lambda \bigl(k^*(\cdot, \BFx') - h^\lambda(\BFx',\cdot) \bigr)\bigr\|_\infty
\nonumber \\
\leq{}& C_{13} \Bigl[ n^{-2}(\log n)^{3(d-1)} +  \lambda^{1/2} (\log n)^d
+ \lambda^{1/2} (\log n)^d
\Bigl(
 1 +
 \sqrt{1 + \lambda^{1/2} (\log n)^d+ \lambda (\log n)^{2d)})}
\Bigr)
\Bigr] \nonumber \\
\leq{} & C_{14}  \Bigl[n^{-2}(\log n)^{3(d-1)}
+ \lambda^{1/2} (\log n)^d \bigl( 1 \vee \lambda^{1/4}(\log n)^{d/2}\vee \lambda^{1/2} (\log n)^d \bigr) \Bigr],
\label{eq:M_2-estimate}
\end{align}
for some positive constants $C_{13}$ and $C_{14}$.

The final piece of the proof is to analyze $M_3(\BFx)$.
Clearly,
\begin{equation}\label{eq:M_3-simple-bound}
M_3(\BFx) = \lambda \BFk^\intercal(\BFx)(\BFK+\lambda \BFI)^{-2}\BFk(\BFx)
\leq \lambda \BFk^\intercal(\BFx)\BFK^{-2}\BFk(\BFx)
= \lambda \BFk^\intercal(\BFx)\BFK^{-1}\BFI\BFK^{-1}\BFk(\BFx).
\end{equation}
Consider the following indicator function
\[I(\BFx,\BFx')=\begin{cases}
1,\quad\text{if }\BFx=\BFx',\\
0,\quad\text{otherwise}.
\end{cases}\]
Note that $\BFk^\intercal(\BFx)\BFK^{-1}\BFI\BFK^{-1}\BFk(\BFx')$ can be obtained by applying the operator $\CalP_n$ twice to $I(\BFx,\BFx')$, first acting on the variable $\BFx$ and then on $\BFx'$.
This implies the following expansion,
\begin{align}
    & \BFk^\intercal(\BFx)\BFK^{-1}\BFI\BFK^{-1}\BFk(\BFx)\nonumber\\
    ={} &\sum_{(\BFl,\BFi):\BFc_{\BFl,\BFi}\in\CalX_n^{\mathsf{TSG}}}\sum_{(\BFl',\BFi'):\BFc_{\BFl',\BFi'}\in\CalX_n^{\mathsf{TSG}}}\frac{\bigl\langle\langle I(\cdot,\cdot),\phi_{\BFl,\BFi}\rangle_{\ScrH_k},\phi_{\BFl',\BFi'}\bigr\rangle_{\ScrH_k}}{ \|\phi_{\BFl',\BFi'}\|^2_{\ScrH_k}\norm{\phi_{\BFl,\BFi}}^2_{\ScrH_k}}\phi_{\BFl,\BFi}(\BFx)\phi_{\BFl',\BFi'}(\BFx)\nonumber\\
    \leq{} & \Biggl(\sup_{(\BFl,\BFi),(\BFl',\BFi'):\BFc_{\BFl,\BFi},\BFc_{\BFl',\BFi'}\in\CalX_n^{\mathsf{TSG}}}\frac{\bigl|\bigl\langle\langle I(\cdot,\cdot),\phi_{\BFl,\BFi}\rangle_{\ScrH_k},\phi_{\BFl',\BFi'}\bigr\rangle_{\ScrH_k}\bigr|}{ \|\phi_{\BFl',\BFi'}\|^2_{\ScrH_k}\norm{\phi_{\BFl,\BFi}}^2_{\ScrH_k}}\Biggr) \cdot
    \Biggl(\sum_{(\BFl,\BFi),(\BFl',\BFi'):\BFc_{\BFl,\BFi},\BFc_{\BFl',\BFi'}\in\CalX_n^{\mathsf{TSG}}}\phi_{\BFl,\BFi}(\BFx)\phi_{\BFl',\BFi'}(\BFx) \Biggr)\nonumber\\
    ={} & \Biggl(\sup_{(\BFl,\BFi),(\BFl',\BFi'):\BFc_{\BFl,\BFi},\BFc_{\BFl',\BFi'}\in\CalX_n^{\mathsf{TSG}}}\frac{\bigl|\bigl\langle\langle I(\cdot,\cdot),\phi_{\BFl,\BFi}\rangle_{\ScrH_k},\phi_{\BFl',\BFi'}\bigr\rangle_{\ScrH_k}\bigr|}{ \|\phi_{\BFl',\BFi'}\|^2_{\ScrH_k}\norm{\phi_{\BFl,\BFi}}^2_{\ScrH_k}}\Biggr)
    \cdot
    \Biggl(\sum_{(\BFl,\BFi):\BFc_{\BFl,\BFi}\in\CalX_n^{\mathsf{TSG}}}\phi_{\BFl,\BFi}(\BFx)\Biggr)^2 \label{eq:T-expansion}.
\end{align}
With an argument similar to \eqref{eq:max-estimate-general}, we can show that
\[
\sup_{(\BFl,\BFi),(\BFl',\BFi'):\BFc_{\BFl,\BFi},\BFc_{\BFl',\BFi'}\in\CalX_n^{\mathsf{TSG}}}\frac{\bigl|\bigl\langle\langle I(\cdot,\cdot),\phi_{\BFl,\BFi}\rangle_{\ScrH_k},\phi_{\BFl',\BFi'}\bigr\rangle_{\ScrH_k}\bigr|}{ \|\phi_{\BFl',\BFi'}\|^2_{\ScrH_k}\norm{\phi_{\BFl,\BFi}}^2_{\ScrH_k}}
\leq C_{15},
\]
for some constant $C_{15}>0$ that is independent of $n$.
Moreover, by \eqref{eq:estimate-sum-phi} and \eqref{eq:estimate-sum-phi-2},
\begin{equation}\label{eq:sample-inequality_squared}
\sum_{(\BFl,\BFi):\BFc_{\BFl,\BFi}\in\CalX_n^{\mathsf{TSG}}}\phi_{\BFl,\BFi}(\BFx) \leq C_{16} (\log n)^d,
\end{equation}
for some constant $C_{16}>0$.
Plugging \eqref{eq:T-expansion} and \eqref{eq:sample-inequality_squared} into \eqref{eq:M_3-simple-bound} gives
\[
\BFk^\intercal(\BFx)\BFK^{-1}\BFT\BFK^{-1}\BFk(\BFx) \leq C_{17} (\log n)^{2d},
\]
for all $\BFx\in(0,1)^d$, and some constant $C_{17}>0$. Therefore,
\begin{equation}\label{eq:M_3-estimate}
\sup_{\BFx\in(0,1)^d} |M_3(\BFx)| \leq C_{17} \lambda (\log n)^{2d}.
\end{equation}

Last, it follows from \eqref{eq:misspec-MSE-decomp}, \eqref{eq:M_1-estimate}, \eqref{eq:M_2-estimate}, and \eqref{eq:M_3-estimate}
that
\begin{align*}
& \sup_{\BFx\in(0,1)^d} \E[(\widehat\SFY_n^{\mathsf{mis}}(\BFx) - \SFY^*(\BFx))^2] \\
={}&
\CalO\Bigl(n^{-2}(\log n)^{3(d-1)}
+ \lambda^{1/2} (\log n)^d \bigl( 1 \vee \lambda^{1/4}(\log n)^{d/2}\vee \lambda^{1/2} (\log n)^d \bigr) \Bigr).
\end{align*}
Note that if $\lambda (\log n)^{2d} \to 0$, then
\[
1 > \lambda^{1/4}(\log n)^{d/2} > \lambda^{1/2} (\log n)^d.
\]
Therefore, if $(\max_{1\leq i\leq n} m_i^{-1/2} \sigma(\BFx_i)) = o\left((\log n)^{-d}\right)$, then as $n\to\infty$,
\[
\sup_{\BFx\in(0,1)^d}\E[(\widehat \SFY_n(\BFx) - \SFY(\BFx))^2] = \CalO\left(n^{-2}(\log n)^{3(d-1)} + (\log n)^d \max_{1\leq i\leq n} m_i^{-1/2}\sigma(\BFx_i)\right). \Halmos
\]
\endproof

\section{Kernel Matrix Inversion in One Dimension}

In this section, we prove Proposition~\ref{prop:inverse}.
The proof relies on explicit calculation using the following basic property of the multivariate normal distribution; see, for example, Theorem~2.3 in \cite{RueHeld05_ec} for its proof.

\begin{lemma}\label{lemma:cond-MVN}
Let $Z=(Z_1,\ldots,Z_n)$ be a multivariate normal random variable with covariance matrix $\BFC\in\Real^{n\times n}$.
Then,  $(\BFC^{-1})_{ii} = 1/\Var[Z_i|Z_{-i}]$ for all $i$ and $(\BFC^{-1})_{ij} = -\Cov[Z_i, Z_j|Z_{-ij}]$ for all $i\neq j$,
where $Z_{-i} = (Z_\ell: \ell \neq i)$ and $Z_{-ij} = (Z_\ell:\ell\notin\{i,j\})$.
\end{lemma}

\proof{Proof of Part (i) of Proposition~\ref{prop:inverse}.}
Let $\SFY$ be a a one-dimensional GP with mean zero and kernel $k(x, x') = p(x\wedge x')q(x\vee  x')$.
Then, $\SFY$ is a Gauss--Markov process by Lemma~\ref{lemma:PD},
so $\{\Psi_i\coloneqq \SFY(x_i):i=1,\ldots,n\}$ form a discrete-time Markov chain, since $x_1<\cdots<x_n$.

Consider two indices $i,j\in\{1,\ldots,n\}$ with $j-i\geq 2$.
Then, $i<j-1<j$, and thus
\[
\pr(\Psi_j\in\cdot \,|\, \Psi_i, \Psi_{-ij}) = \pr(\Psi_j\in\cdot \,|\, \Psi_1,\ldots,\Psi_i,\ldots,\Psi_{j-1}, \Psi_{j+1},\ldots, \Psi_n) = \pr(\Psi_j\in\cdot\,|\, \Psi_{j-1}, \Psi_{j+1}),
\]
where the second inequality follows from the Markov property.
In the same vein, we can show $\pr(\Psi_j\in\cdot \,|\, \Psi_{-ij}) = \pr(\Psi_j\in\cdot\,|\, \Psi_{j-1}, \Psi_{j+1})$,
implying that $\pr(\Psi_j\in\cdot \,|\, \Psi_i, \Psi_{-ij}) = \pr(\Psi_j\in\cdot \,|\, \Psi_{-ij})$.
Therefore, $\Psi_j$ and $\Psi_i$ are conditionally independent given $\Psi_{-ij}$.
It follows immediately from Lemma~\ref{lemma:cond-MVN} that
\[(\BFK^{-1})_{i,j} = -\Cov[\SFY(x_i), \SFY(x_j) \,|\, \SFY(x_\ell):\ell\notin\{i,j\} ] = -\Cov[\Psi_i, \Psi_j\,|\, \Psi_{-ij}] = 0.\]
By symmetry, we also have $(\BFK^{-1})_{i,j}= 0 $ if $i-j\geq 2$.
Thus, $\BFK^{-1}$ is a tridiagonal matrix.

By Lemma~\ref{lemma:cond-MVN}, it suffices to calculate $\Var(\Psi_i\,|\,\Psi_{-i})$ and $\Cov[\Psi_i, \Psi_{i+1}\,|\, \Psi_{-\{i,i+1\}}]$
in order to calculate $(\BFK^{-1})_{ii}$ and $(\BFK_{i,i+1})^{-1}$, respectively.
To this end, we note that by the Markov property,
$\pr(\Psi_i\in\cdot \,|\, \Psi_{-i}) = \pr(\Psi_i\in\cdot \,|\, \Psi_{i-1},\Psi_{i+1})$.
Moreover, $(\Psi_{i-1},\Psi_i, \Psi_{i+1})$ has the multivariate normal distribution with mean $\BFzero$ and covariance matrix with entries $\BFK_{\ell,\ell'}$ for all $\ell,\ell'=i-1,i, i+1$.
Thus,
\begin{align*}
\Var(\Psi_i\,|\,\Psi_{i-1}, \Psi_{i+1})
={}& \BFK_{ii} -
\begin{pmatrix}
\BFK_{i,i-1}  && \BFK_{i,i+1}
\end{pmatrix}
\begin{pmatrix}
\BFK_{i-1,i-1} && \BFK_{i-1,i+1} \\
\BFK_{i+1,i-1} && \BFK_{i+1,i+1}
\end{pmatrix}^{-1}
\begin{pmatrix}
\BFK_{i,i-1} \\
\BFK_{i,i+1}
\end{pmatrix}.
\end{align*}

Likewise, by the Markov property, $\pr((\Psi_i,\Psi_{i+1})\in\cdot \,|\, \Psi_{-\{i,i+1\}}) = \pr((\Psi_i,\Psi_{i+1})\in\cdot \,|\, \Psi_{i-1},\Psi_{i+2})$, so
\begin{align*}
&
\begin{pmatrix*}[l]
\Cov[\Psi_i,\Psi_i\,|\, \Psi_{i-1},\Psi_{i+2}] &&  \Cov[\Psi_{i},\Psi_{i+1}\,|\, \Psi_{i-1},\Psi_{i+2}]  \\
\Cov[\Psi_{i+1},\Psi_i\,|\, \Psi_{i-1},\Psi_{i+2}] &&
\Cov[\Psi_{i+1},\Psi_{i+1}\,|\, \Psi_{i-1},\Psi_{i+2}]
\end{pmatrix*}\\
={}&
\begin{pmatrix*}[l]
\BFK_{i,i} && \BFK_{i,i+1} \\
\BFK_{i+1,i} && \BFK_{i+1,i+1}
\end{pmatrix*}
-
\begin{pmatrix*}[l]
\BFK_{i,i-1} && \BFK_{i,i+2} \\
\BFK_{i+1,i-1} && \BFK_{i+1,i+2}
\end{pmatrix*}
\begin{pmatrix}
\BFK_{i-1,i-1} && \BFK_{i+2,i+2} \\
\BFK_{i-1,i-1} && \BFK_{i+2,i+2}
\end{pmatrix}^{-1}
\begin{pmatrix}
\BFK_{i,i-1} && \BFK_{i+1,i-1} \\
\BFK_{i,i+2} && \BFK_{i+1,i+2}
\end{pmatrix}
\end{align*}

Then, for all $i\leq j$, plugging $\BFK_{ij}=k(x_i,x_j) = p(x_i)q(x_j) =\SFp_i\SFq_j$, we can verify via direct but tedious calculation that the formula \eqref{eq:K-inverse} holds.

\proof{Proof of Part (ii) of Proposition~\ref{prop:inverse}.}
Let $x\in[x_{i^*}, x_{i^*+1})$. For each $i=1,\ldots,d$,
\begin{align}
&(\BFK^{-1}\BFk(x))_i
\nonumber
\\ ={}& \sum_{j=1}^n (\BFK^{-1})_{ij}k(x,x_j) \nonumber\\
={}& \frac{(\SFp_{i+1}\SFq_{i-1} - \SFp_{i-1}\SFq_{i+1}) k(x,x_{i}) - (\SFp_{i+1}\SFq_i - \SFp_i\SFq_{i+1})k(x,x_{i-1}) - (\SFp_i\SFq_{i-1}-\SFp_{i-1}\SFq_i)k(x,x_{i+1}) }{(\SFp_i\SFq_{i-1} - \SFp_{i-1}\SFq_i) (\SFp_{i+1}\SFq_i - \SFp_i \SFq_{i+1}) }.\label{eq:K-inver-k-i}
\end{align}
If $i\leq i^*-1$, then $x_{i-1}<x_i<x_{i+1}\leq x$, so $k(x_j,x) = \SFp_j q(x)$ for all $j=i-1,i,i+1$.
Thus, the numerator of \eqref{eq:K-inver-k-i} equals $q(x)$ multiplied by
\[
(\SFp_{i+1}\SFq_{-1} - \SFp_{i-1}\SFq_{i+1}) \SFp_i - (\SFp_{i+1}\SFq_i - \SFp_i\SFq_{i+1})\SFp_{i-1} - (\SFp_i\SFq_{i-1}-\SFp_{i-1}\SFq_i)\SFp_{i+1} = 0,
\]
so $(\BFK^{-1}\BFk(x))_i=0$ for $i\leq i^*-1$.
Likewise, we may show $(\BFK^{-1}\BFk(x))_i=0$ for $i\geq i^*+2$, in which case $k(x_j,x)= p(x) \SFq_j$ for $j=i-1,i,i+1$.

If $i=i^*$, then $x_{i-1}<x_i\leq x < x_{i+1}$, so
$k(x_{j},x) = \SFp_{j}q(x)$ for $j=i-1,i$, and $k(x_{i+1},x) = p(x)\SFq_{i+1}$.
Thus, the numerator of \eqref{eq:K-inver-k-i} equals
\begin{align*}
& (\SFp_{i+1}\SFq_{i-1} - \SFp_{i-1}\SFq_{i+1}) \SFp_{i-1}q(x) - (\SFp_{i+1}\SFq_i - \SFp_i\SFq_{i+1})\SFp_{i}q(x) - (\SFp_i\SFq_{i-1}-\SFp_{i-1}\SFq_i)p(x)\SFq_{i+1} \\
={}& (\SFp_i\SFq_{i-1}-\SFp_{i-1}\SFq_i) (\SFp_{i+1} q(x)-p(x)\SFq_{i+1}).
\end{align*}
Applying \eqref{eq:K-inver-k-i},
\[
(\BFK^{-1}\BFk(x))_i = \frac{\SFp_{i+1} q(x)-p(x)\SFq_{i+1}}{\SFp_{i+1}\SFq_i - \SFp_i \SFq_{i+1}},
\]
proving the formula \eqref{eq:K-inverse-times-k} for $i=i^*$.
The case of $i=i^*+1$ can be verified similarly.
\Halmos
\endproof

\section{Kernel Matrix Inversion on Truncated Sparse Grids}

In this section, we prove Theorem~\ref{theo:K-inverse}.
Prior to the proof,
we first discuss the number of nonzero entries in the inverse of the kernel matrix on a classical SG.
The following result will be used for proving part (iii) of Theorem~\ref{theo:K-inverse}.

\begin{proposition}\label{prop:nnz}
Let $k(\BFx,\BFx')=\prod_{j=1}^dk_j(x_j,x'_j)$ be a TM kernel that satisfies Assumption~\ref{assump:SL}, $\BFK=k(\CalX_\tau^{\mathsf{SG}}, \CalX_\tau^{\mathsf{SG}})$, and $\BFk(\BFx) = k(\CalX_\tau^{\mathsf{SG}}, \{\BFx\})$ for some $\BFx\in(0,1)^d$.
Then,
\[
\nnz(\BFK^{-1}) = \CalO(|\CalX_\tau^{\mathsf{SG}}|)\quad\mbox{and}\quad \nnz(\BFK^{-1}\BFk(\BFx)) = \CalO(\log(|\CalX_\tau^{\mathsf{SG}}|)^{d-1}),
\]
where $\nnz(\cdot)$ denotes the number of nonzero entries of a vector or matrix.
\end{proposition}
\proof{Proof of Proposition~\ref{prop:nnz}.}
Since the design involved here is a classical SG, we can use Algorithm~\ref{alg:TM-SG} to compute both $\BFK^{-1}$ and $\BFK^{-1}\BFk(\BFx)$.

We first analyze $\nnz(\BFK^{-1})$.
In Algorithm~\ref{alg:TM-SG}, $\BFK^{-1}$ is initialized as a zero matrix.
Then, for each iteration $\BFl$ for which $\tau\leq \abs{\BFl}\leq \tau+d-1$,
the recursion \eqref{eq:update-A} updates
the submatrix of $\BFK^{-1}$ with indices $\{((\BFl',\BFi'),(\BFl'',\BFi'')):\BFc_{\BFl',\BFi'},\BFc_{\BFl'',\BFi''}\in \CalX_\BFl^{\mathsf{FG}}\}$ are updated, where $\CalX_\BFl^{\mathsf{FG}}$ denotes the full grid $\bigtimes_{j=1}^d \CalX_{j,l_j}$ and $\CalX_{j,l_j}$'s are defined in \eqref{eq:dyadic};
moreover, this submatrix is a multiple of $\BFK_\BFl^{-1} = (k(\CalX_\BFl^{\mathsf{FG}},\CalX_\BFl^{\mathsf{FG}}))^{-1}$.
Therefore, the number of nonzero entries of $\BFK^{-1}$ is at most the sum of the number of nonzero entries of $\BFK_\BFl^{-1}$ over all the iterations, namely,
\begin{equation}\label{eq:nnz-K-inv}
\nnz(\BFK^{-1}) \leq \sum_{\tau \leq |\BFl|\leq \tau+d-1} \nnz(\BFK_\BFl^{-1})
= \sum_{\tau \leq |\BFl|\leq \tau+d-1} \prod_{j=1}^d \nnz\bigl(k_j(\CalX_{j,l_j}, \CalX_{j,l_j})^{-1}\bigr),
\end{equation}
where the equality holds because $\BFK_\BFl = \bigotimes_{j=1}^d k_j(\CalX_{j,l_j}, \CalX_{j,l_j})$.
Further, since $\CalX_{j,l_j}$ is a one-dimensional grid, we may apply
\eqref{eq:K-inverse} in Proposition~\ref{prop:inverse} to deduce that
$k_j(\CalX_{j,l_j}, \CalX_{j,l_j})^{-1}$ is a tridiagonal matrix, so
\begin{equation}\label{eq:nnz-K-inv-2}
\nnz\bigl(k_j(\CalX_{j,l_j}, \CalX_{j,l_j})^{-1}\bigr) =  \CalO(|\CalX_{j,l_j}|).
\end{equation}
Combining \eqref{eq:nnz-K-inv} and \eqref{eq:nnz-K-inv-2} yields
\[
\nnz(\BFK^{-1}) \leq \sum_{\tau \leq |\BFl|\leq \tau+d-1} \prod_{j=1}^d \CalO(|\CalX_{j,l_j}| ) = \sum_{\tau \leq |\BFl|\leq \tau+d-1} \CalO(2^{l_j}) = \sum_{\tau \leq |\BFl|\leq \tau+d-1} \CalO(2^\BFl).
\]
Note that
\[
\sum_{\tau \leq \abs{\BFl}\leq \tau+d-1} 2^{\abs{\BFl}} = \sum_{\ell=d\vee\tau}^{\tau+d-1} 2^\ell \binom{\ell-1}{d-1}
= \sum_{\ell=0\vee(\tau-d)}^{\tau-1} 2^{\ell+d} \binom{\ell+d-1}{d-1}
\leq 2^d \cdot\abs{\CalX_\tau^{\mathsf{SG}}},
\]
where the inequality follows from \eqref{eq:SSGNumpt}.
Therefore, $\nnz(\BFK^{-1}) = \CalO(|\CalX_\tau^{\mathsf{SG}}|)$.

We can apply a similar argument, albeit with the use of the recursion \eqref{eq:update-b} and the formula \eqref{eq:K-inverse-times-k} in Proposition~\ref{prop:inverse}, to show that $\nnz(\BFK^{-1}\BFk(\BFx)) = \CalO(\log(|\CalX_\tau^{\mathsf{SG}}|)^{d-1})$.
\Halmos\endproof

\proof{Proof of Part (i) of Theorem~\ref{theo:K-inverse}.}

Let $\{\SFY(\BFx):\BFx\in(0,1)^d\}$ be a zero mean TMGP with kernel $k$.
Then, $(\SFY(\BFc_{\BFl,\BFi}):\BFc_{\BFl,\BFi}\in \CalX_n^{\mathsf{TSG}})$ is a multivariate normal random variable with
covariance matrix $\BFK$.

We cast $\BFK$ into the form of a block matrix,
\begin{equation}\label{eq:K-block-form}
\BFK^ =
\begin{pNiceMatrix}[first-row, columns-width=auto]
\scriptstyle{|\CalX_\tau^{\mathsf{SG}}|\;\,\mathrm{dim.}} & \scriptstyle{\Tilde{n} \;\,\mathrm{dim.}} \\[2pt]
\BFA & \BFF \\[3pt]
\BFF^\intercal &  \BFG
\end{pNiceMatrix},
\end{equation}
where $\BFA = k(\CalX_\tau^{\mathsf{SG}},\CalX_\tau^{\mathsf{SG}})$,  $\BFF = k(\CalX_\tau^{\mathsf{SG}}, \CalA_{\tilde n})$, and $\BFG = k(\CalA_{\tilde n}, \CalA_{\tilde n})$.
Then, the formula for inverting $2\times 2$ block matrices asserts that
\begin{equation}\label{eq:K-inverse-block-form}
\BFK^{-1} =
\begin{pNiceMatrix}
\BFA^{-1}+ \BFA^{-1}\BFF \BFD \BFF^\intercal \BFA^{-1} & -\BFA^{-1}\BFF \BFD\\[3pt]
-\BFD\BFF^\intercal \BFA^{-1} & \BFD
\end{pNiceMatrix},
\end{equation}
where $\BFD = (\BFG - \BFF^\intercal \BFA^{-1} \BFF)^{-1}$.

Let $\BFD_{(\BFl',\BFi'),(\BFl'',\BFi'')}$ denote an arbitrary entry of $\BFD$ indexed by $((\BFl',\BFi'),(\BFl'',\BFi''))$ for which $\BFc_{\BFl',\BFi'}, \BFc_{\BFl'',\BFi''}\in \CalA_{\tilde n}$.
By Lemma~\ref{lemma:cond-MVN},
\[
\BFD_{(\BFl',\BFi'),(\BFl'',\BFi'')} =
\left\{
\begin{array}{ll}
1/\Var[\SFY(\BFc_{\BFl',\BFi'})\,|\, \SFY(\BFx),\BFx\in \CalX_n^{\mathsf{TSG}} \setminus \{\BFc_{\BFl',\BFi'} \}]     &  \mbox{ if } (\BFl',\BFi') = (\BFl'',\BFi''), \\[1ex]
- \Cov[\SFY(\BFc_{\BFl',\BFi'}), \SFY(\BFc_{\BFl'',\BFi''}) \,|\,\SFY(\BFx),\BFx\in \CalX_n^{\mathsf{TSG}} \setminus \{\BFc_{\BFl',\BFi'}, \BFc_{\BFl'',\BFi''} \}],
     & \mbox{ otherwise}.
\end{array}
\right.
\]
We then apply Proposition~\ref{prop:cond_var} to deduce that
\[
\BFD_{(\BFl',\BFi'),(\BFl'',\BFi'')} =
\left\{
\begin{array}{ll}
\|\phi_{\BFl',\BFi'}\|_{\ScrH_k}^2, & \mbox{ if } (\BFl',\BFi') = (\BFl'',\BFi''), \\[1ex]
0, &  \mbox{ otherwise}.
\end{array}
\right.
\]
In light of the expression \eqref{eq:phi-norm}, it is straightforward to see that
\eqref{eq:K-inverse-block-form} yields \eqref{eq:cond-var}.
\Halmos\endproof

\proof{Proof of Part (ii) of Theorem~\ref{theo:K-inverse}.}
We analyze the three cases separately.

\textbf{Case (a.)}
This case  follows immediately from that $\BFD$ is a diagonal matrix.

\smallskip
\textbf{Case (b.)}
Fix indices $(\BFl',\BFi')$ and $(\BFl'',\BFi'')$ such that
$\BFc_{\BFl',\BFi'}\in \CalX_\tau^{\mathsf{SG}}$, $\BFc_{\BFl'',\BFi''}\in \CalX_n^{\mathsf{TSG}}$, and
$\BFc_{\BFl'',\BFi''}\notin \CalR_{\BFl',\BFi'}$.
Then, $(\BFK^{-1})_{(\BFl',\BFi'),(\BFl'',\BFi'')}=-(\BFA^{-1}\BFF \BFD)_{(\BFl',\BFi'),(\BFl'',\BFi'')}$,
belonging to the top-right block of $\BFK^{-1}$ in \eqref{eq:K-inverse-block-form}.
It suffices to show that
\begin{equation}\label{eq:upper-right-block-2}
\bigl(\BFA^{-1}\BFF \bigr)_{(\BFl',\BFi'),(\BFl'',\BFi'')} = 0,
\end{equation}
because $\BFD$ is diagonal, and
multiplying a matrix by a diagonal matrix will retain the zero entries of the former.

Consider the indicator function $f(x) = \ind(\BFx = \BFc_{\BFl',\BFi'}) = \prod_{j=1}^d \ind(x_j = \SFc_{l'_j,i'_j})$.
Then,
\[
\bigl(\BFA^{-1}\BFF \bigr)_{(\BFl',\BFi'),(\BFl'',\BFi'')}
= k(\BFc_{\BFl'',\BFi''}, \CalX_\tau^{\mathsf{SG}}) \bigl(k(\CalX_\tau^{\mathsf{SG}}, \CalX_\tau^{\mathsf{SG}})\bigr)^{-1} f(\CalX_\tau^{\mathsf{SG}}),
\]
where $f(\CalX_\tau^{\mathsf{SG}})$ denotes the vector having entries $f(\BFc_{\BFl,\BFi})$ for all $\BFc_{\BFl,\BFi}\in\CalX_\tau^{\mathsf{SG}}$.

Note that $k(\cdot, \CalX_\tau^{\mathsf{SG}}) (k(\CalX_\tau^{\mathsf{SG}}, \CalX_\tau^{\mathsf{SG}}))^{-1} f(\CalX_\tau^{\mathsf{SG}})$ can be expressed as a linear combination of $\{k(\cdot, \BFc_{\BFl,\BFi}):\BFc_{\BFl,\BFi}\in\CalX_\tau^{\mathsf{SG}}\}$, so it lies in the span of $\{k(\cdot,\BFc_{\BFl,\BFi}):\BFc_{\BFl,\BFi}\in\CalX_\tau^{\mathsf{SG}}\}$,
which is identical to the span of $\{\phi_{\BFl,\BFi}: \BFc_{\BFl,\BFi}\in\CalX_\tau^{\mathsf{SG}}\}$
by virtue of the orthogonal expansion \eqref{eq:TMK_expansion}.
Therefore, it follows from \eqref{eq:g-decomp} and \eqref{eq:prod-neighbors} that
\begin{equation}\label{eq:A-inv-F-decomp}
\bigl(\BFA^{-1}\BFF \bigr)_{(\BFl',\BFi'),(\BFl'',\BFi'')}  =
\sum_{\abs{\BFl}\leq \tau+d-1}\sum_{\BFi\in\rho(\BFl)}
\beta_{\BFl,\BFi}
\phi_{\BFl,\BFi}(\BFc_{\BFl'',\BFi''}),
\end{equation}
where
\begin{align*}
\beta_{\BFl,\BFi}
={}&  \biggl\langle f,\frac{\phi_{\BFl,\BFi}}{\|\phi_{\BFl,\BFi}\|^2_{\ScrH_k}}\biggr\rangle_{\ScrH_k}
\nonumber \\
={}& \biggl\langle \ind(\cdot = \BFc_{\BFl',\BFi'}),\prod_{j=1}^d\biggl(k_j(\SFc_{l_j,i_j},\cdot)+\frac{\eta_{l_j,i_j-1}}{\eta_{l_j,i_j}}k_j(\SFc_{l_j,i_j-1},\cdot)+\frac{\eta_{l_j,i_j+1}}{\eta_{l_j,i_j}}k_j( \SFc_{l_j,i_j+1},\cdot)\biggr)\biggr\rangle_{\ScrH_k} \nonumber  \\
={} & \prod_{j=1}^d \Bigl\langle \ind(\cdot = \SFc_{l'_j,i'_j}), k_j(\SFc_{l_j,i_j},\cdot)+\frac{\eta_{l_j,i_j-1}}{\eta_{l_j,i_j}}k_j(\SFc_{l_j,i_j-1},\cdot)+\frac{\eta_{l_j,i_j+1}}{\eta_{l_j,i_j}}k_j( \SFc_{l_j,i_j+1},\cdot) \Bigr\rangle_{\ScrH_{k_j}}. \label{eq:coeff-beta}
\end{align*}
Hence, $\beta_{\BFl,\BFi} \neq 0$ only if
$l_j=l'_j$ and
$i_j\in \{i'_j-1, i'_j, i'_j+1\}$ for all $j=1,\ldots,d$.
Further, note that $i_j$ and $i'_j$ are both odd numbers by definition, since $\BFc_{\BFl,\BFi},\BFc_{\BFl',\BFi'}\in\CalX_\tau^{\mathsf{SG}}$.
Thus, $\beta_{\BFl,\BFi}\neq 0$ only if $(\BFl,\BFi)=(\BFl',\BFi')$.
We may simplify \eqref{eq:A-inv-F-decomp} to
\begin{equation}\label{eq:A-inv-F-decomp-2}
\bigl(\BFA^{-1}\BFF \bigr)_{(\BFl',\BFi'),(\BFl'',\BFi'')}  =
\beta_{\BFl',\BFi'}
\phi_{\BFl',\BFi'}(\BFc_{\BFl'',\BFi''}).
\end{equation}
Note that
$\supp(\phi_{\BFl',\BFi'})=\CalR_{\BFl',\BFi'}$ by by the definition \eqref{eq:basis-func-tensor}.
The condition $\BFc_{\BFl'',\BFi''}\notin \CalR_{\BFl',\BFi'}$
then implies that $\phi_{\BFl',\BFi'}(\BFc_{\BFl'',\BFi''}) =0$, proving \eqref{eq:upper-right-block-2}.

\smallskip

\textbf{Case (c.)}
Fix indices $(\BFl',\BFi')$ and $(\BFl'',\BFi'')$ such that $\BFc_{\BFl',\BFi'},
\BFc_{\BFl'',\BFi''}\in\CalX_\tau^{\mathsf{SG}}$,
\begin{equation}\label{eq:no-overlap-1-repeat}
\Bigl|i_j'\cdot 2^{l_j'\vee l_j''-l_j'} - i_j''\cdot 2^{l_j'\vee l_j'' - l_j''}\Bigr| \geq 2,\quad\mbox{for some } j=1,\ldots,d,
\end{equation}
and
\begin{equation}\label{eq:no-overlap-2-repeat}
\BFc_{\BFl,\BFi}\notin  \CalR_{\BFl',\BFi'}\cap\CalR_{\BFl'',\BFi''},\quad \mbox{for all }\BFc_{\BFi,\BFl}\in\CalA_{\tilde n}.
\end{equation}

Then, $(\BFK^{-1})_{(\BFl',\BFi'),(\BFl'',\BFi'')}$
belongs to the top-right block of $\BFK^{-1}$ in \eqref{eq:K-inverse-block-form}, and
it suffices to prove
\begin{align}
& (\BFA^{-1})_{(\BFl',\BFi'),(\BFl'',\BFi'')} = 0, \label{eq:top-left-1} \\
& (\BFA^{-1}\BFF\BFD\BFF^\intercal \BFA^{-1})_{(\BFl',\BFi'),(\BFl'',\BFi'')} =0. \label{eq:top-left-2}
\end{align}

Since $\BFA$ is the kernel matrix defined on the classical SG $\CalX_\tau^{\mathsf{SG}}$, its entries can be calculated via  the additive recursion \eqref{eq:update-A} in Algorithm~\ref{alg:TM-SG}.
In particular, $(\BFA^{-1})_{(\BFl',\BFi'),(\BFl'',\BFi'')}$ is updated in each iteration $\BFl$ such that $\tau\leq \abs{\BFl}\leq \tau+d-1$, if and only if both $\BFc_{\BFl',\BFi'}$ and $\BFc_{\BFl'',\BFi''}$ belong to the full grid $\CalX_{\BFl}^{\mathsf{FG}} = \bigtimes_{j=1}^d \CalX_{j,l_j}$,
where $\CalX_{j,l_j}$'s are defined in \eqref{eq:dyadic}.
In this case, this entry of $\BFA^{-1}$ is increased by $C\cdot \bigl(\bigl(k(\CalX_\BFl^{\mathsf{FG}},\CalX_\BFl^{\mathsf{FG}})\bigr)^{-1}\bigr)_{(\BFl',\BFi'),(\BFl'',\BFi'')} $, where
$C\coloneqq (-1)^{\tau+d-1-\abs{\BFl}}\binom{d-1}{\tau+d-1-\abs{\BFl}}$.
Note that $k(\CalX_\BFl^{\mathsf{FG}},\CalX_\BFl^{\mathsf{FG}})$ can be expressed as a Kronecker product.
Hence,
by Algorithm~\ref{alg:TM-lattice},
\begin{align}
\Bigl(\bigl(k(\CalX_\BFl^{\mathsf{FG}},\CalX_\BFl^{\mathsf{FG}})\bigr)^{-1}\Bigr)_{(\BFl',\BFi'),(\BFl'',\BFi'')}={}& \biggl(\bigotimes_{j=1}^d \bigl(k_j({\CalX_{j, l_j}, \CalX_{j, l_j}} ) \bigr)^{-1}\biggr)_{(\BFl',\BFi'),(\BFl'',\BFi'')} \nonumber \\
={}& \prod_{j=1}^d \bigl(k_j({\CalX_{j, l_j}, \CalX_{j, l_j}} ) \bigr)^{-1}_{(l_j',i_j'),(l_j'',i_j'')}.\label{eq:inverse-tensor-product}
\end{align}

Consider the $j$-th dimension for which the condition \eqref{eq:no-overlap-1-repeat} holds.
Since $\BFc_{\BFl',\BFi'},\BFc_{\BFl'',\BFi''} \in \CalX_\BFl^{\mathsf{FG}}$,
we know that
both $\SFc_{l_j', i_j'}$ and $\SFc_{l_j'', i_j''}$ are design points in the one-dimensional grid $\CalX_{j,l_j}$, and particularly, $l_j',l_j''\leq l_j$.
Hence, $\SFc_{l_j',i_j'}$ and $\SFc_{l_j'',i_j''}$ can be rewritten as
\[
\SFc_{l_j',i_j'} = (i_j'\cdot 2^{l_j-l_j'})\cdot 2^{-l_j}\quad\mbox{and}\quad \SFc_{l_j'',i_j''} = (i_j''\cdot 2^{l_j-l_j''})\cdot 2^{-l_j}.
\]
It is then straightforward  to see from \eqref{eq:no-overlap-1-repeat} that
$\SFc_{l_j',i_j'}$ and $\SFc_{l_j'',i_j''}$ are neither identical nor adjacent points in
$\CalX_{j,l_j}$.
It then follows from \eqref{eq:K-inverse} in Proposition~\ref{prop:inverse} that $(k_j({\CalX_{j, l_j}, \CalX_{j, l_j}} ) )^{-1}_{(l_j',i_j'),(l_j'',i_j'')} = 0$, which implies
$((k(\CalX_\BFl^{\mathsf{FG}},\CalX_\BFl^{\mathsf{FG}}))^{-1})_{(\BFl',\BFi'),(\BFl'',\BFi'')}= 0$ by \eqref{eq:inverse-tensor-product}.
Since this is true for each iteration $\BFl$ that is involved for computing $(\BFA^{-1})_{(\BFl',\BFi'),(\BFl'',\BFi'')}$, we conclude that \eqref{eq:top-left-1} holds.

Let $\BFB \coloneqq \BFA^{-1} \BFF$. Since $\BFD$ is a diagonal matrix,
\begin{align}
(\BFA^{-1}\BFF\BFD\BFF^\intercal \BFA^{-1})_{(\BFl',\BFi'),(\BFl'',\BFi'')}
={}&  (\BFB\BFD\BFB^\intercal)_{(\BFl',\BFi'),(\BFl'',\BFi'')} \nonumber \\
={}& \sum_{(\BFl,\BFi):\BFc_{\BFl,\BFi}\in \CalA_{\tilde n}}  \BFB_{(\BFl',\BFi'),(\BFl,\BFi)} \BFD_{(\BFl,\BFi),(\BFl,\BFi)}\BFB_{(\BFl'',\BFi''),(\BFl,\BFi)} \nonumber\\
={}& \sum_{(\BFl,\BFi):\BFc_{\BFl,\BFi}\in \CalA_{\tilde n}} \beta_{\BFl',\BFi'} \beta_{\BFl'',\BFi''} \BFD_{(\BFl,\BFi),(\BFl,\BFi)}  \phi_{\BFl',\BFi'}(\BFc_{\BFl,\BFi})  \phi_{\BFl'',\BFi''}(\BFc_{\BFl,\BFi}), \label{eq:top-left-2-decomp}
\end{align}
where the last step follows from \eqref{eq:A-inv-F-decomp-2}.
Note that
$\phi_{\BFl',\BFi'}(\BFc_{\BFl,\BFi})  \phi_{\BFl'',\BFi''}(\BFc_{\BFl,\BFi}) \neq 0$ if and only if $\BFc_{\BFl,\BFi} \in \supp(\phi_{\BFl',\BFi'}) \cap \supp(\phi_{\BFl'',\BFi''}) = \CalR_{\BFl',\BFi'}\cap\CalR_{\BFl'',\BFi''}$.
It then follows from the condition \eqref{eq:no-overlap-2-repeat} that
each term in the summation \eqref{eq:top-left-2-decomp} is zero, proving \eqref{eq:top-left-2}.
\Halmos\endproof

\proof{Proof of Part (iii) of Theorem~\ref{theo:K-inverse}.}

Recall that $\BFA=k(\CalX_\tau^{\mathsf{SG}}, \CalX_\tau^{\mathsf{SG}})$,
and that
each column of $\BFF$ is $k(\CalX_{\tau}^{\mathsf{SG}}, \{\BFc_{\BFl'',\BFi''}\})$ for $\BFc_{\BFl'',\BFi''}\in\CalA_{\tilde n}\subset \CalX_{\tau+1}^{\mathsf{SG}}\setminus \CalX_\tau^{\mathsf{SG}}$.
It follows from  Proposition~\ref{prop:nnz} that
each column of $\BFA^{-1}\BFF$, which is of the form $\BFA^{-1} k(\CalX_{\tau}^{\mathsf{SG}}, \{\BFc_{\BFl'',\BFi''}\})$, has
$\CalO(\log(|\CalX_\tau^{\mathsf{SG}}|)^{d-1})$ nonzero entries.
Hence,
$\nnz(\BFA^{-1}\BFF)=\CalO(\tilde n \log(|\CalX_\tau^{\mathsf{SG}}|)^{d-1})$.
Since $\BFD$ is a diagonal matrix, we conclude that
\[\nnz(\BFA^{-1}\BFF\BFD) = \nnz(\BFA^{-1}\BFF) = \CalO(\tilde n \log(|\CalX_\tau^{\mathsf{SG}}|)^{d-1}) = \CalO(n (\log n)^{d-1}).\]

We now consider the top-left block of $\BFK^{-1}$.
Note that an entry of $\BFE^{-1}$ is nonzero if either the same entry of $\BFA^{-1}$ is nonzero or that of $\BFA^{-1} \BFF \BFD \BFF^\intercal\BFA^{-1}$ is nonzero.
Hence,
\begin{equation}\label{eq:nnz-E}
\nnz(\BFE)\leq \nnz(\BFA^{-1}) + \nnz(\BFA^{-1} \BFF \BFD \BFF^\intercal\BFA^{-1}).
\end{equation}

By Proposition~\ref{prop:nnz},
\begin{equation}\label{eq:nnz-A-inv}
\nnz(\BFA^{-1}) = \CalO(|\CalX_\tau^{\mathsf{SG}}|).
\end{equation}

On the other hand, the analysis that follows \eqref{eq:top-left-2-decomp} asserts that
$(\BFA^{-1} \BFF \BFD \BFF^\intercal\BFA^{-1})_{(\BFl',\BFi'),(\BFl'',\BFi'')} = 0$ if
$\BFc_{\BFl,\BFi} \notin \CalR_{\BFl',\BFi'} \cap\CalR_{\BFl'',\BFi''}$ for all $\BFc_{\BFl,\BFi}\in\CalA_{\tilde n}$.
Therefore,
\begin{align}
& \nnz(\BFA^{-1} \BFF \BFD \BFF^\intercal\BFA^{-1}) \nonumber \\
\leq {}&
|\{((\BFl',\BFi'),(\BFl'',\BFi'')): \BFc_{\BFl',\BFi'}, \BFc_{\BFl'',\BFi''}\in\CalX_\tau^{\mathsf{SG}}, \,\BFc_{\BFl,\BFi} \in  \CalR_{\BFl',\BFi'} \cap\CalR_{\BFl'',\BFi''}\mbox{ for some } \BFc_{\BFl,\BFi}\in\CalA_{\tilde n} \} | \nonumber    \\
\leq{}& \sum_{(\BFl,\BFi):\BFc_{\BFl,\BFi}\in\CalA_{\tilde n}}
|\{((\BFl',\BFi'),(\BFl'',\BFi'')): \BFc_{\BFl',\BFi'}, \BFc_{\BFl'',\BFi''}\in\CalX_\tau^{\mathsf{SG}}, \, \BFc_{\BFl,\BFi} \in  \CalR_{\BFl',\BFi'} \cap\CalR_{\BFl'',\BFi''} \} | . \label{eq:NNZ-top-left}
\end{align}

Fix an arbitrary $\BFc_{\BFl,\BFi}\in\CalA_{\tilde n}$.
For each $\BFl'\in\NatInt^d$ with $|\BFl'|\leq \tau+d-1$, there exists a unique $i^*(\BFl')\in \rho(\BFl')$ such that $\BFc_{\BFl,\BFi}\in \CalR_{\BFl',\BFi^*(\BFl)}$, because $\{\CalR_{\BFl',\BFi'}: \BFi'\in\rho(\BFl')\}$ forms a partition of the design space $(0,1)^d$.
Hence, for each $\BFl'\in\CalX_\tau^{\mathsf{SG}}$, the number of $\CalR_{\BFl',\BFi'}$'s that cover $\BFc_{\BFl,\BFi}$ is exactly 1.
This implies that
\begin{align*}
|\{(\BFl',\BFi'): \BFc_{\BFl',\BFi'}\in\CalX_\tau^{\mathsf{SG}}, \, \BFc_{\BFl,\BFi}\in\CalR_{\BFl',\BFi'}\}|
={}& |\{\BFl'\in\NatInt^d: |\BFl'|\leq \tau+d-1 \}| = \sum_{\ell=d}^{\tau+d-1}\binom{\ell-1}{ d-1} =  \CalO( \tau^{d}),
\end{align*}
where the last step follows from a derivation similar to  \eqref{eq:estimate-sum-phi-2}.
Thus,
\begin{align}
& |\{((\BFl',\BFi'),(\BFl'',\BFi'')): \BFc_{\BFl',\BFi'}, \BFc_{\BFl'',\BFi''}\in\CalX_\tau^{\mathsf{SG}}, \, \BFc_{\BFl,\BFi} \in  \CalR_{\BFl',\BFi'} \cap\CalR_{\BFl'',\BFi''} \} | \nonumber \\
={}& |\{(\BFl',\BFi'): \BFc_{\BFl',\BFi'}\in\CalX_\tau^{\mathsf{SG}}, \, \BFc_{\BFl,\BFi} \in  \CalR_{\BFl',\BFi'} \} | \times
|\{(\BFl'',\BFi''): \BFc_{\BFl'',\BFi''}\in\CalX_\tau^{\mathsf{SG}}, \, \BFc_{\BFl,\BFi} \in \CalR_{\BFl'',\BFi''} \} | \nonumber \\
={}& \CalO(\tau^{2d}). \label{eq:NNZ-top-left-2}
\end{align}
It follows from \eqref{eq:NNZ-top-left} and \eqref{eq:NNZ-top-left-2} that
\[
\nnz(\BFA^{-1} \BFF \BFD \BFF^\intercal\BFA^{-1}) = \CalO(\tilde n \tau^{2d}),
\]
which, in conjunction with \eqref{eq:nnz-E} and \eqref{eq:nnz-A-inv}, implies that
\[\nnz(\BFE) = \CalO(|\CalX_\tau^{\mathsf{SG}}| + \tilde n \tau^{2d})
= \CalO(n(\log n)^{2d}),\]
since $|\CalX_\tau^{\mathsf{SG}}| =\CalO(n)$, $\tilde n =\CalO(n)$, and $\tau = \CalO(\log (n))$.
Consequently, the density of $\BFK^{-1}$ is
\begin{align*}
\frac{1}{n^2}\left(\nnz(\BFE) + 2\times \nnz(\BFA^{-1}\BFF \BFD) + \nnz(\BFD)\right)
={}& \frac{1}{n^2} \left(\CalO(n(\log n)^{2d}) + \CalO(n(\log n)^{d-1}) + \CalO(n)\right) \\
={}& \CalO\bigl(n^{-1}(\log n)^{2d}\bigr). \Halmos
\end{align*}
\endproof

\end{document}